\documentclass[reprint,
superscriptaddress,
 amsmath,amssymb,
 notitlepage,
10pt,
]{revtex4-2}

\usepackage[utf8]{inputenc}
\usepackage[caption=false]{subfig}
\usepackage{graphicx}% Include figure files
\usepackage{dcolumn}% Align table columns on decimal point
\usepackage{bm}% bold math
\usepackage{xcolor}
\usepackage{verbatim}
\usepackage{bbold}
\usepackage[ruled,vlined]{algorithm2e}
\usepackage{bm}

\usepackage[colorlinks=True, citecolor=blue, urlcolor=blue]{hyperref}
\usepackage{booktabs}
\usepackage{natbib}

\providecommand{\braket}[1]{\mathinner{\langle{#1}\rangle}} %Solve conflicting braket definitions
\usepackage{tikz}
\usetikzlibrary{quantikz}
\DeclareMathOperator{\sinc}{sinc}
\SetKw{KwIn}{in}

\begin{document}
\title{Adiabatic spectroscopy and a variational quantum adiabatic algorithm}% 
\author{Benjamin~F.~Schiffer}
\email[Corresponding author: ]{Benjamin.Schiffer@mpq.mpg.de}
\affiliation{Max-Planck-Institut f\"ur Quantenoptik, Hans-Kopfermann-Str.~1, D-85748 Garching, Germany}%
\affiliation{Munich Center for Quantum Science and Technology (MCQST), Schellingstr.~4, D-80799 Munich, Germany}%
\author{Jordi~Tura}%
\affiliation{Instituut-Lorentz, Universiteit Leiden, P.O. Box 9506, 2300 RA Leiden, The Netherlands}%
\affiliation{Max-Planck-Institut f\"ur Quantenoptik, Hans-Kopfermann-Str.~1, D-85748 Garching, Germany}%
\affiliation{Munich Center for Quantum Science and Technology (MCQST), Schellingstr.~4, D-80799 Munich, Germany}%
\author{J.~Ignacio~Cirac}%
\affiliation{Max-Planck-Institut f\"ur Quantenoptik, Hans-Kopfermann-Str.~1, D-85748 Garching, Germany}%
\affiliation{Munich Center for Quantum Science and Technology (MCQST), Schellingstr.~4, D-80799 Munich, Germany}%
\date{\today}

\begin{abstract}
Preparing the ground state of a Hamiltonian is a problem of great significance in physics with deep implications in the field of combinatorial optimization. The adiabatic algorithm is known to return the ground state for sufficiently long preparation times which depend on the a priori unknown spectral gap. Our work relates in a twofold way. First, we propose a method to obtain information about the spectral profile of the adiabatic evolution. Second, we present the concept of a variational quantum adiabatic algorithm (VQAA) for optimized adiabatic paths. We aim at combining the strengths of the adiabatic and the variational approaches for fast and high-fidelity ground state preparation while keeping the number of measurements as low as possible. Our algorithms build upon ancilla protocols which we present that allow to directly evaluate the ground state overlap. We benchmark for a non-integrable spin-1/2 transverse and longitudinal Ising chain with $N=53$ sites using tensor network techniques. Using a black box, gradient-based approach, we report a reduction in the total evolution time for a given desired ground state fidelity by a factor of ten, which makes our method suitable for the limited decoherence time of noisy-intermediate scale quantum devices. 
\end{abstract}
\maketitle

\section{Introduction}
Remarkable scientific progress in recent years has led to the first noisy intermediate-scale quantum (NISQ)~\cite{preskill2018quantum} devices. Current NISQ devices are still limited by the number of qubits available, their gate fidelity and the maximum circuit depth. Much research effort is put into the investigation of algorithms for digital NISQ devices that could hold the promise of a (practical) quantum advantage~\cite{arute2019quantum, Zhongeabe8770}. Nevertheless, simulating large quantum many-body systems on these devices still remains challenging within the next years. Analog quantum simulators, however, are able to implement some quantum dynamics very efficiently. Furthermore, analog quantum simulators are especially well suited for probing universal features of quantum many-body systems. Several powerful concepts and experimental realizations of digital NISQ devices and analog quantum simulators have been put forward~\cite{wendin2017quantum, cirac2012goals, gross2017quantum}.  \\
One particularly important problem is preparing the ground state of quantum many-body systems, relevant for a wide range of physics applications and closely related to combinatorial optimization tasks. Quantum computers already save an exponential memory cost compared to a classical computer in the representation of the quantum state, making a quantum device the natural choice to compute desired quantum states. Moreover, ground states are also of great importance for optimization problems as the solutions to combinatorial problems can be naturally encoded into a classical Hamiltonian.\\
The ground state finding problem is known to be QMA-complete~\cite{kempe2006complexity}, which translates roughly to the analogue of NP-complete for a quantum computer. However, two different promising heuristic approaches have been established. First and most straightforward, there is the adiabatic approach. In order to prepare the ground state of a target Hamiltonian $H_T$ with a quantum adiabatic algorithm (QAA)~\cite{farhi2000quantum}, one takes an adiabatic path $H(s) = (1-s) H_0 + s H_T$. Starting from the ground state of a trivial Hamiltonian $H_0$ at $s(0)=0$, the parameter $s(t)$ is changed with time until the final value $s(T)=1$. Given a non-degenerate ground state along the path, the adiabatic algorithm is known to return the ground state for a sufficiently long preparation time $T$. However, $T$ is a function of the spectral energy gap $\Delta(s)$ between the ground state and the first excited state along the adiabatic path and not known a priori~\cite{messiah1962quantum, amin2009consistency, jansen2007bounds, elgart2012note}.
Due to the limited decoherence times of current NISQ devices and analog quantum simulators, the preparation time $T$ is of great relevance for the feasibility of the adiabatic approach. Different adiabatic paths can be constructed and a linear time schedule is generally not optimal in the sense of a minimal $T$. This is especially relevant when the gap $\Delta(s)$ becomes very small as it occurs in the presence of a quantum phase transition.\\
Optimal adiabatic paths have been the subject of intensive research efforts both in the framework of shortcuts to adiabaticity as well as optimal control theory \cite{guery2019shortcuts, albash2018adiabatic}. For the problem of an unstructured search, Grover-type speed-ups have been shown where the full spectral information about the problem is available~\cite{roland2002quantum, rezakhani2009quantum}.\\
Another approach to prepare the ground state is the quantum approximate optimization algorithm (QAOA)~\cite{farhi2014quantum} which seeks to overcome the limitations of the QAA. It generalizes the QAA by splitting the total time into chunks, $\{T_i\}$, where one alternates between $H_0$ and $H_T$, and takes $\{T_i\}$ as the variational parameters. The QAOA includes the QAA in the sense that there exist $T_i=iT/L$, where $i=\{1,\ldots,L\}$ and $L$ sufficiently large, corresponds to the Trotter evolution of the QAA. However, it is expected that the QAOA can provide a speedup by choosing the parameters $\{T_i\}$ larger than in trotterized QAA. \\
In fact, the QAOA belongs to a wider class of variational quantum algorithms (VQA) in the spirit of the variational principle which is widely used in physics. While the quantum computer is used to prepare the states and perform the measurements, the optimization of the parameters is carried out classically. In practice, the performance of VQA can be curbed since the number of measurements necessary to estimate an objective function may scale unfavorably, and due to the presence of plateaus in the energy landscape, including noise-induced barren plateaus ~\cite{Wang2020, franca2020limitations, cerezo2020variational, bittel2021training}. \\
In this work, we propose the concept of a variational quantum adiabatic algorithm (VQAA) to find optimal adiabatic paths for high-fidelity ground state preparation, thereby combining the strengths of adiabatic state preparation and VQA. Akin to the QAOA, the times $\{T_i\}$ are treated as the variational parameters. However, the evolution in the different chunks is performed adiabatically, similarly to the QAA. Here, the VQAA differs from the fixed Hamiltonians found in the QAOA. The VQAA allows for a significant acceleration compared to the QAA with a linear adiabatic path, yet requires fewer parameters and measurements than the QAOA.\\
We discuss different approaches to find such a parametrized optimal adiabatic path. The approaches are suited to different resource requirements, some making use of ancilla protocols to estimate the ground state overlap. These protocols rely on
controlled unitary evolution which leads to many benefits in the quantum computing setting. Besides being central to the canonical quantum phase estimation algorithm~\cite{nielsen2002quantum}, it also allows to access overlaps between initial and final states for a given unitary transformation by looking at the ancilla only and enables spectral projections on the state of the system if one enables postselection on the ancilla~\cite{chen2020quantum}. \\
The ancilla techniques presented here may prove to be useful tools by themselves for other variational algorithms. This is especially true for our protocol with two ancilla qubits, as this constitutes a case of non-trivial distributed quantum computing. Interconnecting multiple quantum devices using a coherent link is a promising path forward for the field of quantum computing~\cite{magnard2020microwave}. \\
We benchmark the different algorithms with a quantum Hamiltonian and up to $N=100$ qubits, which is what is expected for the new generation of NISQ devices or analog quantum simulators. We take a Hamiltonian that is non-trivial (non-integrable) but for which we can simulate the action of a quantum computer classically using tensor network techniques. 
In the case of a small gap in the adiabatic path, in the presence of a phase transition, we report significant reductions in the required evolution time to reach a given desired ground state fidelity. For a chain of $N=53$ qubits and a target fidelity of 90\%, the VQAA is able to reduce the total evolution time by a factor of 10 compared to a linear adiabatic path.\\
Having mentioned before that the spectral gap $\Delta(s)$ is a priori unknown, obtaining knowledge about the spectral gap $\Delta(s)$ can be as hard a problem as finding the ground state itself. However, due to the interconnection between the spectral gap and the performance of adiabatic state preparation, we are able to propose a form of adiabatic spectroscopy to find spectral gap properties.
Performing adiabatic sweeps on the system and being equipped with either the ancilla protocols for ground state estimation or backward-time evolution, a measure for both the position and the smallness of the spectral gap can be obtained.
A related scheme using backward-time evolution, albeit initializing the evolution in a superposition, has been suggested in \cite{matsuzaki2021direct}.
Knowledge about the spectral gap may be applied to gain important insights about quantum many-body physics, obtain the phase diagram of a physical system or formulate the optimal path for adiabatic state preparation. \\
The structure of this paper is as follows. In Section~\ref{sec:results} we outline our main results. Then, we discuss our approach to adiabatic spectroscopy in Sec.~\ref{sec:AdSpec}, and in Sec.~\ref{sec:protocols}, we present two protocols for estimating the ground state at a given point along the adiabatic path using one or two ancillas. In Sec.~\ref{sec:vqaa}, we introduce the general concept of a variational quantum adiabatic algorithm for ground state preparation and discuss different specific algorithms with different resource requirements (Sec.~\ref{sec:algos}). After that, in Sec.~\ref{sec:benchmarkin}, the model for benchmarking our algorithms is described and we present results. Finally, we comment on the number of measurements necessary in our approach~(Sec.~\ref{sec:mmt}) and the impact of noise on our algorithms~(Sec.~\ref{sec:noise}). In the Appendices, we give the necessary background to make the paper self-contained: a theoretical description of the adiabatic algorithm and QAOA as well as Bayesian inference for Beta-Bernoulli models which are relevant for performing hypothesis testing.
\section{Main results}\label{sec:results}
Our main results include \emph{(1)} protocols for eigenstate closeness estimation, \emph{(2)} a proposal for adiabatic spectroscopy, as well as a \emph{(3)} concept for variational quantum adiabatic algorithms including a black box gradient-based method.\\
\emph{(1)} Provided the possibility to implement controlled unitary evolution on a quantum state, where the dynamics are controlled by a single ancilla qubit, information about the closeness to the next eigenstate can be extracted from the ancilla. We show that for a quantum state $\ket{\psi} = \sum_j \psi_j \ket{\phi_j}$ and Hamiltonian $H = \sum_j E_j \ket{\phi_j}\bra{\phi_j}$, we need to obtain
$\alpha := \sum_j |\psi_j|^2 \exp(-i E_j \tau)$ by measuring the ancilla. Then, for suitable $\tau$, we can deduce that the state $\ket{\psi}$ is an eigenstate of $H$ only if $|\alpha|=1$. By a self-consistent argument and suitable construction, the closest eigenstate may be identified with the ground state.\\
\emph{(2)} The profile of the spectral gap $\Delta(s)$ between the ground state and first excited state energy of an interpolating Hamiltonian $H(s)$ is closely connected to the evolution time $T$ required for adiabatic state preparation.
\begin{figure}[htbp]
    \includegraphics[width=.9\linewidth]{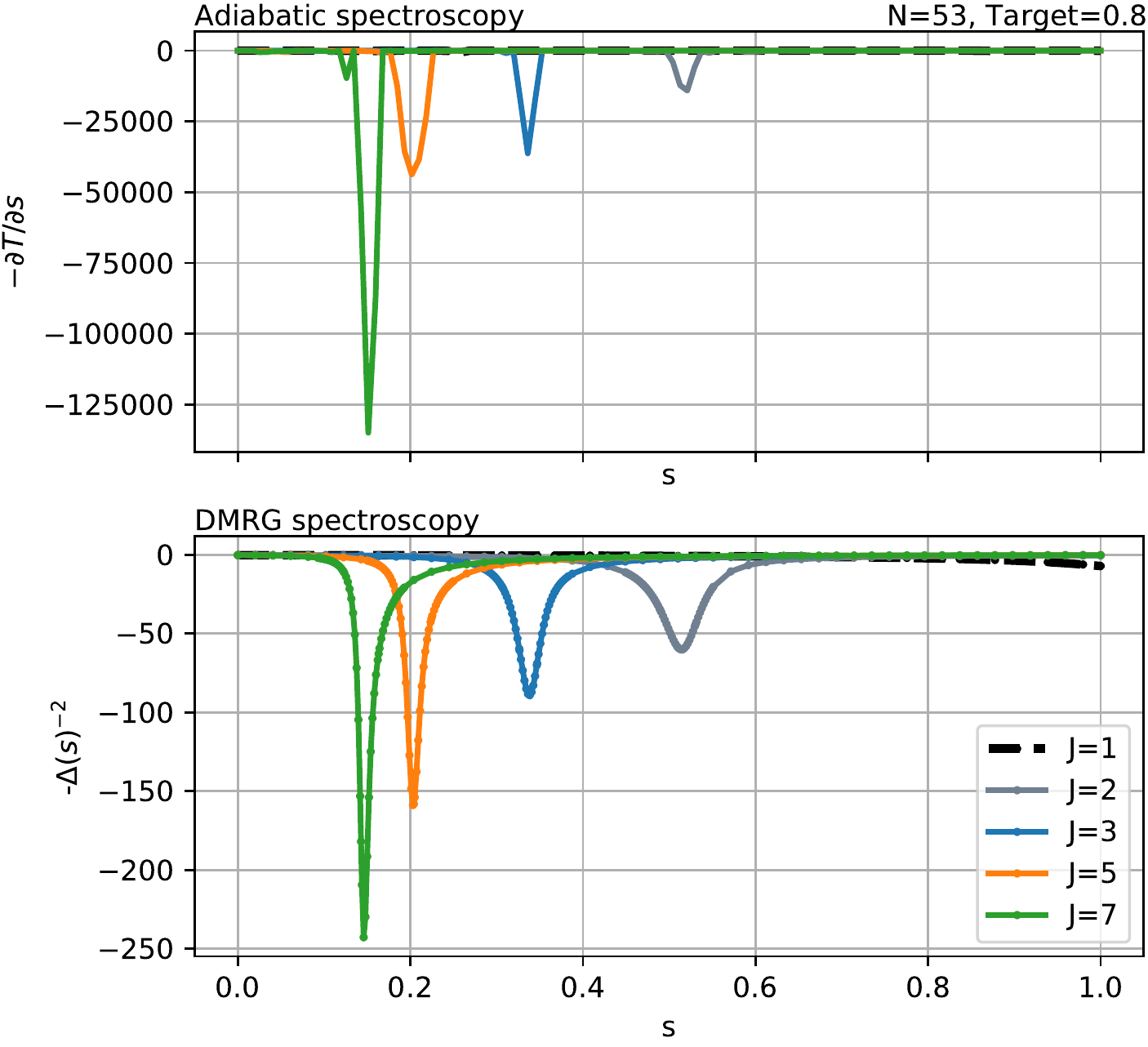}
    \caption{Adiabatic spectroscopy for a chain of $N=53$ qubits and a target ground state fidelity of 0.7. For the Hamiltonian $H(J,g,h)=\sum_i (J\sigma_i^z \sigma_{i+1}^z + h\sigma_i^x + g\sigma_i^z)$, the adiabatic path is a linear interpolation from $H(0,1,0)$ to $H(J,1,1)$. The minima of the different curves correspond to position and smallness of the respective spectral gap. In the lower plot, $-\Delta(s)^{-2}$ is shown as a comparison, obtained with DMRG. The match with the Landau-Zener scaling is not exact as higher order corrections from adiabatic perturbation theory are playing a role.}
    \label{fig:AdSpec}
\end{figure}
We analyze the evolution time $T(s)$ necessary to prepare the ground state of $H(s)$ with a given target fidelity. The ground state fidelity is computed either by time-evolving both forward and backward and measuring the fidelity with the initial product state, or by making use of the ancilla protocol which is generally expected to give superior results. After obtaining the data points for $T(s)$, we interpolate the curve and compute the derivative.
Corresponding to physical intuition, the curve of the evolution time features a strong increase when a small gap is crossed. The derivative $\partial T/\partial s$ can then provide the position and also a measure for the smallness of the spectral gap (Fig.~\ref{fig:AdSpec}).\\
\emph{(3)} In our work, we present the concept of the VQAA and give specific algorithms which attempt to improve over QAA. The main idea is to optimize the adiabatic path $s(t)$ by performing a moderate number of measurements. 
\begin{figure}[htbp]
    \includegraphics[width=.9\linewidth]{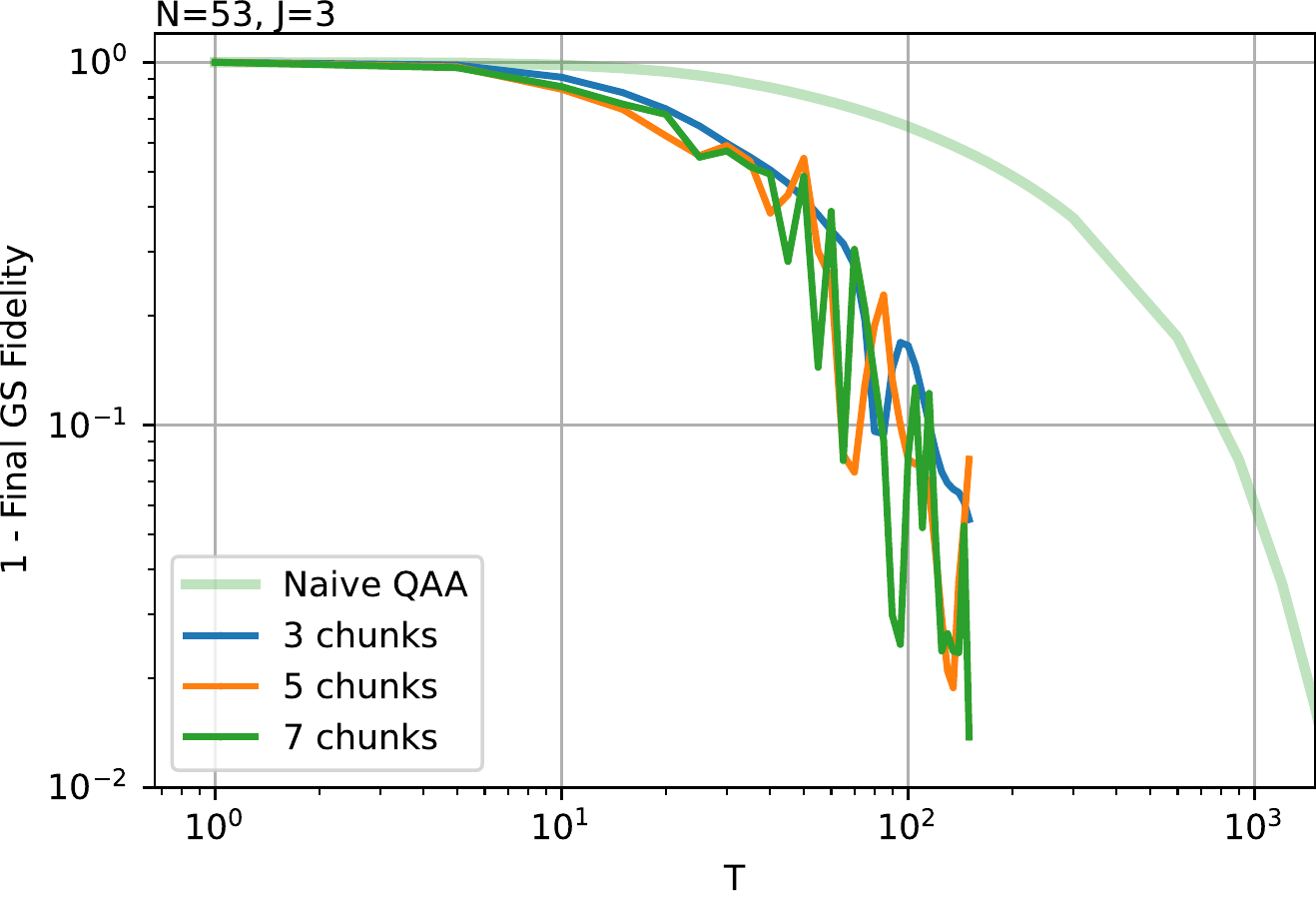}
    \caption{Black box VQAA results for 53 qubits and the ZZXZ model for the case of a crossed phase transition ($J=3$). Results for 3, 5 and 7 chunks, respectively, are shown. More chunks do not improve the results as for one avoided level crossing an optimized path can be approximated already with very few chunks. The quantum-classical feedback loop converges to optimized adiabatic paths that prepare the ground state with $>90\%$ fidelity for $T\lesssim 100$. In order to achieving this ground state fidelity with (non-optimized) naive QAA would have required an evolution time about 10 times larger.}
    \label{fig:BBBPT}
\end{figure}
In our setup, $s(t)$ depends on a set of parameters, which are chosen through optimization procedures. The key ingredient is to use the overlap with the ground state as the figure of merit. The algorithms presented in this work make use of different protocols to estimate this overlap, and are also distinguished in the way the optimization is performed: either by intending to remain in the ground state at intermediate steps in the adiabatic path,
or by optimizing for a maximum ground state overlap at the end of the state preparation (like in QAOA). The main feature of the method presented in this paper is that it requires fewer measurements. Our approach is well suited to be used in NISQ devices or analog quantum simulators by reducing the required preparation time and thus avoiding decoherence. 
Without knowing the adiabatic spectrum beforehand, the optimal adiabatic paths yield the desired ground state with high fidelity at only a fraction of the total evolution time of a non-optimized adiabatic algorithm (Fig.~\ref{fig:BBBPT}). 
\section{Adiabatic spectroscopy}\label{sec:AdSpec}
We seek to gain insights about the adiabatic spectrum of the Hamiltonian
\begin{align}
    H(s) = (1-s) H_0 + s H_T.
\end{align}
The Hamiltonian $H(s)$ describes the adiabatic evolution from the ground state of a trivial Hamiltonian $H_0$ at $s=0$ to the ground state of $H_T$ at $s=1$ where $s=t/T$ is parametrized time. The probability that the system transitions out of the ground state in the course of the evolution is intimately connected with the size of the spectral gap $\Delta(s) = E_1(s) - E_0(s)$ between the ground state and the first excited state energy.
For each $s_i$ in a set of data points $\{s_i\}$, $i\in \{1,\dots,r\}$ and $r$ being the resolution of the spectroscopy, an efficient root finding algorithm is used (e.g.~using a bisection algorithm on the time variable $T_i$) to find the evolution time $T_i$ which reaches a given target ground state overlap $O_T$ (see Suppl.~for a simple algorithmic description).\\
If a small gap is crossed in the adiabatic path, e.g.~at $s^*$, this will correspond to a large increase of the $\{T_i\}$ around and after this value $s^*$. Hence, if we observe such a rise in the required evolution times $\{T_i\}$, we conclude that a local minimum in the spectral gap $\Delta(s^*)$ must be present. In this manner, the position of a small spectral gap can be obtained.
Moreover, the smallness of $\Delta(s^*)$ is related to the steepness of the increase in the $\{T_i\}$ around $s^*$. For the Landau-Zener model, we establish the relation $\partial T(s)/\partial s \sim  1/\Delta(s)^2$ around the minimal gap (App.~\ref{App:LZ}). Since the Landau-Zener approach is a toy model to qualitatively study the property of the spectral gap in adiabatic algorithms, we presume that a similar scaling could persist in a more general sense.\\
In order to obtain a measure for the ground state overlap for given $s_i$, we propose two different approaches. The first approach is not making use of the ancilla protocol and is thus very simple to implement. The initial ground state $\ket{\psi_0}$ at $s=0$ is adiabatically time-evolved forward with evolution time $\widetilde T_j$, implemented by the unitary operator $U_{0\rightarrow s_i}(\widetilde T_j)$. Here, the $\widetilde T_j$ denote the probing values in the search. We seek to determine the forward time so that we obtain the evolution time $T_i$ which succeeds in reaching the given overlap $O_T$. Next, the interactions are reversed in a backward-time evolution from $s=s_i$ to $s=0$ implemented by $U_{0\leftarrow s_i}(\widetilde T^B_j)$. The backward time $\widetilde T^B_j \gg \widetilde T_j$ is chosen to be larger than the forward time evolution. This allows for a trivial measurement of the ground state overlap at $s=0$
\begin{align}
    \widetilde O_j &= \left|\braket{\psi_0|U_{0\leftarrow s_i}(\widetilde T^B_j) U_{0\rightarrow s_i}(\widetilde T_j)|\psi_0}\right|
\end{align}
as the state $\ket{\psi_0}$ is a product state. Once a $\widetilde T_j$ is found for which $\widetilde O_j \approx O_T$, we set $T_i := \widetilde T_j$ and proceed with the next data point $s_{i+1}$. Note that it is also sensible to use a large constant evolution time for the backward path.\\
We extend this approach by noting that if a small spectral gap $\Delta(s^*)$ has been found and we continue to probe for those $s_i$ for which $s_i>s^*$, it is economical to make use of the spectral information already obtained. In the spirit of the VQAA, the adiabatic schedule used in the spectroscopy should be altered so that the evolution around $s^*$ is performed very slowly. In this manner, multiple gaps in the adiabatic spectrum could be investigated. This ancilla-free approach provides the least stringent requirements on the experimental setup in a NISQ framework.\\
However, with some additional effort, there is the possibility with to directly estimate the ground state overlap without going back to the initial state at $s=0$. This second approach leverages on the ability to obtain a measure for the eigenstate closeness through an ancilla protocol. For every data point in $\{s_i\}$, we obtain the ground state overlap for different $\widetilde T_j$ directly:
\begin{align}
    \widetilde O'_j &= \left|\braket{\text{GS}_{s_i}|U_{0\rightarrow s_i}(\widetilde T_j)|\psi_0}\right|,
\end{align}
with $|\text{GS}_{s_i}\rangle$ the ground state at $s_i$. Now just like in the first approach, we search for the $\widetilde T_j$ such that $\widetilde O'_j \approx O_T$ and set $T_i := \widetilde T_j$.\\
Hence, this form of adiabatic spectroscopy can provide a tool in order to experimentally gain knowledge about the adiabatic spectrum. We simulate this technique classically with $N=53$ qubits for different spectral profiles and present the results in (Fig.~\ref{fig:AdSpec}) and Sec.~\ref{ssec:ResAdSpec}.
\section{Protocol for ground state closeness estimation}\label{sec:protocols}
A very important ingredient of our algorithms is the ability to estimate the overlap with a non-degenerate ground state. In this section we introduce two suitable protocols using ancillas and comment briefly on their benefits. The ancilla protocols only require the ability to implement unitary evolution controlled on a single ancilla and to perform measurements on the ancilla. Controlled unitary evolution is a valuable ingredient for quantum computing, e.g.~in the well-known quantum phase estimation algorithm~\cite{nielsen2002quantum}. Using ancilla measurements for ground state preparation has been investigated previously and can be used to construct spectral projection operators similar to the quantum Zeno effect if one allows for postselection on the ancilla~\cite{chen2020quantum}. In our work, we utilize the ancilla in order to extract information about the eigenstate closeness which can then be used in the different applications of the VQAA in a highly versatile manner.
\subsection{Single-ancilla protocol} \label{ssec:singleancillaprotocol}
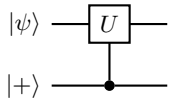
\begin{figure}[htbp]
    \centering
    \begin{tikzpicture}
        \node[scale=1] {
            \begin{quantikz}
                \lstick{\ket{\psi}} & \gate{U} &  \qw \\
                \lstick{\ket{+}}  &  \ctrl{-1} &  \qw %\meter{}  
            \end{quantikz}
        };
    \end{tikzpicture}
    \caption{Quantum circuit for unitary dynamics controlled by a single ancilla qubit in the $\ket{+}$ state.}
    \label{fig:singleancilla}
\end{figure}

\begin{figure}[htbp]
    \centering
    \includegraphics[width=.75\linewidth]{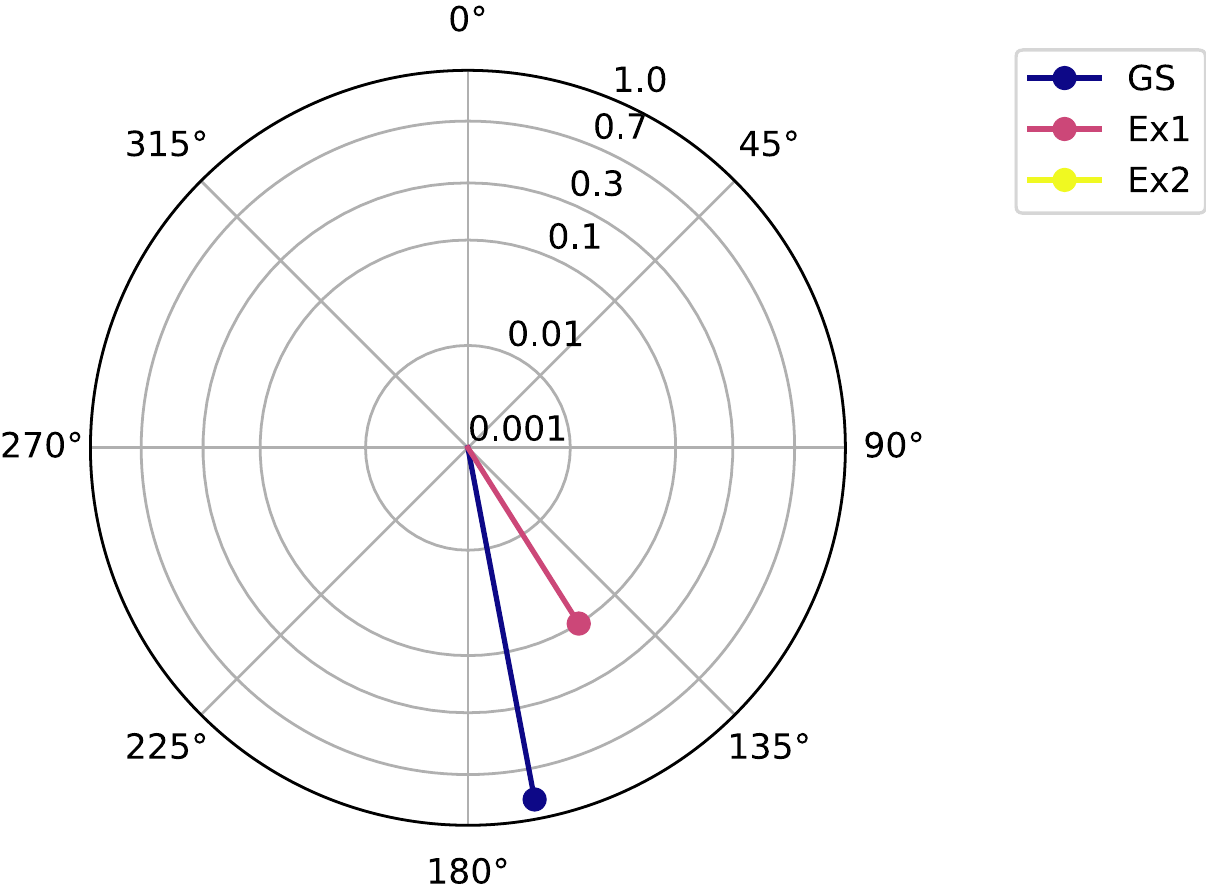}
    \caption{Exemplary eigenstate clock featuring the complex summands of $\alpha$ for $\tau$=1. Here, the term corresponding to the ground state $|\psi_0|^2 \exp(-i E_0 \tau)$ and the first excited state term $|\psi_1|^2 \exp(-i E_1 \tau)$ lie quite closely together in phase which leads to problematic constructive interference. However, as the pointers rotate with their respective eigenenergy, they will be well-separated for suitable larger $\tau$. Note that in this instance, the summands corresponding to higher excited eigenstates are taken very small and not visible.}
    \label{fig:clock}
\end{figure}

We first present a protocol for eigenstate closeness estimation using a single ancilla in the $\ket{+}$~state \footnote{The state $\ket{+}$ is an eigenvector of the Pauli matrix $\sigma^x$ and defined as $\ket{+}=(\ket{0}+\ket{1})/\sqrt{2}$ where $\ket{0}$ and $\ket{1}$ are the computational basis states}. The quantum circuit shown in (Fig.~\ref{fig:singleancilla}) can be used to
compare the overlap of an initial quantum state $\ket{\psi}$ with the state after the unitary evolution $\ket{\psi_\text{evo}}$
\begin{align}
    \braket{\psi | \psi_\text{evo}} = \braket{\sigma_x + i \sigma_y}_\text{ancilla}.
\end{align}
For a fixed Hamiltonian $H = \sum_j E_j \ket{\phi_j}\bra{\phi_j}$ with the unitary 
\begin{align}
    U \ket{\phi_j} = e^{-iH\tau}\ket{\phi_j} = e^{-iE_j\tau}\ket{\phi_j},
\end{align}
we write the normalized state $\ket{\psi} = \sum_j \psi_j \ket{\phi_j}$ in the eigenbasis of $H$. Then, we analyze the quantum circuit for this choice of $U$. We obtain for the density matrix of the ancilla qubit after the controlled unitary evolution (App.~\ref{App:DM})
\begin{align}
    \rho_{\text{a}} = \sum_j \frac{|\psi_j|^2}{2} \begin{pmatrix}
    1 & e^{i E_j \tau}\\
    e^{-i E_j \tau} & 1\end{pmatrix}.
    \label{eqn:ancilladm}
\end{align}
For a quantum state $\ket{\psi}$ that is an eigenstate, the rank of $\rho_\text{a}$ will be 1. Due to the specific structure of $\rho_\text{a}$, only one off-diagonal matrix element is needed in order to determine the rank of $\rho_\text{a}$. We note with (Eqn.~\ref{eqn:overlap}) that 
\begin{align}
    \braket{\psi | \psi_\text{evo}} 
    = \sum_j |\psi_j|^2 e^{-i E_j \tau}  =: \alpha.
    \label{eqn:alpha}
\end{align}
For suitable $\tau$, $\ket{\psi_\text{evo}}$ is an eigenstate only if $|\alpha|=1$. The time $\tau$ needs to be chosen so that the complex summands of $\alpha$ with non-vanishing amplitude do not have approximately equal phases. In this unlikely case of matching phases, we would see constructive interference so that $|\alpha|=1$ could be true even if $\ket{\psi_\text{evo}}$ is not an eigenstate. Visualizing the summands of $\alpha$ on a complex plane (Fig.~\ref{fig:clock}), this becomes rather intuitive. The choice for $\tau$ is related to the spectrum of $H$. For the sake of this argument we assume that the overlap with the ground state and first excited state are the only other non-vanishing overlaps. Then, an arbitrary $\tau$ would correspond to choosing an $l\in \mathbb{Z}$ in $\tau = \pi l/\Delta$ at random~(App.~\ref{App:tau}). For $l\gg 1$, the probability of choosing an odd value of $l$ is approximately 1/2. Therefore, by testing several random values of $\tau \in O(\Delta^{-1})$ it is possible to deduce information about the system whether it is in a mixed state or an eigenstate with high confidence (cf.~the Chernoff-Hoeffding bound in Sec.~\ref{sec:mmt}). Through a self-consistent argument we can conclude that the main contribution to $\alpha$ comes from the ground state,
provided we remain nearly adiabatic throughout the path. We can bound the maximal error in $|\alpha (\tau)|^2$ by computing the average for up to large values of an uniformly distributed $\tau$ as
\begin{align}
    E^2 := \lim_{K\rightarrow\infty} \mathbb{E}_{\tau\: \sim \text{ unif. dist. in } [0, K]} (|\alpha(\tau)|^2)= \sum_i |\psi_i|^4.
\end{align}
Then, we obtain
\begin{align}
    |\psi_0|^2 \geq \frac{1}{2} + \frac{1}{2}\sqrt{2E^2-1}
\end{align}
for the ground state population $|\psi_0|^2$.
In practice, terms in (Eqn.~\ref{eqn:alpha}) corresponding to higher eigenenergies will be rather small and destructively interfere with each other. Therefore, this bound is not tight and much smaller errors are expected in an experiment (cf. App.~\ref{App:Bound}).\\
Our protocol takes inspiration from semi-classical approaches to the quantum phase estimation algorithm~\cite{santagati2018witnessing, obrien2020error}. However, these algorithms also seek to determine to energy. As we argued above, the ground state overlap is better suited as the cost function for ground state finding algorithms. Therefore we devised this protocol that is oblivious to the energy value at any given point while being very simple to implement.
\subsection{Entangled-ancillas protocol} \label{sec:entangled-ancilla}
Building upon the single-ancilla protocol, we introduce a protocol using two identical quantum systems with one ancilla each. The protocol is motivated by the intuition that the single-ancilla protocol gathers information about the complex phase of $\braket{\psi | \psi_\text{evo}}$ which is without practical use to us. Instead we would like to estimate $|\alpha|$ directly, which this protocol achieves.\\
We require the possibility to conduct an entangling measurement (i.e.~a Bell measurement) between the two ancillas of the two systems. Such an entangling measurement could be performed with a microwave quantum link between two superconducting circuits~\cite{magnard2020microwave}. This protocol constitutes an instance of distributed quantum computing for such a quantum network. Moreover, it is well-suited to existing NISQ devices lacking an all-to-all connectivity where the adiabatic evolutions could be implemented on separate but connected sub-graphs~\cite{boothby2020next}. \\
Our goal is to determine the purity of the ancilla. In general, there is the relation $\rho$ pure $\Leftrightarrow \text{Tr}[\rho^2] = \lambda_1^2 +\lambda_2^2$ for density matrices, with $\lambda_1$ and $\lambda_2$ the eigenvalues of $\rho$. We write the density matrix and its square as
\begin{align}
    \rho_\text{a} = \begin{pmatrix}
    a & b\\ c & d
    \end{pmatrix} \quad \text{and} \quad
    \rho_\text{a}^2 = \begin{pmatrix}
    a^2+bc & \cdot\\ \cdot & bc+d^2
    \end{pmatrix}
\end{align}
where matrix elements irrelevant for our protocol are denoted with a dot. 
With the Bell state $\ket{\Phi^-} = \left( \ket{00}-\ket{11}\right)/\sqrt{2}$ we construct a Bell measurement.
The diagonal matrix elements of $\rho_\text{a}^2$ may then be attained by considering a composite system where the second system has controlled negative time evolution (implementable by changing the sign of $H$). Then, the density matrix of the ancilla of the second system effectively corresponds to the transpose of the density matrix of the first ancilla $\rho_\text{a2}=\rho_\text{a1}^T$. The composite system gives 
\begin{align}
    \rho_\text{a1} \otimes \rho_\text{a2} =  \begin{pmatrix}
    a^2 & \cdot & \cdot & bc\\
    \cdot & \cdot & \cdot & \cdot\\
    \cdot & \cdot & \cdot & \cdot\\
    bc & \cdot & \cdot & d^2
    \end{pmatrix} = 
    \frac{1}{4}\begin{pmatrix}
    1 & \cdot & \cdot & |\alpha|^2\\
    \cdot & \cdot & \cdot & \cdot\\
    \cdot & \cdot & \cdot & \cdot\\
    |\alpha|^2 & \cdot & \cdot & 1
    \end{pmatrix}.
\end{align}
If $\ket{\psi}$ is in an eigenstate, the density matrix of the ancilla $\rho_\text{a}$ is pure and the expectation of the $\ket{\Phi^-}$ measurement is 
\begin{align}
    \langle\rho_\text{a1} \otimes \rho_\text{a2}\rangle_{\ket{\Phi^-}} = \frac{1}{4} (1-|\alpha|^2) = 0,
\end{align}
allowing for very low variance measurements when $\ket{\psi}$ is in the vicinity of the ground state. This protocol is especially well-suited for hypothesis testing (cf.~Suppl.).
\section{Variational quantum adiabatic algorithms} \label{sec:vqaa}
Preparing the ground state of a Hamiltonian is a problem of great significance in physics with deep implications in the field of combinatorial optimization. While adiabatic state preparation is known to return the ground state for sufficiently long preparation times only, variational quantum algorithms require a very large number of measurements in the training phase.  We present a toolbox for variational quantum adiabatic algorithms (VQAA). Our objective is to prepare the ground state of a problem Hamiltonian $H_T$ with high fidelity while keeping the number of measurements in the process as low as possible. We aim at finding an optimized profile for the adiabatic evolution $H(s)$ from the ground state of $H_0$ to the ground state of $H_T$. For reasons of completeness, we refer to the Appendix (see~Suppl.) for a treatment of the adiabatic algorithm and its implementation. \\
In order to find an optimized adiabatic evolution, we choose a positive resolution $L\in \mathbb{N}$ for this velocity profile and split up the adiabatic path into $L$ chunks. 
\begin{figure}[htbp]
    \centering
    \includegraphics[width=.9\linewidth]{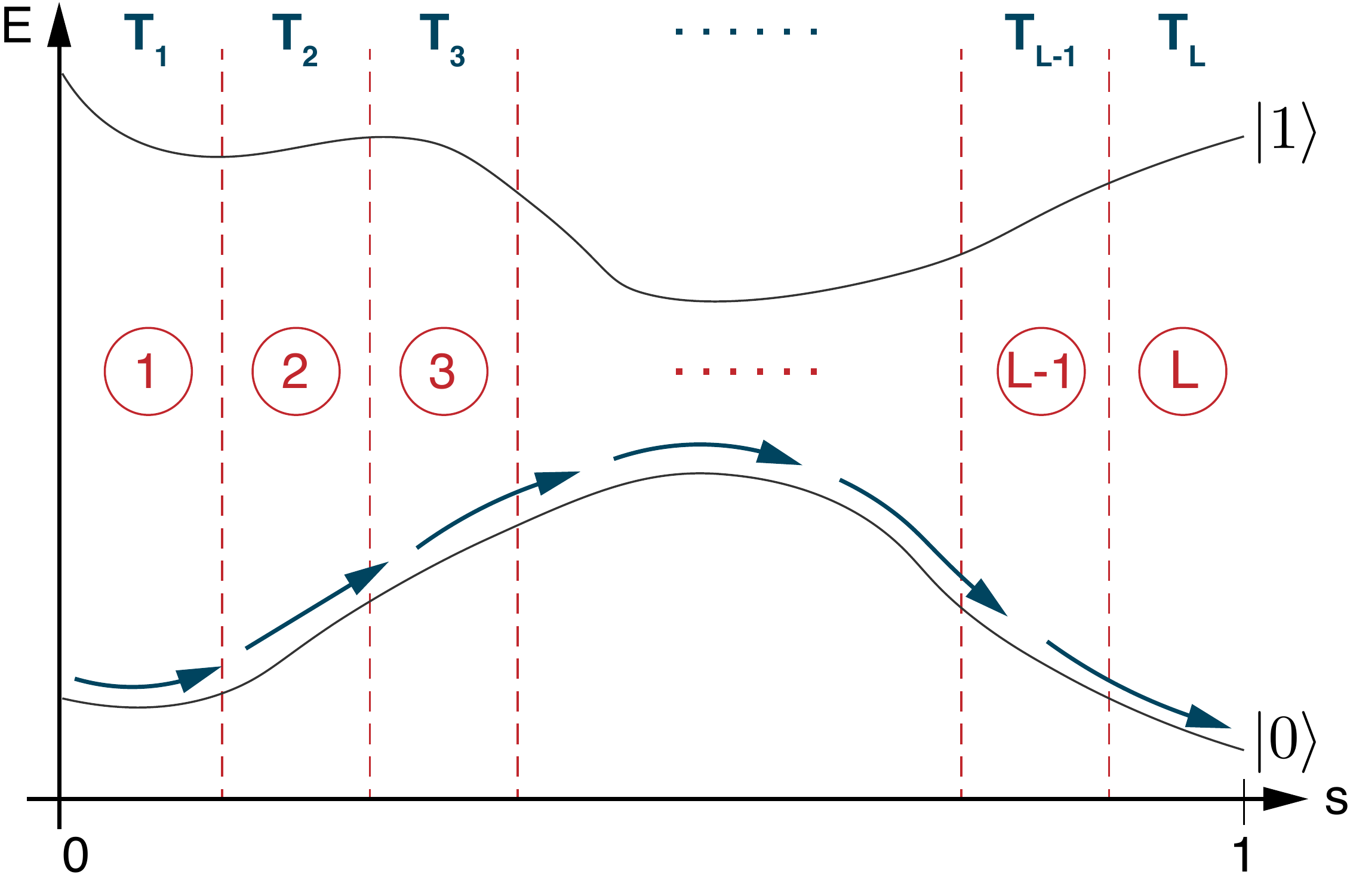}
    \caption{Illustration of a VQAA by splitting the adiabatic path into $L$ chunks. The ground state and higher excited state energies of $H(s)$ are shown here, being separated by a finite spectral gap. In an adiabatic algorithm, the ground state is prepared by following a path from a trivial Hamiltonian ground state at $s=0$ to the target Hamiltonian at $s=1$.}
    \label{fig:vqaa}
\end{figure}
Then, for every chunk $i\in\{1,\ldots, L\}$ an optimal adiabatic evolution time $T_i$ needs to be determined~(Fig.~\ref{fig:vqaa}).\\
In every optimization task, keeping the number of parameters to be optimized as low as possible is paramount. This is because every new parameter yields an additional cost in the number of repetitions necessary to make all the estimations which are required in the optimization of that parameter. In a VQAA, two different options are possible. One could distribute the total time budget $T$ for the evolution from $s=0$ to $s=1$ onto the $L$ chunks (e.g.~evenly spaced) and optimize the chunk lengths. Or alternatively, one could distribute the chunks in a given way and optimize the $T_i$. For both options, we present suitable specific algorithms with different resource requirements. An essential ingredient for the algorithms is the ability to obtain information about the closeness of the quantum state at a given point in the adiabatic path with the ground state.
\section{Presentation of the algorithms}\label{sec:algos}
The concept for the VQAA allows flexibility in the question of whether the chunk lengths or the chunk evolution times $T_i$ are the parameters to be optimized. Also, different specific classical algorithms can be used for the optimization process. Here, we present three different algorithms for finding the chunk lengths for fixed total evolution time and one algorithm for finding the chunk evolution times for flexible total $T$. These algorithms have different resource requirements and can make use of different ground state closeness protocols. The results stated in the abstract have been obtained using the gradient-based black box optimization for fixed total time (Sec.~\ref{sec:BB}).
\subsection{Fixed total time optimization}
Our proposal for the VQAA allows to set a maximum total time for the adiabatic evolution, which is well-suited for the limitations of current quantum devices. Here, the total adiabatic evolution time $T$ is allocated evenly between the chunks of the adiabatic path, so that
\begin{align}
    T_i=T/L \quad \forall i\in\{1, \ldots, L\}.
\end{align}
The chunk lengths become the variational parameters to be optimized, effectively controlling the density of adiabatic steps.
\subsubsection{Ancilla-free optimization for fixed total time} \label{sssec:ancillafreeopt}
In order to optimize $\left|\braket{\psi(s=1)|\text{GS}}\right|$, we try to keep the loss in fidelity in the ground state overlap along the adiabatic path as small as possible. Chunk lengths are initialized with equal lengths so that the end positions of each chunk are at 
\begin{align}
    s_i=i/L \quad \forall i\in\{1, \ldots, L\},
\end{align}
reproducing what we call naive QAA. The chunk lengths are $\bar s_i = s_i - s_{i-1}$, with $s_0=0$. Then, an adiabatic evolution is performed from the initial trivial product state $\ket{\psi_0}=\ket{\psi(s=0)}=\ket{-}^{\otimes N}$ up to the end points of each chunk. This adiabatic evolution is then time-reversed (at the same speed) by changing the sign in the unitaries. The total forward and backward-time evolution between $s=0$ and $s=s_i$ are described by $W_{0\rightarrow s_i}$ and $W_{0\leftarrow s_i}$, respectively, with the backward evolution being slower than the forward evolution. By going back to the initial product state $\ket{-}^{\otimes N}$, the implementation of this protocol is rather simple and allows for low variance ground state overlap measurements without ancillas. We denote the individual adiabatic evolution operators for chunk $j$ from $s_{j-1}$ to $s_j$ in time $T_j$ with the unitaries $V_{s_{j-1}, s_j}(T_j)$, so that we write
\begin{align}
    W_{0\rightarrow s_i} = \prod_{j=1}^i V_{s_{j-1}, s_j}(T_j),\quad  W_{0\leftarrow s_i} = \prod_{j=i}^1 V_{s_{j}, s_{j-1}}(T^B_j).
\end{align}
Now, we compute
\begin{align}
    O_i &= \left|\braket{\psi_0|W_{0\leftarrow s_i} W_{0\rightarrow s_i}|\psi_0}\right|
\end{align}
for all $i\in\{1,\ldots, L\}$ with $O_0=1$. The consecutive ratios of the overlap are
\begin{align}
    R_i = \frac{O_i}{O_{i-1}} \quad \forall i\in\{i,\ldots, L\}.
\end{align}
These $R_i$ correspond to the drop in ground state overlap with each next chunk along the adiabatic path. In order to find chunk lengths which correspond to a smooth decrease of the ground state overlaps, chunks $i$ where the drop in ground state overlap is larger than the average
\begin{align}
    R_i > \frac{1}{L}\sum_{i=1}^L R_i
\end{align}
are made smaller and vice-versa (where $R_j$ is below average, the $j$th chunk is made larger). The sum of the chunk lengths is kept normalized to one ($\sum_{i=1}^L \bar s_i = 1$). Then, new values $O_i$ are computed and the procedure repeats until convergence.\\
Clearly, what is optimized here is not exactly the instantaneous ground state overlap because the reversed adiabatic evolution will accumulate extra phases distorting the results slightly. Nevertheless, this method can be a useful compromise between a protocol which is very simple to execute and can still yield improved results of an optimized adiabatic routine~(Sec.~\ref{ssec:resultsfixed}).
\subsubsection{Forward only evolution for fixed total time} \label{ssec:forwardonly}
Improving over the ancilla-free algorithm, this algorithm makes use of the ancilla-based ground state closeness protocol  (Sec.~\ref{ssec:singleancillaprotocol}) in order to estimate the ground state overlap at the point $s_i$ in the adiabatic path
\begin{align}
    O'_i &= \left|\braket{\text{GS}_{s_i}|W_{0\rightarrow s_i}|\psi_0}\right|
\end{align}
without backward-time evolution. Here, $\ket{\text{GS}_{s_i}}$ is the instantaneous ground state at point $s_i$ of the adiabatic path.  The rest of the procedure is analogue to the previous algorithm. This algorithm enables us to try to keep quantum state along the adiabatic path close to the instantaneous ground state for fixed total adiabatic runtime.
\subsubsection{Black box optimization for fixed total time} \label{sec:BB}
\begin{figure}[htbp]
    \includegraphics[width=.9\linewidth]{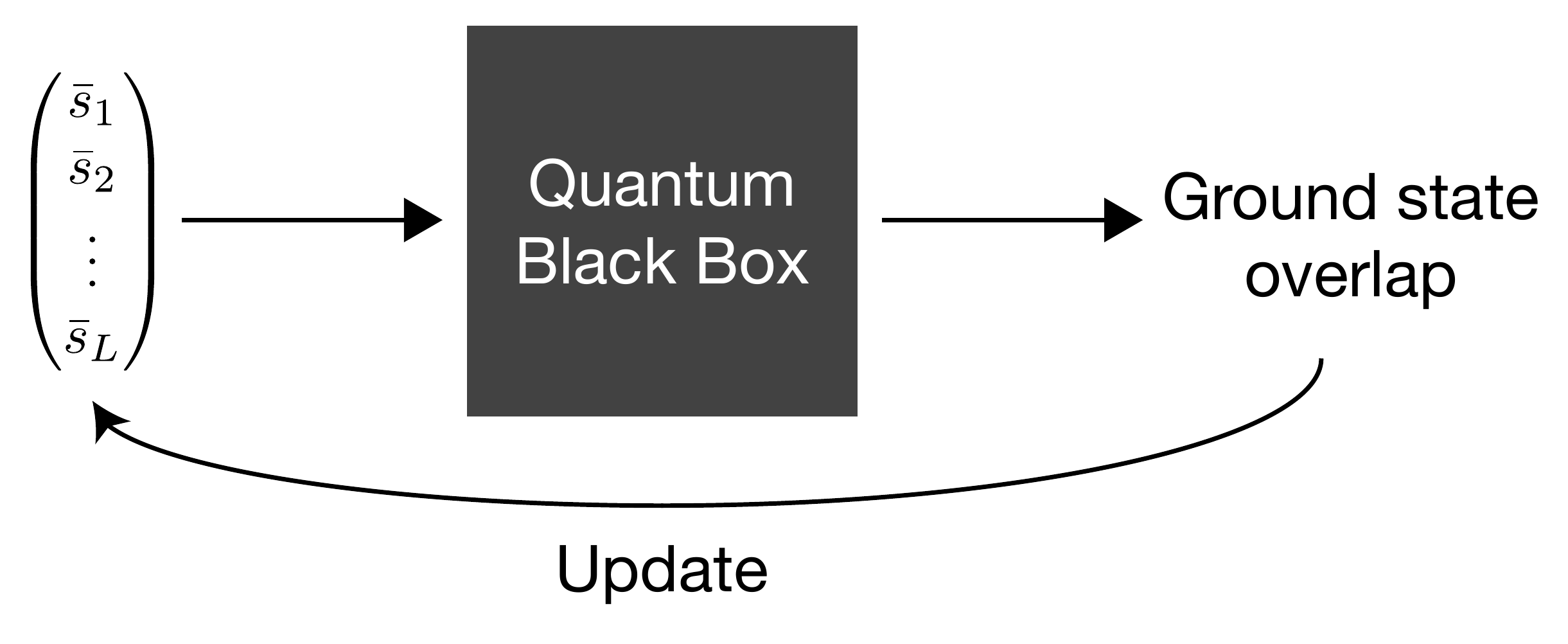}
    \caption{Quantum classical feedback loop for optimizing the chunk lengths of an adiabatic evolution while keeping the total time fixed. A parameter vector containing the chunk lengths is used as input for a quantum black box. The black box implements an optimized adiabatic evolution accordingly and outputs the ground state overlap. Making use of a classical optimizer, the chunk lengths vector is updated in order to maximize the ground state overlap.}
    \label{fig:BBalgorithm}
\end{figure}

We present a black box optimizer routine which makes use of the gradient to find optimized chunk lengths $\{\bar s_i\}$. Here, the chunk lengths are optimized in a quantum-classical feedback loop similar to typical variational quantum algorithms. In our approach, the vector containing the chunk lengths is used as the input for a quantum black box (Fig.~\ref{fig:BBalgorithm}). The chunk lengths remain normalized to 1. Within the quantum black box, an optimized adiabatic evolution is implemented according to the current chunk length vector, and the output of the black box is the final ground state overlap at $s=1$. The ground state overlap may be obtained, e.g.~by making use of our proposed one-ancilla protocol~(Sec.~\ref{ssec:singleancillaprotocol}). This value of the ground state overlap is fed into a classical optimizer which updates the input vector. Our cost function is the ground state overlap which we seek to maximize. We use the (quasi-Newtonian) bounded limited memory BFGS (L-BFGS-B) algorithm~\cite{byrd1995limited} or the gradient-free Nelder-Mead  (or downhill simplex) algorithm~\cite{nelder1965simplex} for the classical optimization. To make the setting of the optimizer more realistic for an experimental set-up, we fix the relative step size in L-BFGS-B to be larger than $1\%$ of the chunk lengths $\{\bar s_i\}$. The feedback loop is repeated until convergence or until another suitable termination criterion is reached, e.g.~a desired ground state fidelity.
\subsection{Target fidelity profile with flexible total time}
The concept of the VQAA also allows for flexible total time optimization. Here, the goal is to find an optimized adiabatic evolution which gives a final fidelity as close as possible to a given threshold 
\begin{align}
\left|\braket{\psi(s=1)|\text{GS}}\right| \gtrsim \theta_L,    
\end{align}
the adiabatic path is split up into $L$ chunks of length $s/L$ with respective evolution times $T_i$. The state $\ket{\text{GS}}$ is the ground state at the end of the adiabatic path at $s=1$. By smoothly interpolating from $\theta_0$ at $s=0$ to $\theta_L$ at $s=1$ the intermediate target thresholds $\theta_i$ are set.\\
Starting with the first chunk, we now search for the evolution time $T_1$ which suffices so that $\left|\braket{\psi(s_1=1/L)|\text{GS}}\right| \gtrsim \theta_1$. Here, we can make use of hypothesis testing which is highly efficient in the number of measurements (cf.~Suppl.). Search algorithms such as bisection methods feature exponentially fast convergence. The $\{T_i\}$ define the optimized adiabatic evolution. This algorithm is then able to follow a target fidelity profile with resolution $L$ as closely as possible.
We note that due to the limited decoherence time of NISQ devices, a maximal value for the $T_i$ can be fixed.
\section{Benchmarking and results}\label{sec:benchmarkin}
For benchmarking purposes, we choose a non-trivial problem where we are able to simulate the evolution of a quantum computer with and without noise on a classical computer. Simulating the dynamics of a large quantum system on a classical computer is, in general, very hard as the number of coefficients necessary for the classical description increases exponentially with the system size $N$. For this reason, we choose a gapped one-dimensional Hamiltonian as our system. In the algorithms we propose, the state is always close to the ground state, i.e. has few excitations only. The ground states of gapped one-dimensional systems can be described efficiently using a matrix product state (MPS) ansatz. We give a brief outline of the tensor network techniques~\cite{cirac2020matrix} we use to simulate our algorithms classically in the Appendix (cf.~Suppl.).
Therefore, even though MPS cannot approximate time evolution in general, it is ideally suited for our problem since our states have few excitations.
\subsection{Model}
\begin{figure}[htbp]
    \centering
    \includegraphics[width=.8\linewidth]{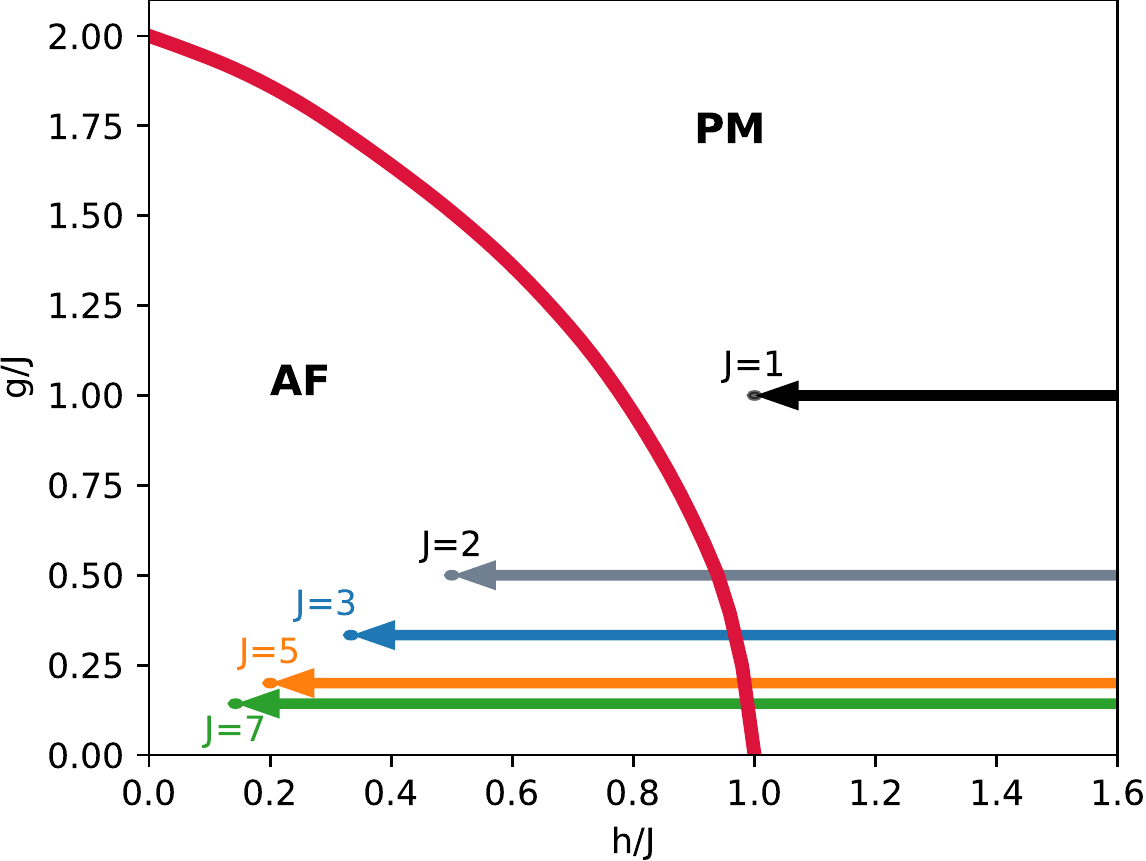}
    \caption{Zero-temperature phase diagram of the ZZXZ model in the thermodynamic limit. The antiferromagnetic phase (AF) is separated from the paramagnetic phase (PM) by a second-order phase transition (red line). At the multi-critical point at $(h,g)=(0,2)$ the model becomes classical resulting in a first-order phase transition. For different interaction strengths $J$ the adiabatic evolution follows different paths in the phase diagram. Phase diagram after~\cite{ovchinnikov2003antiferromagnetic}, note that their different Hamiltonian formulation results in an appropriately rescaled phase diagram.}
    \label{fig:PD}
\end{figure}

For benchmarking a simple, yet non-integrable quantum model, we use the translationally invariant Ising model with transverse and longitudinal fields, hereafter referred to as the ZZXZ model
\begin{align}
    H_T=\sum_{i=1}^N (J\sigma_i^z \sigma_{i+1}^z + h\sigma_i^x + g\sigma_i^z)
\end{align}
as a finite system with open boundary conditions~\footnote{The Pauli matrices are defined as
$\sigma_x = \big(\begin{smallmatrix}0 & 1\\1 & 0\end{smallmatrix}\big)$, $\sigma_y = \big(\begin{smallmatrix}0 & -i\\i & 0\end{smallmatrix}\big)$ and $\sigma_z = \big(\begin{smallmatrix}1 & 0\\0 & -1\end{smallmatrix}\big)$.}. For $J>0$ we are considering the antiferromagnetic ZZXZ model. A choice of the coefficient $J=1$ places the model in the paramagnetic phase (Fig.~\ref{fig:PD}) and 
an adiabatic evolution from the trivial Hamiltonian $H_0=\sum_i h\sigma_i^x $ stays entirely within the paramagnetic phase, therefore no phase transition is crossed~\cite{novotny1986zero, ovchinnikov2003antiferromagnetic}. This is different for choices, e.g.~$J=\{2,3,5,7\}$, where the adiabatic paths crosses a second-order phase transition from the paramagnetic phase into the antiferromagnetic phase. The adiabatic spectrum following a linear interpolation from $H_0$ to $H_T$ is discrete for the lowest energy eigenstates (for a discussion on smooth reparametrizations of the path, see Suppl.). 
\subsection{Results for adiabatic spectroscopy} \label{ssec:ResAdSpec}
\begin{figure}[htbp]
    \centering
    \includegraphics[width=.9\linewidth]{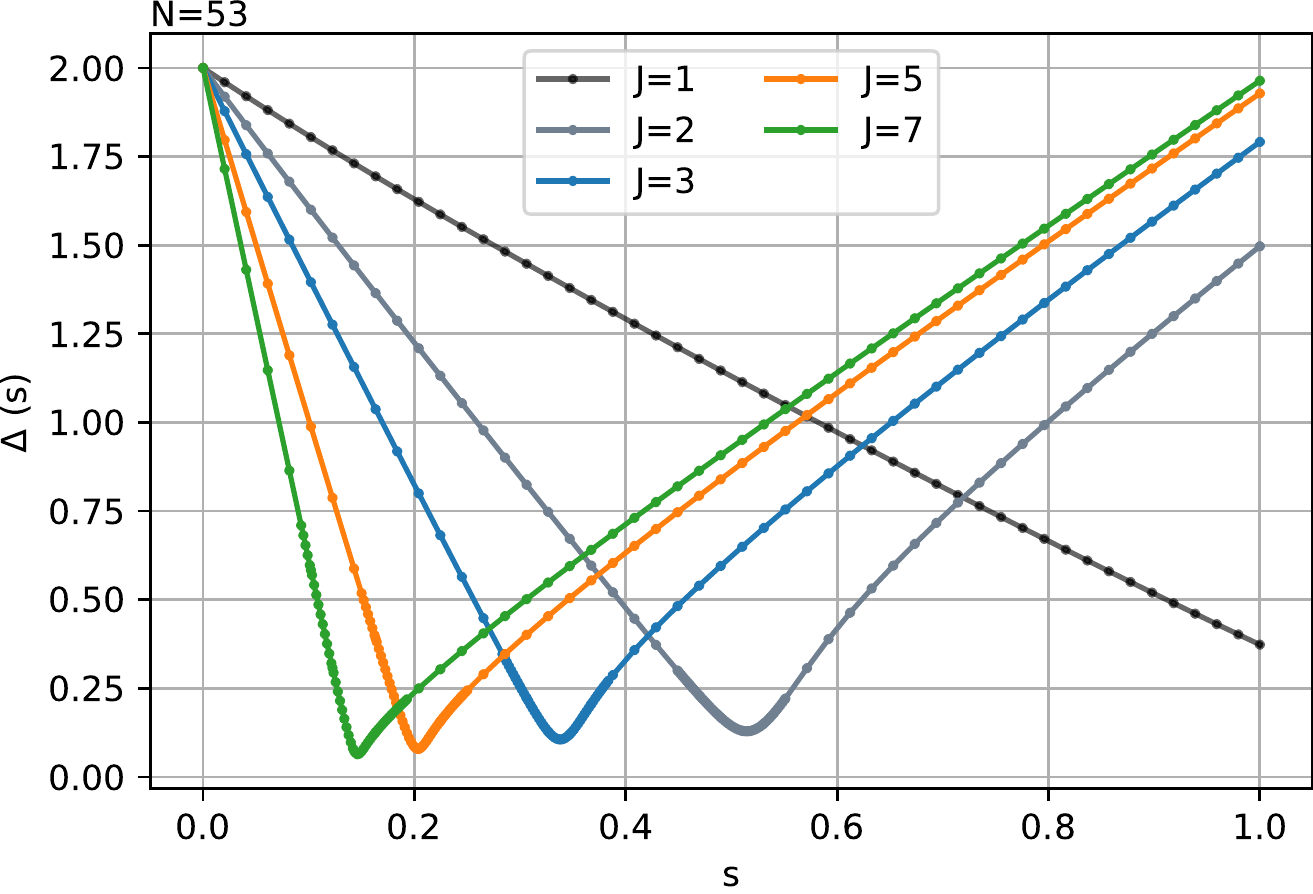}
    \caption{Spectral gap $\Delta(s)$ for the ZZXZ model $N=53$ qubits and different values of $J$, obtained with DMRG methods. The energy difference between the ground state and the first excited state is plotted along the adiabatic path in $s$. The minimal spectral gap along the adiabatic path is smallest for large interaction strength $J$ and the phase transition is crossed earlier in parametrized time $s$.}
    \label{fig:DMRG}
\end{figure}

\begin{figure}[htbp]
    \centering
    \includegraphics[width=.9\linewidth]{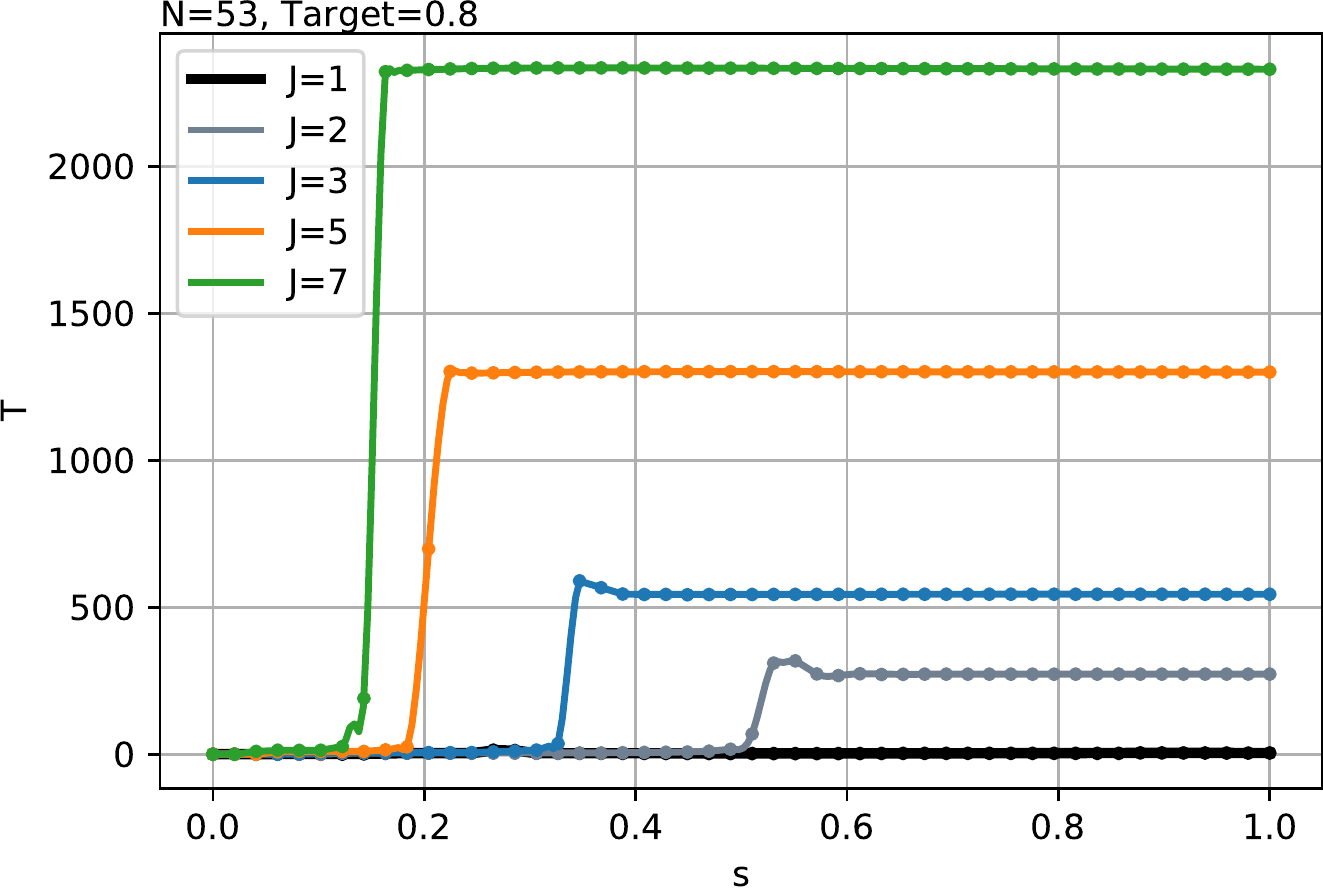}
    \caption{Adiabatic spectroscopy for $N=53$ qubits and a given target ground state fidelity of 0.8, we perform adiabatic sweeps from 0 to $s$ and find the evolution time $T$ required to reach the target fidelity. We obtain data for 50 values of $s$ linearly distributed along the x-axis. When a phase transition needs to be overcome, the $T$s rise to much larger values. The steepness of curve at this point provides a measure for the smallness of the spectral gap.}
    \label{fig:AdSpecCumulative}
\end{figure}

We benchmark the adiabatic spectroscopy for the ZZXZ model on a qubit chain with $N=53$ sites. For our model, the spectral gap $\Delta(s)$ has been obtained using DMRG methods (Fig.~\ref{fig:DMRG}).
In our simulations, we compute the ground state fidelity directly using tensor contractions, however, in an experiment, this information would be gathered using our ancilla protocols (Sec.~\ref{sec:protocols}). In an experimental setting, every time we would like to make a measurement to estimate $|\alpha(\tau)|$, we can choose to evolve uniformly at random for different values of $\tau$ between 0 and some very large $K$. Under these assumptions a simple analytical formula for the expectation value $E^2$ of $|\alpha(\tau)|^2$ can be given (App.~\ref{App:Bound}) and a target ground state fidelity of $|\psi_0| = 0.8$ translates to
\begin{align}
    E^2 := \lim_{K\rightarrow\infty} \mathbb{E}_{\tau\: \sim \text{ unif. dist. in } [0, K]} (|\alpha(\tau)|^2) \geq 0.54.
\end{align}
In the adiabatic spectroscopy, we obtain the evolution times $T(s)$ required to reach this target for a given value of $s$ (Fig.~\ref{fig:AdSpecCumulative}).
Clearly, there is a strong increase in $T$ around the position of the minimal spectral gap for respective $J$. By computing the derivative of the (cubic) splines, we extract the position and smallness of the gap which correspond to the position and steepness of the $T$-increase in the $T(s)$ plot. For closer resemblance with the actual spectral gap profile, we plot $-\partial T (s) / \partial s$ (and clip values above zero) in (Fig.~\ref{fig:AdSpec}). 
\subsection{Results for VQAA}
Here we compare the benchmarking results obtained for the different algorithms and interpret their respective behavior. In particular we would like to highlight that there are two regimes where an optimized adiabatic path improves over non-optimized (or naive) QAA.
When a phase transition is crossed in an adiabatic evolution, as intuition would suggest, optimal adiabatic paths concentrate the majority of the evolution time around the position of the spectral gap. We also examine the case without a phase transition ($J=1$) and benchmark for systems with up to $N=100$ qubits. Here, we observe that rotations in the low-energy eigenspace can significantly improve the performance of the quasi-adiabatic evolution. Moreover, we find that leaving the instantaneous ground state can pay off in finding a better final ground state overlap. This seems to be especially significant for short total evolution times. 
\subsubsection{Black box optimization for fixed total time} \label{ssec:BBresults}
\begin{figure}[hbtp]
    \centering
    \includegraphics[width=.9\linewidth]{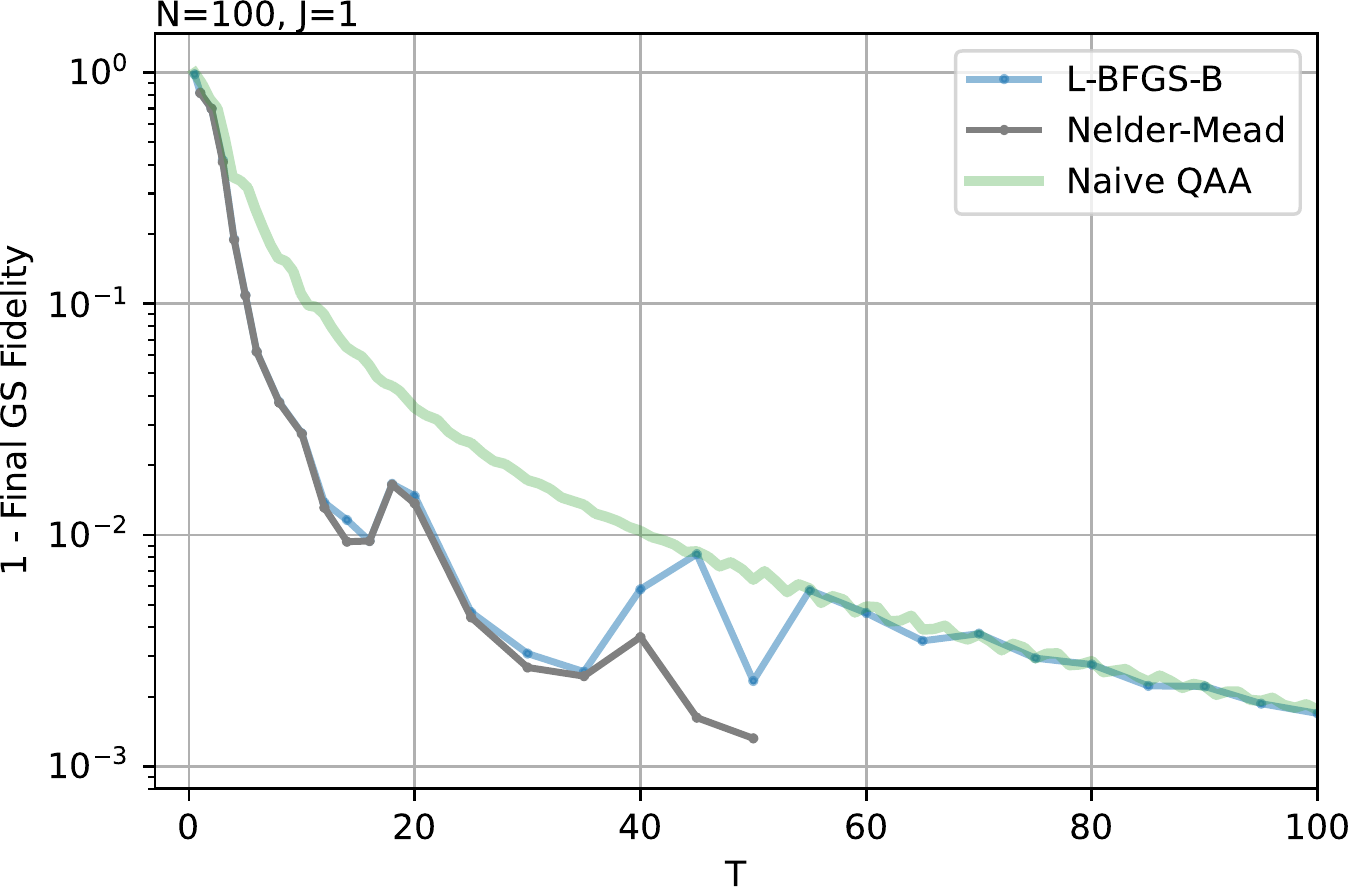}
    \caption{Large $T$ black box optimizer benchmark for the two classical optimization algorithms L-BFGS-B and Nelder-Mead (NM). Data for was obtained only up to $T=50$ and requires, in general, more measurements than gradient-based methods. However, the optimization with L-BFGS-B became increasingly unstable for large $T>40$ and NM was observed to be more robust. The instability of L-BFGS-B was because a very large amount of fine-tuning of the initial step size for the different T would have been required.}
    \label{fig:LargeBenchmark}
\end{figure}

\begin{figure}[hbtp]
    \centering
    \includegraphics[width=.9\linewidth]{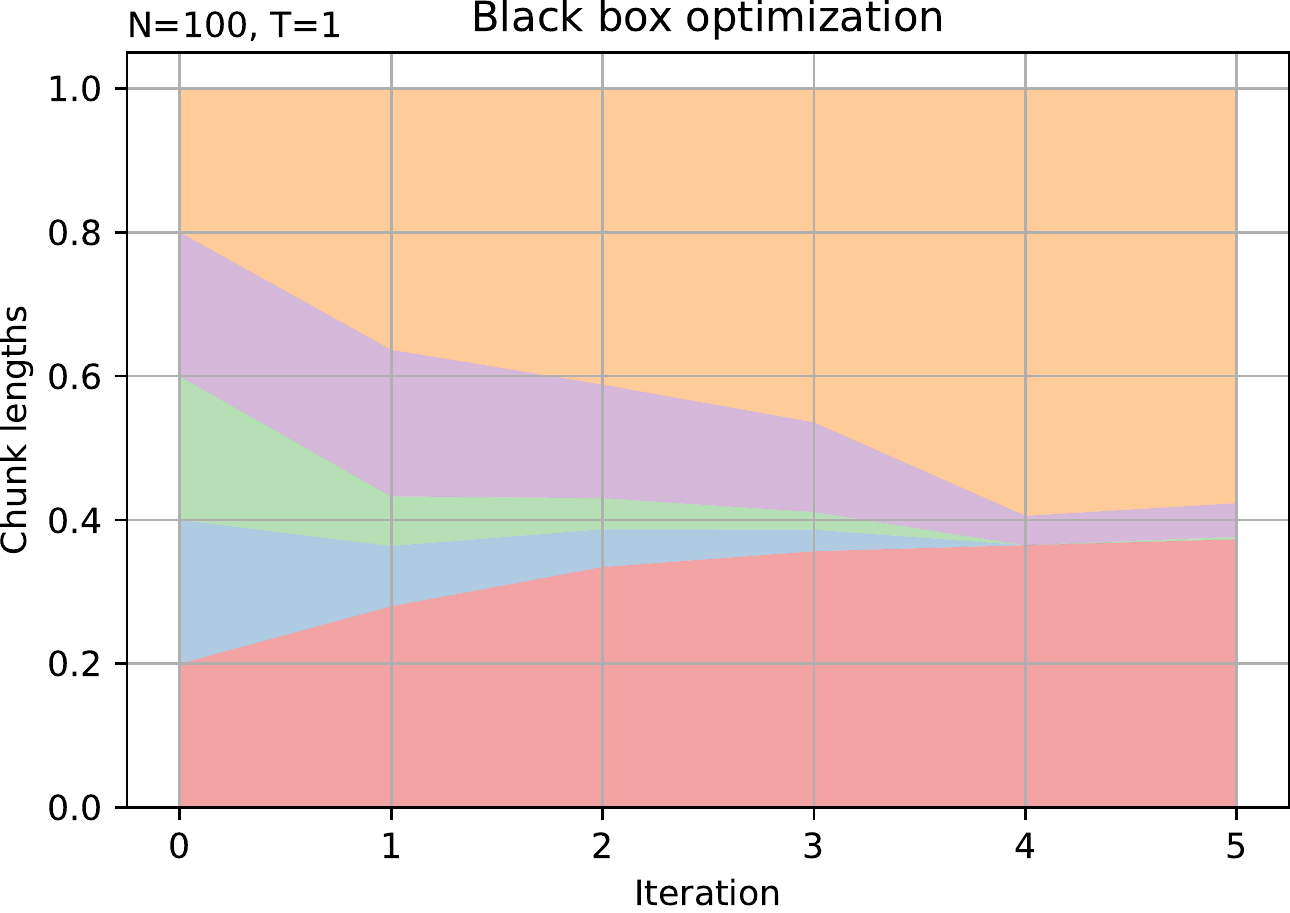}
    \caption{Stacked chunk lengths adding to $s=1$ display the evolution of an instance of a black box optimizing routine. Chunks are initialized with equal lengths and their lengths optimized using a L-BFGS-B optimizer with regard to a maximal final fidelity with the ground state of the ZZXZ Hamiltonian. A confluence of chunks around $s\approx0.4$ can be observed. 
    \label{fig:EvolutionChunksBB}}
\end{figure}

\begin{figure}[hbtp]
    \centering
    \includegraphics[width=.9\linewidth]{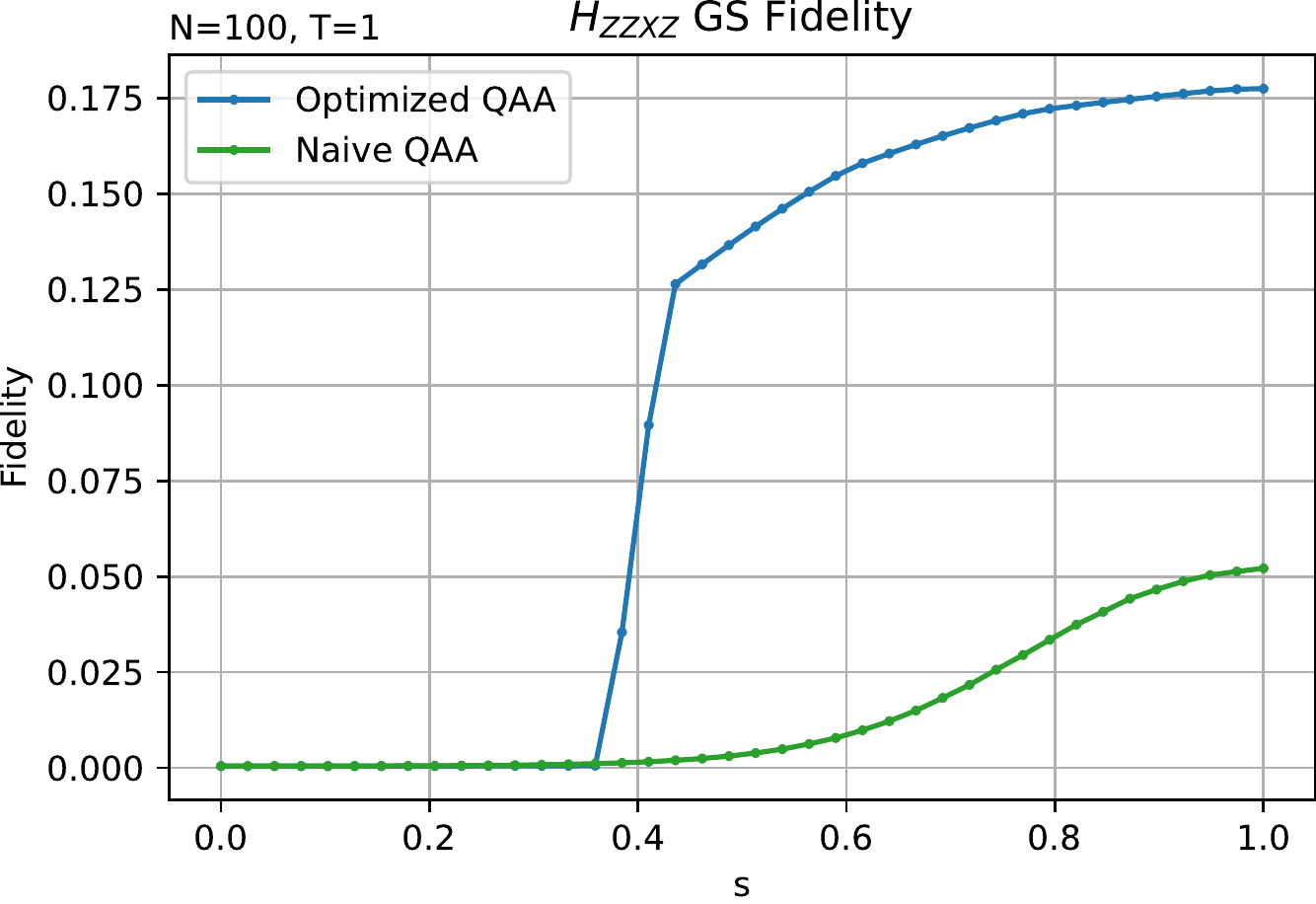}
    \includegraphics[width=.9\linewidth]{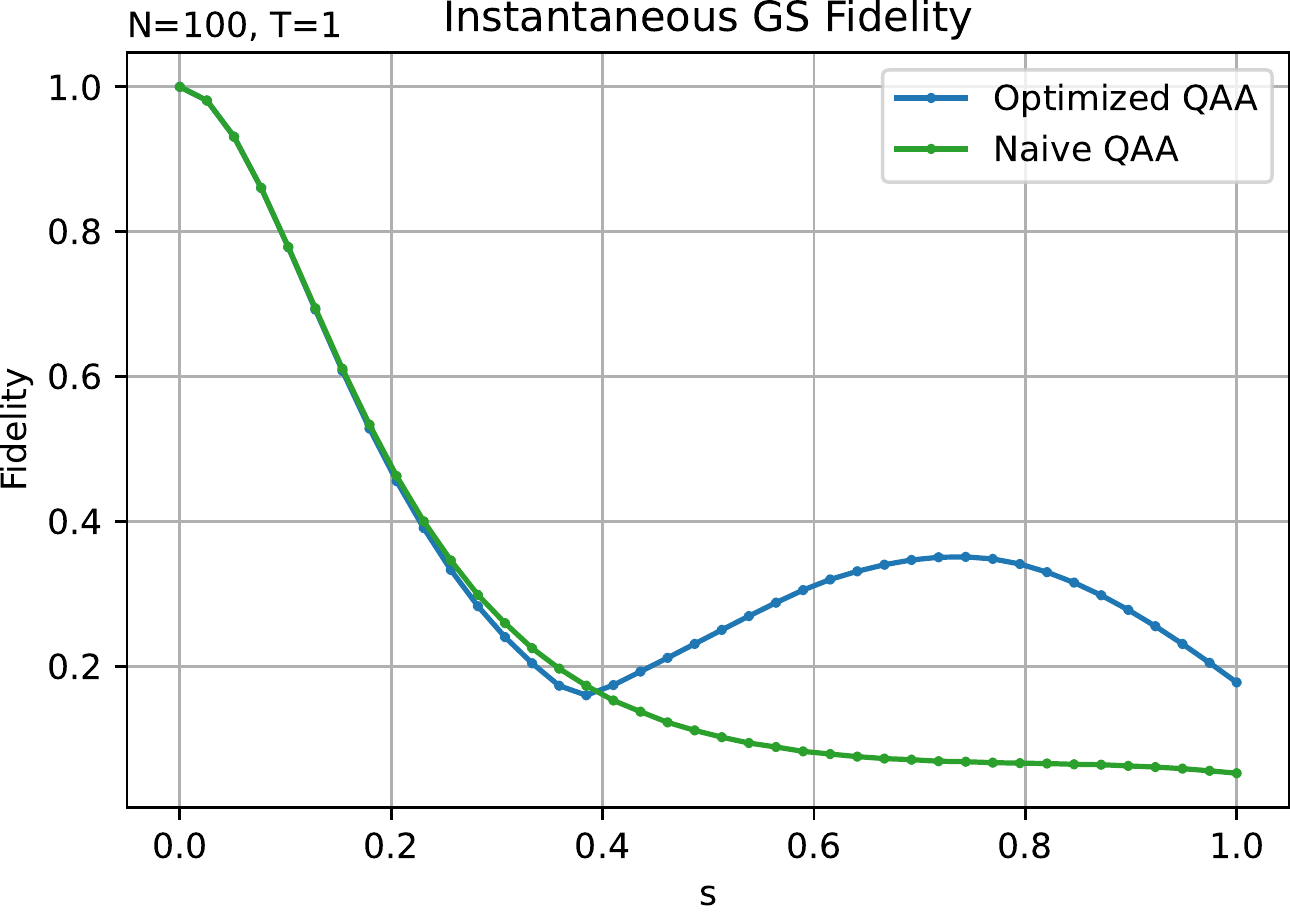}
    \caption{The ground state fidelity is plotted along the adiabatic path for different values of $s$. The results are for $N=100$ and fixed total time $T=1$. The optimized QAA curve (blue) differs significantly from naive QAA (green curve) in the region of very small chunks around $s\approx0.4$. Note that the spectral gap between the ground state and the first excited state is not minimal in this region, but decreases strictly monotonically with $s$. At the end of the evolution, optimized QAA achieves a ZZXZ model ground state fidelity of $17.8\%$ while naive QAA results in a fidelity of $5.2\%$.}
    \label{fig:GSoverlapBB}
\end{figure}

The best results for fixed time are achieved with the black box algorithm. Even for large system sizes, this algorithm can be able to improve significantly over naive QAA for fixed total time. Good results are usually achieved for a handful of chunks already and the performance of the algorithm does not improve much further for more chunks (which would correspond to a higher resolution in the adiabatic velocity profile). This is the case both when a small spectral gap needs to be crossed, but also in the absence a phase transition. In the latter case, the black box algorithm very often returns an adiabatic path including one or even several very small chunks.\\
These rotations can be understood in the following way. The set-up of the VQAA with fixed time allows some chunk length $\bar s_i$ to become very small. In these small chunks, however, there is still an amount of time $T_i$ spent. The evolution in such a chunk in the limit of $\bar s_i \rightarrow 0$ is implemented by a unitary
\begin{align}
    U_\text{rot}(T_i, s_i) = e^{-i T_i H(s_i)}
\end{align}
which effectively changes the local phases of $\ket{\psi(s_i)}$ thereby physically changing the quantum state. As $\ket{\psi(s_i)}$ in our algorithm is expected to have a considerable ground state population and also small populations in the lowest excited states, the change in the local phases of $\ket{\psi(s_i)}$ corresponds to a rotation in the low energy eigensector. Some rotations eventually become beneficial for the performance of the adiabatic routine. Rotations in the low energy eigensector have also been considered in the context of the Eigenstate Thermalization Hypothesis~\cite{wurtz2020emergent}.\\
In the case of a \emph{phase transition}, the improvement in the evolution time $T$ between naive QAA and an optimized adiabatic evolution is much larger than without a small gap. For very few chunks only, target fidelities of over 90\% in a system of 53 qubits were reached at around $T\approx 100$. With a non-optimized adiabatic path, this would have required very long ($\approx$ factor 10 longer) preparation times as depicted in (Fig.~\ref{fig:BBBPT}). The classical optimizer used for obtaining the data in this figure is the Nelder-Mead algorithm. In our simulations, Nelder-Mead proved to be more robust at large $T$ at the expense of requiring more measurements. Being a gradient-free method, we expect it to behave better than the gradient-based L-BFGS-B for noisy systems. \\
In
\begin{figure}[hbtp]
    \centering
    \includegraphics[width=.9\linewidth]{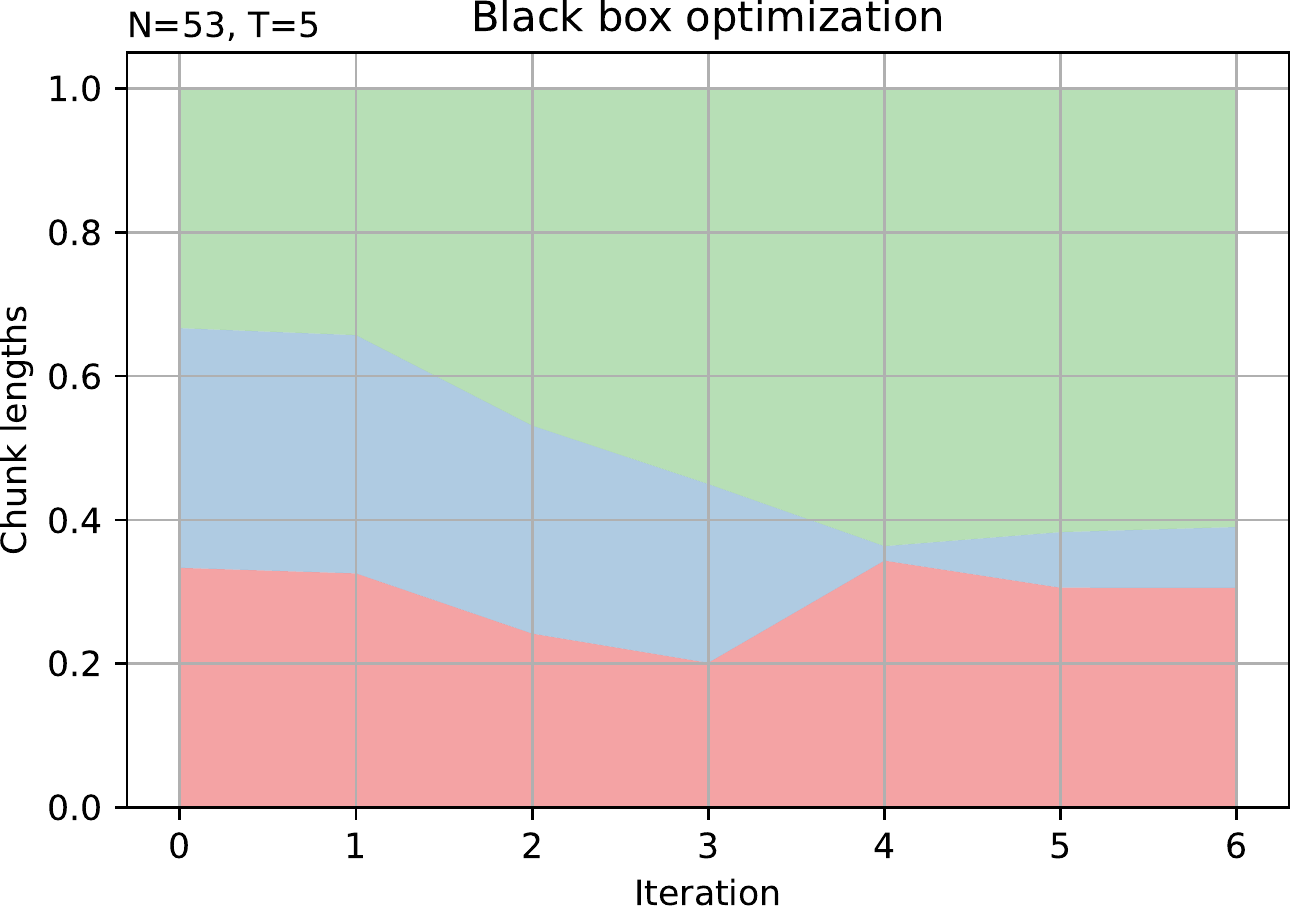}
    \caption{Instance of a black box optimizer for the case when crossing a phase transition, here for $J=3$ and $N=53$. The stacked chunk lengths are shown for the first nine iterations as the middle chunk becomes smaller. This leads to the evolution time being spend effectively around the smallest gap around $s\approx0.34$. In this instance, the final fidelity is nearly seven times larger using the VQAA compared with a linear sweep.}
    \label{fig:BBPT}
\end{figure}
(Fig.~\ref{fig:BBPT}), the evolution of the chunk positions for an instance of the black box algorithm is shown for 53 qubits and $J=3$. Here, the total time is $T=5$ and L-BFGS-B was chosen as the classical optimizer due to its faster convergence in the classical simulations. The adiabatic spectrum for $J=3$ features a small gap around $s=0.34$. This can be approximately captured with three chunks only. Over a few iterations, the center chunk becomes smaller around the position of the gap, so that more time in the adiabatic evolution is spend there.\\
In the case \emph{without a phase transition}, for $J=1$, we find reductions of the total evolution time by a factor of over 3 when comparing with naive QAA. Here, we benchmark the black box optimization routine for large values of $T$ and 100 qubits (Fig.~\ref{fig:LargeBenchmark}). 
We note that for $T>40$, the classical L-BFGS-B algorithm was in most cases not able to find an optimized adiabatic path. It was made sure that the reason for this was not due to memory-limitations in the optimizer, the relative step size or tolerance values for the termination of the algorithm. Increasing the number of chunks does not generally help in finding better optimized paths. However, sequential initializations of the optimizers can improve the performance. Instead of starting with naive QAA, the optimizer then begins with the chunk lengths of the previous, shorter-time optimizer instance. \\
A typical black box algorithm instance (for $J=1$, i.e.~no small gap) is shown in (Fig.~\ref{fig:EvolutionChunksBB}) where the small chunks can be easily observed around $s\approx0.4$. Indeed, the ground state fidelity is maximized and the energy minimized in five iterations only. In (Fig.~\ref{fig:GSoverlapBB}), the fidelity between $\ket{\psi(s)}$ and the ZZXZ ground state or the instantaneous ground state of $H(s)$, respectively, is shown. The exact ground states have been obtained using DMRG methods. A discontinuity in the path of the optimized QAA is clearly visible for $s\approx0.4$ due to the implemented rotation. As the black box optimizer only has access to the ground state overlap at $s=1$, it is agnostic to the actual curve of the optimized QAA for values below $s<1$.
It is interesting to observe that in fact leaving the instantaneous ground state can lead to better final results at $s=1$, also discussed for adiabatic evolutions in~\cite{benseny2020all} and in a related way in the context of diabatic transitions in QAOA in~\cite{zhou2018quantum}. 
\begin{figure}[hbtp]
    \centering
    \includegraphics[width=.9\linewidth]{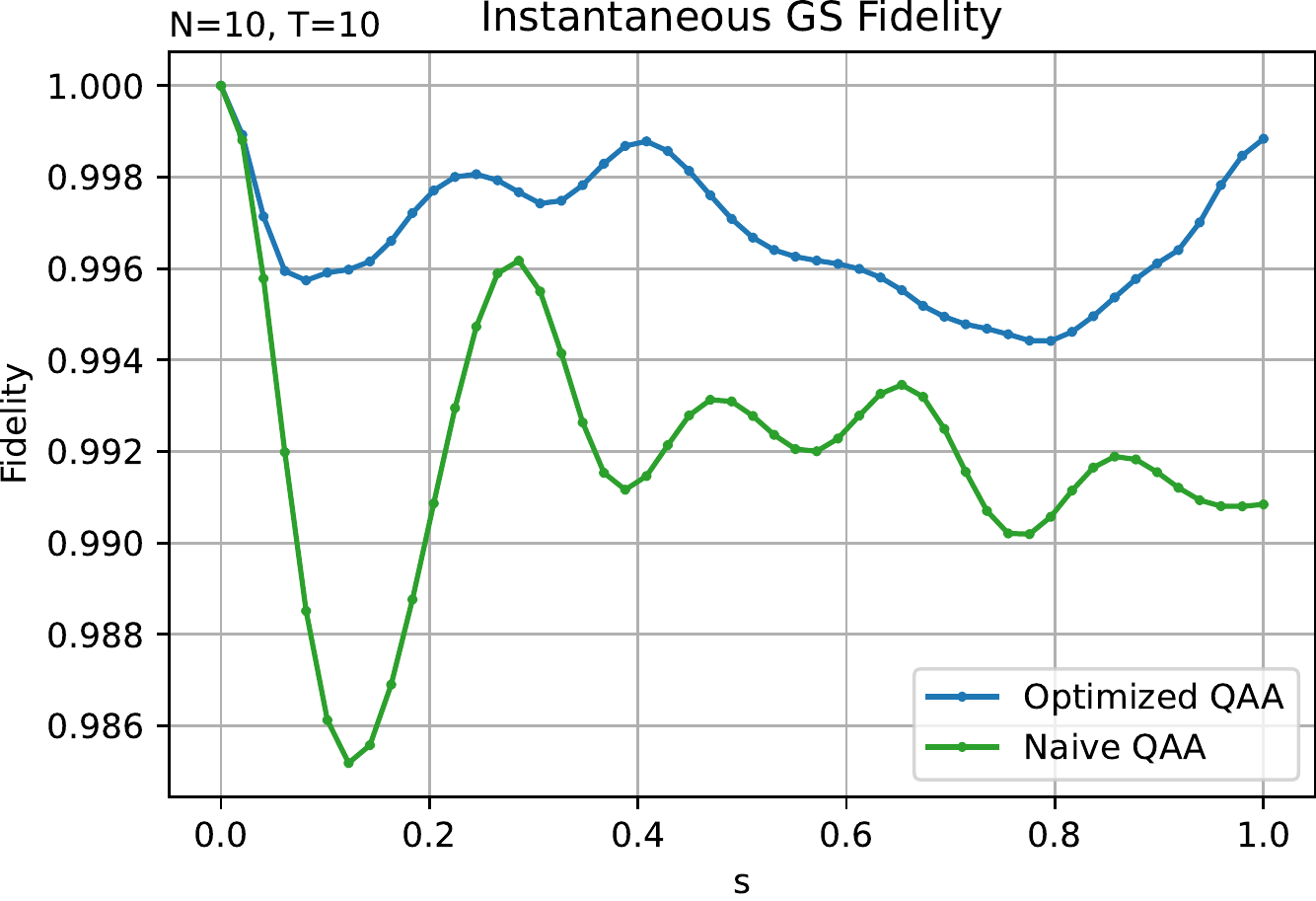}
    \caption{In this instance, albeit for $N=10$, we see near perfect ground state preparation for $T=10$ without increasing the total time budget. Even though the instantaneous ground state fidelity with $H(s)$ along the adiabatic path is smaller, the final ground state fidelity is 0.999.}
    \label{fig:Backto1}
\end{figure}
This feature can also be observed in a black box optimizer instance for a smaller system in (Fig.~\ref{fig:Backto1}) where the optimized QAA curve gives a final fidelity of approximately 0.999 and recovers most of its ground state fidelity in the last 20\% of the adiabatic evolution path only.
\subsubsection{Converging fidelity ratios for fixed total time} \label{ssec:resultsfixed}
Besides the black box approach to VQAA, we have presented other algorithms which aim to stay close to the instantaneous ground state along the adiabatic path. Both the ancilla-free and the one-ancilla method seek for convergence in the fidelity ratios between consecutive chunks. The one-ancilla method is cleaner in the sense that it directly uses the overlap and does not accumulate extra transitions and phases on the backward path (Sec.~\ref{ssec:forwardonly}). The one-ancilla method can achieve a smooth adiabatic path which remains as close as possible to the ground state at all times for fixed total evolution time. 
However, the ancilla-free method finds adiabatic paths which effectively implement a rotation in the low energy eigensector. This property is also observed in the gradient-based black box method~(Sec.~\ref{ssec:BBresults}) and seems to be very relevant for preparing the ground state on NISQ devices. Therefore, which method will perform better is likely quite model-dependent for these two algorithms. 
\subsubsection{Target fidelity profile for flexible total time}
\begin{figure}[htbp]
    \centering
    \includegraphics[width=.9\linewidth]{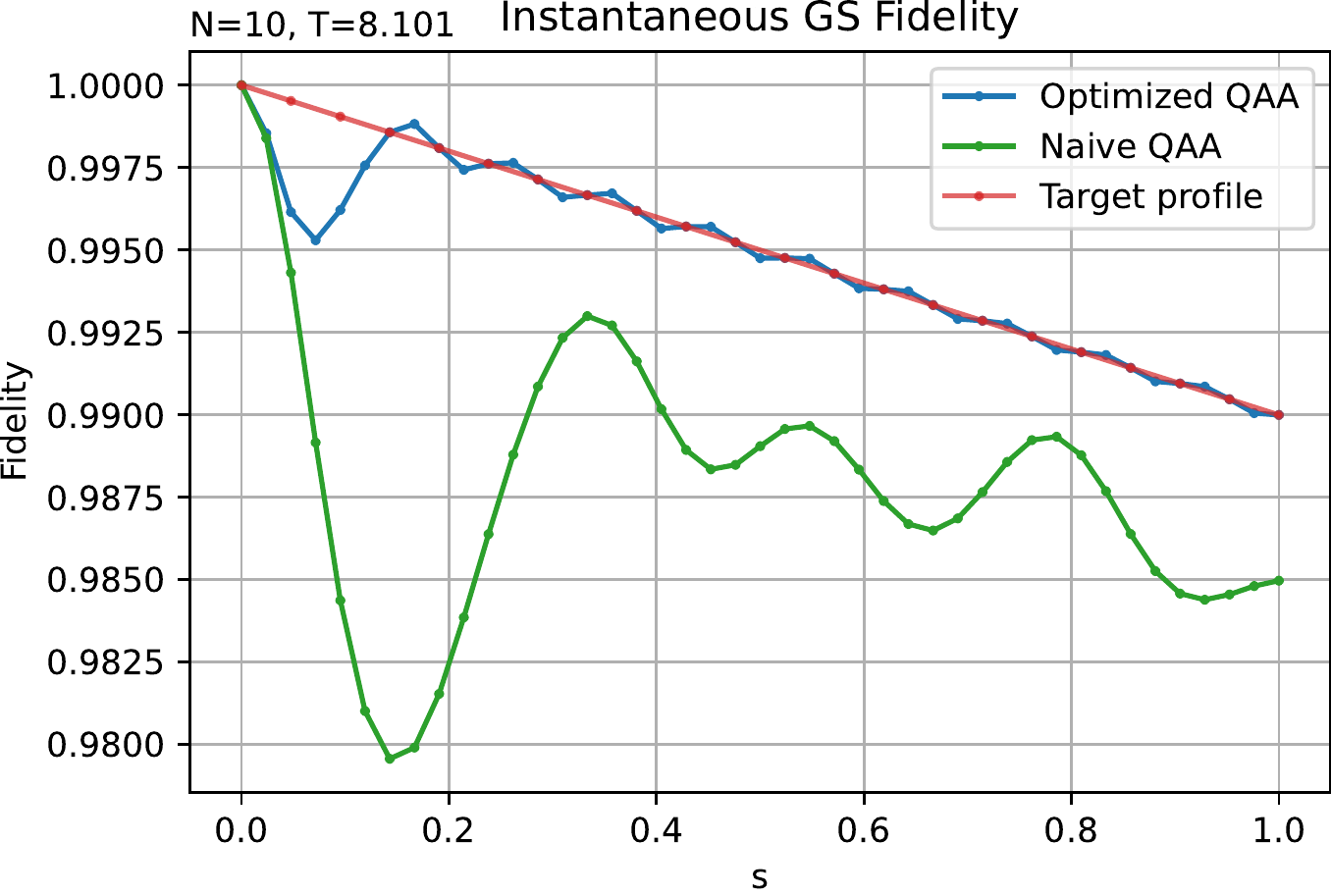}
    \caption{Instantaneous ground state fidelity for QAA optimized with the flexible total time algorithm compared to naive QAA with the same time budget. The target fidelity at $s=1$ was set to $\theta=0.99$ and linearly interpolated along the adiabatic path. The search interval for the time spent in each chunk was upper bounded such that $T_i \leq 20$ $\forall i$.}
    \label{fig:flexT}
\end{figure}

The flexible total time algorithm with a given target fidelity with the ground state at the end of the adiabatic evolution is well-suited in the kind of instances in which staying close to the ground state is desired. As we observe in (Fig.~\ref{fig:flexT}) the naive QAA will occasionally leave the ground state leading to oscillations in the instantaneous ground state fidelity.
\noindent Considering the adiabatic path split-up into several chunks, due to the changing Hamiltonian $H(s)$ different evolution times $\{T_i\}$ for each chunk are necessary in order to achieve a given fidelity with the ground state at the end of the adiabatic evolution. Values for the $\{T_i\}$ strongly depend on the spectral gap $\Delta(s)$ between the ground state and the excited state as well as on the Berry connections~(cf.~Suppl.).\\
We observe that our algorithm is able to reduce the total adiabatic evolution time compared to naive QAA in this toy example. This is because it uses the time available in a more economic way by spending much of the evolution time only when required to stay close to the ground state. 
One useful property of this algorithm is the fact that it is self-verifying in a sense that the hypothesis testing at the end of every chunk guarantees with high confidence that the ground-state fidelity is larger than a given value.
Further developments of this algorithm can be envisioned where the $\{T_i\}$ are optimized to follow a more complicated profile. Moreover, the results of this algorithm serve as a good initial point for the gradient-based black box algorithm for fixed total evolution time.\\
Here, we have set the chunk lengths $\{\bar s_i\}$ to equal values. Adaptions of this algorithm with unevenly spaced chunks are also possible. Including additional chunks with very small chunk length could possibly provide a better performance of the algorithm by making it possible to implement rotations in the low energy eigensector at the expense of an increased number of measurements~(Sec.~\ref{ssec:resultsfixed}).
\subsection{Cost of implementation}\label{sec:mmt}
In every variational algorithm, the number of measurements needed in order to obtain satisfactory results is of upmost importance as it directly determines the feasibility of the approach. Here, we discuss the number of measurements necessary for each respective algorithms presented in this paper. We focus on the number of ground state overlap evaluations necessary in our classical simulations. To obtain the ground state overlap from $|\alpha|$, a small overhead is required~(Sec.~\ref{ssec:singleancillaprotocol}).\\
Both the \textit{ancilla-free and one-ancilla fixed time algorithms} converge with very few iterations even for large system sizes ($N=100$). The resolution of the velocity profile in the adiabatic evolution is the number of chunks chosen. In our case, the number of ground state overlap evaluations was fairly low with
\begin{align}
    \# \text{GS overlaps} = \# \text{iterations} \cdot \# \text{chunks} \lesssim 100
\end{align}
which sufficed for our simulations. The number of chunks are the number of parameters to be estimated and for the \textit{flexible total time algorithm} this corresponds to the number of bisection searches required. As these search algorithms converge exponentially fast, roughly 10-20 ground state overlap estimations per search are usually sufficient for highest accuracies of the optimized $\{T_i\}$ values. In this algorithm, the two-ancilla algorithm is used which is suitable for hypothesis testing. Hypothesis testing converges exponentially fast as well, asking for about another 10 measurements for each ground state overlap estimation. Exemplary numbers for hypothesis testing are given in the Appendix~(cf.~Suppl.).\\
The \textit{black box algorithm} depends on ground state overlap evaluations in order to estimate the gradient. The number of ground state overlap evaluations necessary  in the optimization process is directly related to the number of chunks. Already for very few iterations, good results can be obtained with this method, so that for five chunks, we achieved good results for $N=100$ and $T=1$ with significantly less than 50 ground state overlap evaluations in total~(App.~Fig.~\ref{fig:measment}).
For a random variable $X$ with values in $[a,b]$, $\bar \Delta=b-a$ and independent and identically distributed samples, the Chernoff-Hoeffding inequality~\cite{hoeffding1994probability} gives an upper bound on the probability to find the measurement deviating more than $\epsilon$ from its expectation value $\mu$
\begin{align}
    \text{Pr}(|X-\mu| \geq \epsilon) \leq e^{-2m\epsilon^2/\bar\Delta^2} =: \eta
\end{align}
For Pauli measurements, we can have either $+1$ or $-1$ as results, so $\bar\Delta=2$ and the number of measurements 
\begin{align}
    m \geq \frac{2}{\epsilon^2} \log\left(\frac{1}{\eta}\right)
\end{align}
depends on the $\eta$ and $\epsilon$ needed. Setting the precision $\epsilon$ to $(1 - (\text{GS fidelity}))/ 20$ and the failure probability so that 1 in 2 experiments is successful with all estimations within the deviation $\epsilon$, we estimate the number of measurements necessary to be of the order of $10^3$.\\
The number of ground state fidelities required to reach very high fidelities $> 0.9$ at larger $T$ is naturally larger than what stated above for $T=1$. In (Fig.~\ref{fig:BBBPT}), the maximum number of ground state fidelity evaluations was set to 1000, while typically few hundred evaluations sufficed to find an adiabatic path with maximal ground state fidelity up to $10^{-8}$ relative accuracy (i.e. the relative accuracy of the optimization process). Without doubt, this accuracy will be unattainable in current experiments and much less ground state evaluations give already very good results. In fact, we observed that for $N=53$ in the case of a phase transition and $T\approx100$, when obtaining less than 200 ground state evaluations, the Nelder-Mead method will yield results of practically the same quality. These estimations suggest considerably low number of measurements even for optimizing the adiabatic evolution of large quantum systems.\\
While the number of measurements seems to be the most relevant figure of merit to assess the cost of the methods presented here, we also include a short discussion of the number of gates required on different architectures of quantum devices. On analogue quantum simulators, for instance, the ancilla-free optimization method can be implemented natively with the only overhead being the additional parametrized adiabatic sweeps to find optimized adiabatic paths. In a gate-model architecture, for a qubit-chain with $N=53$ sites and only allowing for next-neighbour interaction, we upper bound the total number of CNOT gates for one unit of time to be around 120 CNOTs/$N$. Here, we consider a decomposition of the unitary gates in the trotterized evolution of the numerical simulation. Actual gate counts will be significantly lower because the circuit will be optimized for the respective experimental hardware platform. In the single-ancilla protocol, an upper bound on the number of CNOTs per unit of time ($\tau=1$) including the required SWAPs for the chain topology is at around 2100 CNOTs/$N$. We note that with $\tau$ scaling as the inverse of the spectral gap $\Delta(s)$, the cost of the protocol is not excessively demanding, especially at the end of the adiabatic sweep ($s=1$) where the gap is large. A more detailed discussion of the gate count is provided in the Supplement.
\section{Noise} \label{sec:noise} 
\subsection{Noise in adiabatic quantum computation}
The noise in current quantum devices severely limits the performance of many quantum algorithms~\cite{Zhou2020}. Therefore, we discuss some important properties of noisy quantum adiabatic algorithms. A general inherent robustness of adiabatic evolution has already been established for some time~\cite{childs2001robustness}, here we focus on a few points that are especially important to our method.\\
In a gate-model quantum algorithm without error correction, a flipped qubit will in the worst case render the whole quantum computation nonsensical. This is quite different in an adiabatic algorithm as for physical instances low energy spectral lines are rare~(App.~Fig.~\ref{fig:DOS}). While a flipped qubit in the preparation of the ground state in an adiabatic algorithm can also lead to a quantum state orthogonal to the ground state, the energy, however, of this orthogonal excited state will still be a very good approximation to the ground state energy. Intuitively, one bit flip corresponds to a single excitation of the system. We can therefore assume that errors increase the energy only by $\mathcal{O}(1)$ for fixed time, when flipped qubits are rare. 
Also, the position $s_0$ in the adiabatic path, where a flipped qubit occurs, is not critical.~(App.~Fig.~\ref{fig:adiabnoise}). This may seem somehow surprising, but it can be explained because a perfect adiabatic evolution suppresses all transitions between the eigenstates. It does not only apply to the ground state but also to excited states. Therefore, a noise-induced excitation will in the regime of an adiabatic evolution not lead to further deviations from the ground state energy.\\
However, for time evolutions that are faster than an adiabatic evolution, which is in general the case on quantum devices limited in coherence time, we would generally expect light-cone spreading of noise through the spin-chain. Yet, in our simulations this was not observed to be problematic for the performance of the VQAA. In general, the noise behavior of adiabatic algorithms is encouraging as it suggests very benign noise features in these kind of algorithms making them a suitable candidate for NISQ devices.
\subsection{Impact of noise in the presented algorithms}
Here, we include a qualitative discussion about the expected performance of the algorithms presented in this paper in the presence of noise. We expect the \emph{adiabatic spectroscopy} to be quite robust to noise as the information obtained using this method relies on multiple data points and a rather distinctive feature in the $T(s)$ curve resulting from a small spectral gap. Noise effects will become stronger towards the end of the adiabatic evolution as noise accumulates in the circuit. However, as the results of this spectroscopy are quite pronounced, a qualitative description of the gap is likely only slightly impaired by moderate noise in the circuit. \\
Regarding the \emph{VQAA algorithms}, when considering noise, there are two main points to consider. First, noise can substantially impede the training phase of the algorithm, when the parameters of an optimal adiabatic path are being searched for. Second, in order to prepare the ground state with a desired high fidelity, an adiabatic evolution time $T$ that is only a fraction of the $T$ required for naive QAA suffices with an optimal adiabatic path. This may help strongly in suppressing errors.\\
Following a \emph{target profile} is especially tricky when noise comes into play. This is because noise strongly alters the required target profile. Therefore, concerning the target profile method, we do not expect this method to be very robust to noise, especially when finite size effects are playing a role. In the presence of noise, optimizing with regard to the final ground state overlap instead, is thus advisable.\\
For the \emph{black box method}, in classical simulations with noise, we observed the gradient-based training to be not too well-behaving. Convergence to optimized paths that improve over naive QAA was often impossible even when only few bit flips occurred in the quantum circuit. Classical optimization routines which are more robust towards noise seem to be asked for here, i.e. a classical optimizer that is combined with an error mitigation technique so that noisy outputs of the quantum black box can be corrected.
\begin{figure}[htbp]
    \centering
    \includegraphics[width=.9\linewidth]{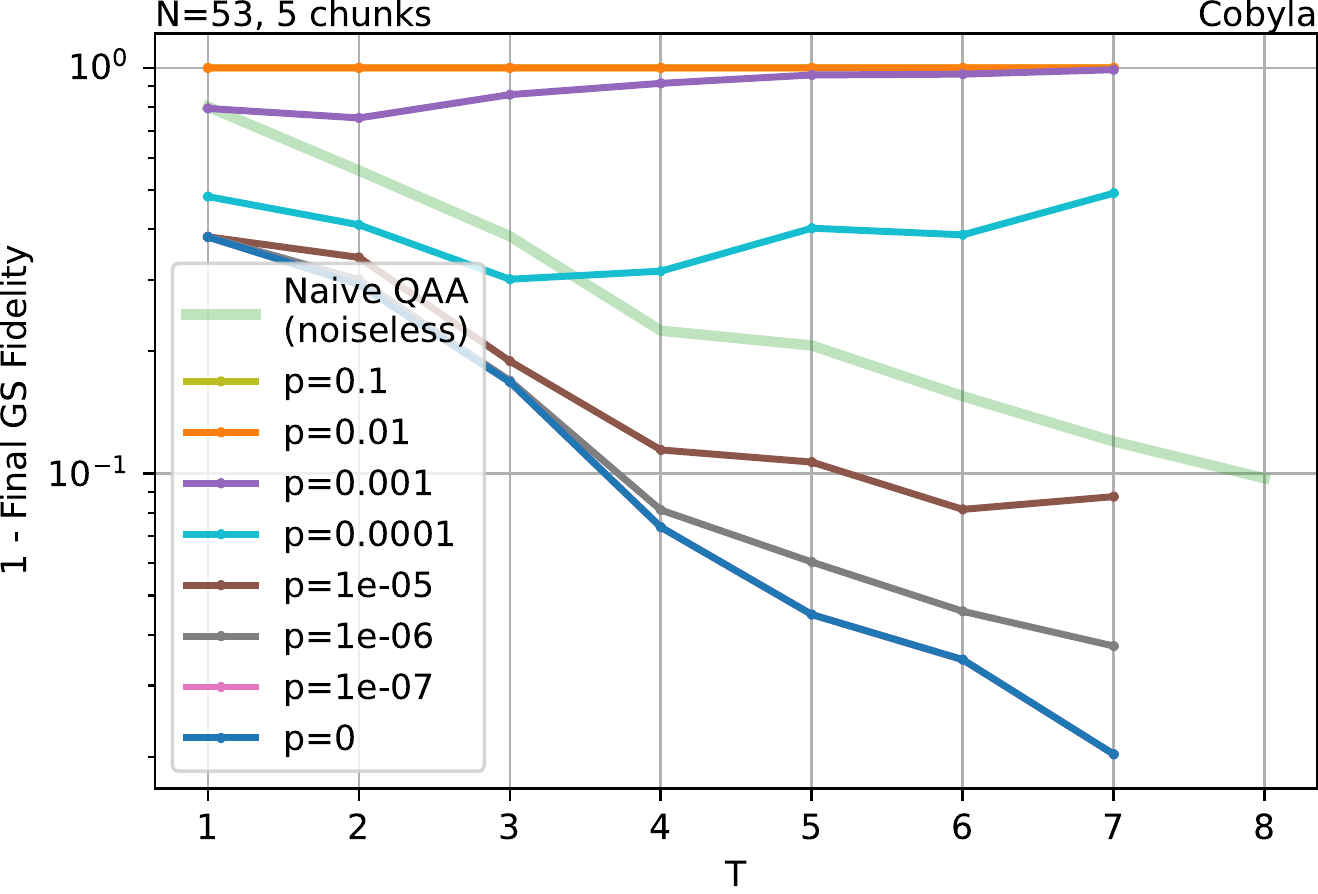}
    \caption{Noisy benchmark of the black box algorithm for $N=53$, $J=1$ and a maximum of 7 optimizer iterations. The noise strength $p$ determines the expected number of noisy qubits. Noise is applied in the circuit both in training as in the testing phase. The pink curve ($p=10^{-7}$) lies behind the blue curve due to the very rare noise events (cf.~Suppl.). No measurement noise was considered and naive QAA is given as a reference without noise. Cobyla was used as the classical optimizer.}
    \label{fig:Noise}
\end{figure}
We note, that gradient-free optimization methods are expected to be more robust to noisy environments and could provide better performance in an experimental setup than a gradient-based method. For this reason, we made use of the COBYLA method~\cite{powell1978fast}. In (Fig.~\ref{fig:Noise}), it can be observed that small amounts of noise significantly impair the training process of an optimized adiabatic path. For small noise strengths $p$, however, the results can improve even over naive QAA. For simulating the noisy quantum circuit, 100 MPS samples were taken. We refer to the Appendix for more details on the noise model (cf.~Suppl.).\\
Besides noise in the adiabatic evolution, noise also is present in the measurement process. For both the one-ancilla method and the two-ancilla method, we have benchmarked the black box routine numerically. The shot noise simulation is performed by finite sampling of $m$ independent measurements and taking the average of these measurements. From the numerical data, we can conclude that for a large system around 10.000 measurements can reduce sufficiently reduce the shot noise. This is one order of magnitude larger than the estimations made earlier in (Sec.~\ref{sec:mmt}). The benchmark for the one-ancilla method is shown in the main text, the plot for the two-ancilla method can be found in the Appendix.
\begin{figure}[htbp]
    \centering
    \includegraphics[width=.9\linewidth]{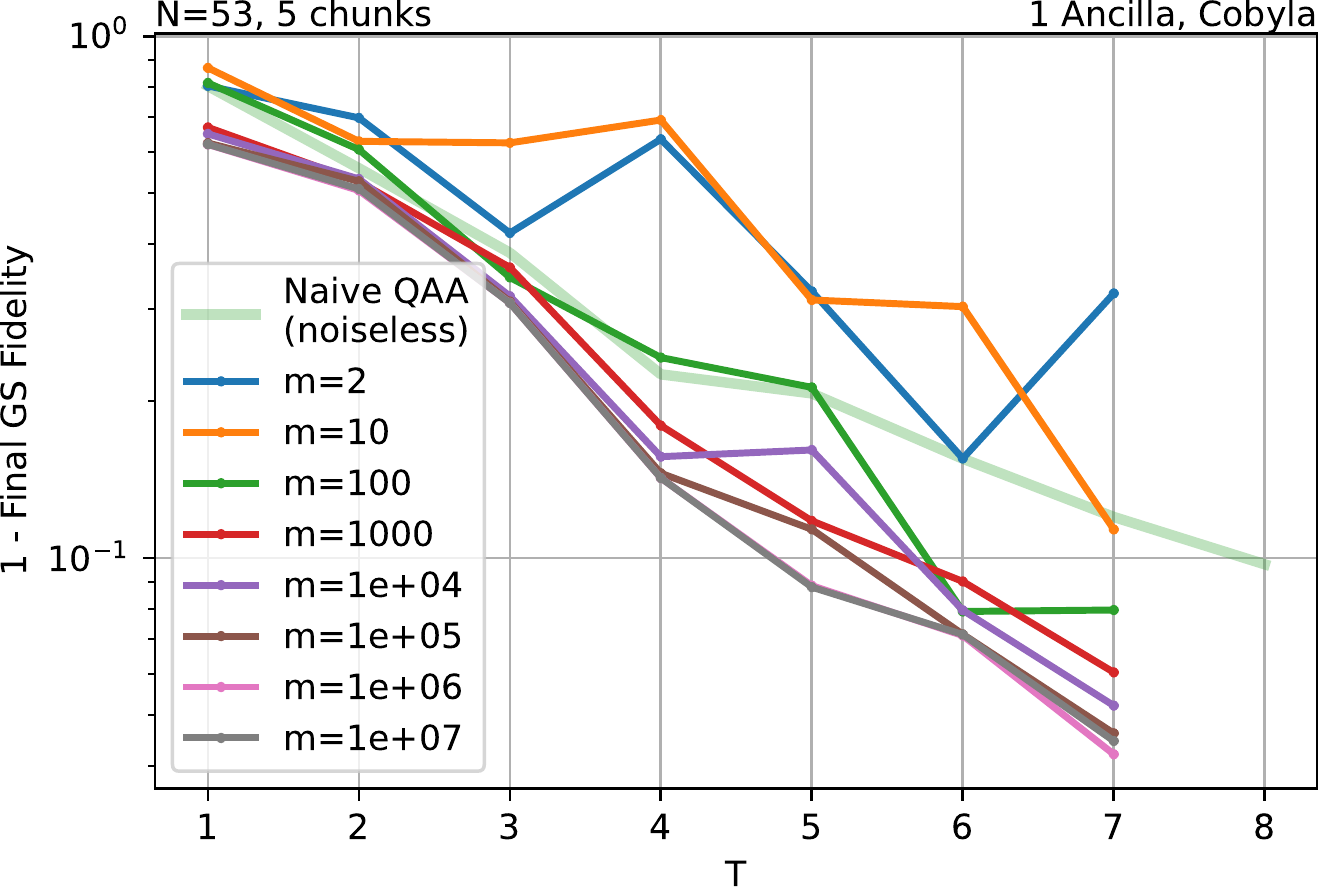}
    \caption{Simulation of shot noise in the black box algorithm for $N=53$, $J=1$ using the one-ancilla method. The number of measurements taken for every ground state overlap estimation is given by $m$. Note that no extra noise is applied during the circuit. Cobyla was used as the classical optimizer. For $m>10^3$ the results begin to converge and shot noise is sufficiently small.}
    \label{fig:Noise_Shot1}
\end{figure} \\
A further comment shall be specifically addressing the noise in the black box algorithm which makes use of a gradient to optimize the adiabatic path. In a recent work, it has been shown that the gradient in a variational quantum algorithm vanishes exponentially in the number of qubits $N$ when the number of layers scales as $\text{poly}(N)$~\cite{Wang2020}. These noise-induced barren plateaus severely hinder the scalability of variational quantum algorithms on NISQ devices. In our work, however, the algorithms either optimize the adiabatic path for fixed total evolution time (which includes the gradient-based black box algorithm), or have a maximum time budget in the case of the target fidelity profile algorithm. Thus, the (maximum) circuit depth is fixed in our approach which makes the results on noise-induced barren plateaus not directly applicable. 
\section{Discussion}
In this paper, we present a toolkit for quantum adiabatic computation. This toolkit includes a proposal for adiabatic spectroscopy, ancilla protocols for estimating the ground state overlap as well as the VQAA as a flexible yet powerful framework for variationally optimized adiabatic paths for high-fidelity ground state preparation.\\
The \emph{adiabatic spectroscopy} offers a straightforward approach to obtain information about an adiabatic spectrum. Our method relies on protocols to evaluate the closeness of an $N$-qubit quantum state $\ket{\psi}$ to an eigenstate using controlled unitary evolution. By remaining sufficiently adiabatic throughout the evolution, a self-consistent argument applies, enabling us to identify this eigenstate with the ground state. We note, that the error in the ancilla protocols presented is smallest when being close to an eigenstate. The protocol which we propose requires the ability to implement controlled time evolution on a single ancilla qubit. Current technology already meets the requisites of our protocol: Conditional dynamics have been explored in the context of trapped ion simulators and Rydberg atom arrays~\cite{lu2020algorithms, urban2009observation} and they can be implemented efficiently in a gate model.\\
A natural requirement for the spectroscopy is that the decoherence time of the quantum device is not a limiting factor to determine the evolution time required in order to reach the target overlap. The applications of the adiabatic spectroscopy go beyond obtaining the spectral gap for a given Hamiltonian $H(s)$. By aggregating spectral information from several adiabatic paths which cut through the phase diagram of a target Hamiltonian $H_T$, rich properties of quantum many-body systems might be acquired. \\
We note that the numerical results presented in (Fig.~\ref{fig:AdSpec}) support our argument that this technique is suitable to derive information about the spectral gap. However, the relation $\partial T(s)/\partial s \sim  1/\Delta(s)^2$, which is obtained from the Landau-Zener model, is only a first order approximation. Improvements of the quantitative validity of the adiabatic spectroscopy are left for future work and might build upon the rich literature on adiabatic perturbation theory~\cite{rigolin2008beyond}.\\
We shall now discuss the \emph{VQAA} from two perspectives. First, the perspective of VQAA as a quantum algorithm for optimal adiabatic paths, and second, VQAA as a variational quantum algorithm which requires only few measurements.\\
Adiabatic quantum computation is known to prepare the ground state of target Hamiltonian $H_T$ for sufficiently long preparation time $T$. However, $T$ scales as a function of the minimal spectral energy gap $\Delta(s)$. For general $H_T$, the best rigorous bound on $T$ has a inverse cubic gap dependence $T=\mathcal{O}((\min_s\Delta(s))^{-3})$~\cite{jansen2007bounds}. Due to the limited decoherence time of NISQ devices, large evolution times necessary for high-fidelity ground state preparation can be a difficulty. If it was possible to reduce $T$ to fit into the coherent time frame of a quantum device, it would become possible to adiabatically prepare ground states, e.g.~the solution of an optimization problem, that have remained unattainable before.\\
The question of finding an optimal path for the adiabatic evolution has been in the focus of research efforts for several years already and for the unstructured search problem, the Grover-type speed-up has been recovered using an optimized adiabatic sweep~\cite{roland2002quantum}. However, the position and size of the spectral gap are, in general, a priori unknown, and obtaining the spectral properties of the adiabatic path can be as hard a problem as preparing the ground state. Therefore, it remains a challenge how an optimized adiabatic path can be obtained when no or only little spectral information is available. One recent work employed techniques from reinforcement learning to find an optimal adiabatic path~\cite{lin2020quantum}. In our work, we phrase the problem of finding an optimized adiabatic path as a problem to be solved through a variational quantum algorithm with step-wise adiabatic velocity profile. \\
Turning now towards a discussion of the VQAA as a variational algorithm, we begin by noting that the ground state overlap can be a suitable cost function for quantum-classical feedback loops. Variational approaches such as the quantum approximate optimization algorithm (QAOA)~\cite{farhi2014quantum} sparked intensive research interest in recent years. The number of measurements necessary to estimate an objective function scales with $\mathcal{O}(\epsilon^{-2})$, where $\epsilon$ is the maximum error that can be tolerated in the optimization process. The proposal of the VQAA aims to reduce the number of measurements by requiring relatively fewer parameters that need to be optimized and by considering a cost function with a low variance.\\
Variational approaches for preparing a ground state generally make use of the energy as a cost function and that energy is estimated via the local observables that compose the Hamiltonian~\cite{bonet2020nearly}. As the actual ground state will generally not be an eigenstate of the Hamiltonian local terms (e.g.~if there is frustration), a low variance in the estimates cannot be guaranteed.
Moreover, the orthogonal eigenstates in the low-energy sector typically yield similar energy values, hindering convergence to the ground state. If we were indeed able to directly measure in the eigenbasis of the Hamiltonian at a given point in the adiabatic path, we would seek to exploit the property of proximity to an eigenstate, which is inherent to adiabatic algorithms. The textbook approach for this problem would be a quantum phase estimation (QPE) algorithm, and direct implementation requires an ancilla overhead~\cite{nielsen2002quantum}. Recent semi-classical approaches are able to use a single ancilla only by utilising post-processing schemes~\cite{santagati2018witnessing, obrien2020error}. 
For the VQAA, we suggest the overlap with the ground state as a figure of merit. 
In the case of the adiabatic algorithm, the optimal value of some other figures of merit, such as the energy, are not directly accessible.
Therefore, we present two protocols to evaluate the closeness to an eigenstate using controlled unitary evolution. The entangled ancilla protocol offers the possibility to perform low-variance measurements by harnessing the power of hypothesis testing when being close to an eigenstate.
We note that in the case of a small spectral gap, e.g.~when a phase transition is crossed, the ground state overlaps are in general very small and special care is needed to extract useful information with the ancilla protocol. \\
In the limit of very large depth, QAOA has the possibility to recover a trotterized adiabatic evolution. Therefore, a black box VQAA algorithm bears some similarities to QAOA. Several key differences are remarked though.
First, analog to the evolution times in a (trotterized) adiabatic evolution, the unitaries in QAOA feature \emph{angles} as parameters.  However, for a quantum cost function $H_T$, the optimized angles could be too large for the decoherence limit of the NISQ device which the algorithms is supposed to be implemented on. On the contrary, limiting the maximum angles could be rather problematic for the performance of QAOA. This issue does not arise in the black box VQAA as the total evolution time has been fixed.
Second, even for large system sizes with 53 or 100 qubits, VQAA significantly improves over naive QAA for a very small number of parameters $L$ only. In QAOA, it is generally expected that deep circuits are necessary to obtain good results for large systems. 
Finally, even for very small $L$, the performance of VQAA is lower bounded by the QAA with a linear time profile. This is true for QAOA only when the angles are initialized akin to a trotterized adiabatic evolution which requires large depth QAOA.
\section{Summary and outlook}
We seek to combine the best from two worlds by combining the strengths of the adiabatic and the variational approaches. We present a toolbox for VQAA building upon ancilla-based methods to evaluate the ground state fidelity at any point in the adiabatic path. Our approach only obtains information about the proximity to the ground state and is deliberately oblivious to the actual value of the energy throughout the adiabatic path.
Due to the small parameter space and the ground state overlap as our cost function, the number of measurements necessary in the optimization of the adiabatic evolution is dramatically lower than for typical variational quantum algorithms such as QAOA. On the whole, our work suggests that a further exploration of NISQ algorithms based on variational adiabatic concepts is indicated. \\
For instance, there seems to be room for sequential VQAA algorithms. In some instances, when going towards larger $T$, the black box VQAA does not improve in a strictly monotonous manner. By reusing information from shorter $T$, the optimized paths for larger $T$ might be improved and obtained more rapidly. In a similar direction, it would be interesting to see how the information gathered through adiabatic spectroscopy will be used for the optimal adiabatic path in an experimental setting. If it was possible to use adiabatic spectroscopy to eliminate or drastically reduce the quantum-classical training phase of VQAA, this would surely be a large advancement in NISQ algorithms for ground state preparation. Also, it remains to be seen, what bounds can be established on how quickly and closely VQAA can find an adiabatic path which is optimal.\\
Furthermore, there is increasing research interest in error mitigation techniques for NISQ devices~\cite{koczor2020exponential, huggins2020virtual}, and a further exploration of how VQAA might benefit from these techniques appears promising. More generally, the protocols for estimating the ground state overlap presented in this work could make a wide range of new, exciting quantum algorithms for ground state preparation possible. This might include the opportunity for an algorithm to find optimized adiabatic paths using techniques from reinforcement learning or a combination of the protocols with techniques such as projected measurements and the quantum Zeno effect~\cite{boixo2009eigenpath}. Besides, in the regime where the time evolution is not strictly adiabatic, high final ground states fidelities might be achieved by not starting in the ground state of the initial Hamiltonian but in an appropriate superposition in the low energy sector of the initial Hamiltonian.
\begin{acknowledgments}
The authors thank M.C. Ba\~{n}uls, T. O'Brien, S. Lu, D. Wild, A. Sauer and F. Pollmann for insightful discussions. We acknowledge support from the ERC Advanced Grant QUENOCOBA under the EU Horizon 2020 program (grant agreement 742102) and from the Deutsche Forschungsgemeinschaft (DFG, German Research Foundation) under the project number 414325145 in the framework of the Austrian Science Fund (FWF): SFB F7104. BS was supported by the German Academic Scholarship Foundation (Studienstiftung des deutschen Volkes). 
\end{acknowledgments}
\vspace{1em}
\emph{Note added.} --- After the completion and submission of this manuscript, some of the variational adiabatic methods presented here have been tested experimentally using large arrays of neutral Rydberg atoms to prepare the ground state of the maximum independent set problem on unit disk graphs with disorder~\cite{ebadi2022quantum}.
\appendix
\section{Adiabatic state preparation in the Landau-Zener model} \label{App:LZ}
We seek to better understand adiabatic spectroscopy as presented in the main text. Therefore, we would like to obtain a qualitative relation between the spectral gap $\Delta(s)$ along the adiabatic path and the evolution time $T(s^*)$ required to prepare the ground state of $H(s^*)$ with a given target fidelity. In order to do so, we turn towards the well-known Landau-Zener (LZ) model which describes a simple two-level system~\cite{landau1932theorie, zener1932non}.
The model Hamiltonian is given as
\begin{align}
    H = \lambda(t) \sigma^z + g \sigma^x = \begin{pmatrix} \lambda(t) & g\\ g & -\lambda(t) \end{pmatrix}.
\end{align}
With $\tan \theta=g/\lambda(t)$, we write the eigenvectors as 
\begin{align}
    \ket{a} = \begin{pmatrix} \sin (\theta / 2) \\ -\cos (\theta / 2)  \end{pmatrix} \quad \text{and} \quad
    \ket{b} = \begin{pmatrix} \cos (\theta / 2) \\ \sin (\theta / 2)  \end{pmatrix}.
\end{align}
The eigenenergies are given as $E_\pm = \pm \sqrt{\lambda(t)^2 +g^2}$ and the coupling is assumed to be a linear function in time $\lambda(t)=\delta t$. This implies that the minimum of the spectral gap (the avoided level-crossing) is found at $t=0$, i.e. an adiabatic evolution parametrized by $s$ from 0 to the position of a small gap $s^*$ is understood to be mapped onto the LZ evolution from very small initial $t$ to $t=0$. \\
A perturbative approach yields the probability
\begin{align}
    |\alpha_+(t_f)|^2 \approx \frac{\delta^2}{16 g^4} \left( \frac{g^6}{(g^2+\lambda(t_i)^2)^3} + \frac{g^6}{(g^2+\lambda(t_f)^2)^3}\right)
\end{align}
to find the system in the excited state at $t_f$ after initializing in the ground state at $t_i$~\cite{de2010adiabatic}. Assuming the beginning of the adiabatic evolution at $t_i=-\infty$, we are given
\begin{align}
    |\alpha_+(t_f)|^2 \approx \frac{\delta^2}{16 g^4} \frac{g^6}{(g^2+\lambda(t)^2)^3} = \frac{\delta^2 g^2}{16 E_+(t)^6}
\end{align}
where we identified $t$ with $t_f$. Now, we set a target fidelity $A_+^2 := |\alpha_+(t)|^2 $ and assume that the total evolution time $T$ scales as $T \sim 1/\delta$:
\begin{align}
    T \sim \frac{1}{\delta} \approx \frac{g}{4A_+E_+(t)^3}. \label{eqn:LZdelta}
\end{align}
As we are interested in the change of $T$, we compute the time-derivative of $T$
\begin{align}
    \dot T \sim -\frac{3}{4}\frac{g \delta}{A_+ E_+(t)^5} = -3\frac{1}{E_+(t)^2} \label{eqn:lasteqn}
\end{align}
where we used (Eqn.~\ref{eqn:LZdelta}) in (Eqn.~\ref{eqn:lasteqn}). Because of $\Delta(t) = 2E_+(t)$ in the LZ model, we obtain $\dot T \sim 1/\Delta(t)^2$ as an approximation to the scaling of $\dot T$ up to coefficients and possible corrections.
\section{Calculations for the one-ancilla and entangled-ancillas protocol}
\subsection{Single-ancilla protocol}
\label{App:DM}
The combined system after the unitary evolution can be written as 
\begin{align}
    \text{C-}U \ket{\psi}\ket{+} = \frac{1}{\sqrt{2}}\left[\ket{\psi}\ket{0} + U\ket{\psi}\ket{1}\right] =: \ket{\zeta}.
\end{align} 
Denoting the quantum state after the unitary evolution as $\ket{\psi_\text{evo}}=U\ket{\psi}$, we note the following relationships regarding Pauli measurements of the ancilla
\begin{align}
    & \braket{\zeta| \mathbb{1} \otimes \sigma_x | \zeta} \\
    = & \frac{1}{2} \left[\bra{0} \bra{\psi}U\ket{\psi}\ket{0} + \bra{1}\bra{\psi}U^\dag \ket{\psi}\ket{1} \right] \\
    = & \frac{1}{2} \left[\braket{\psi | \psi_\text{evo}} + \braket{\psi_\text{evo} | \psi}\right] 
    = \text{Re}(\braket{\psi| \psi_\text{evo}})
\end{align}
and
\begin{align}
   & \braket{\zeta| \mathbb{1} \otimes \sigma_y | \zeta} \\
    = & \frac{i}{2} \left[ \bra{0}\bra{\psi}U\ket{\psi}\ket{0} - \bra{1}\bra{\psi}U^\dag \ket{\psi}\ket{1} \right] \\
    = & \frac{i}{2} \left[\braket{\psi | \psi_\text{evo}} - \braket{\psi_\text{evo} | \psi}\right] 
    = -\text{Im}(\braket{\psi| \psi_\text{evo}})
\end{align}
so that 
\begin{align}
\braket{\psi | \psi_\text{evo}} = \braket{\sigma_x - i \sigma_y}_\text{ancilla} = \left\langle\begin{pmatrix}
0 & 0\\
2 & 0
\end{pmatrix}\right\rangle_\text{ancilla}.
\label{eqn:overlap}
\end{align}
For a fixed Hamiltonian $H = \sum_j E_j \ket{\phi_j}\bra{\phi_j}$ with the unitary $U \ket{\phi_j} = e^{-iH\tau}\ket{\phi_j} = e^{-iE_j\tau}\ket{\phi_j}$, we write the state $\ket{\psi} = \sum_j \psi_j \ket{\phi_j}$ in the eigenbasis of $H$ with $\ket{\psi}$ normalized ($\sum_j |\psi_j|^2 = 1$). The total quantum system can be expressed as
\begin{align}
    \ket{\zeta} =& \frac{1}{\sqrt{2}}\left[\ket{\psi}\ket{0} + U\ket{\psi}\ket{1}\right] \\
    %=& \frac{1}{\sqrt{2}}\left[\left(\sum_j \psi_j \ket{\phi_j}\right)\ket{0} + U\left(\sum_j \psi_j \ket{\phi_j}\right)\ket{1}\right] \\
    =& \frac{1}{\sqrt{2}}\left[\left(\sum_j \psi_j \ket{\phi_j}\right)\ket{0} + \left(\sum_j e^{-iE_j\tau} \psi_j \ket{\phi_j}\right)\ket{1}\right] \\
    =& \frac{1}{\sqrt{2}}\left[\sum_j \psi_j \ket{\phi_j} \left(\ket{0} +  e^{-iE_j\tau} \ket{1}\right)\right] 
\end{align}
which we now use to calculate the density matrix of the ancilla qubit
\onecolumngrid
\begin{align}
    \rho_{\text{ancilla}} = \text{Tr}_\text{non-ancilla} \left(\ket{\zeta}\bra{\zeta}\right)
    =& \frac{1}{2} \sum_{j, m} \text{Tr}_\text{non-ancilla} \left( \psi_j \psi_m^*  \underbrace{\left[\ket{\phi_j}\bra{\phi_m}\right]}_{\delta_{j, m}} \left[\ket{0} + e^{-iE_j\tau}\ket{1}\right] \left[\bra{0} + e^{iE_m\tau}\bra{1}\right] \right) \\
    =& \sum_j \frac{|\psi_j|^2}{2} \left[ \ket{0} + e^{-i E_j \tau} \ket{1} \right] \left[ \bra{0} + e^{i E_j \tau} \bra{1} \right] 
    = \sum_j \frac{|\psi_j|^2}{2} \begin{pmatrix}
    1 & e^{i E_j \tau}\\
    e^{-i E_j \tau} & 1\end{pmatrix}.  \label{eqn:rho_anc}
\end{align}
\twocolumngrid
\subsection{Entangled-ancillas protocol}
Our goal is to determine the purity of the ancilla. In general, there is the relation $\rho$ pure $\Leftrightarrow \text{Tr}[\rho^2] = \lambda_1^2 +\lambda_2^2$ for density matrices, with $\lambda_1$ and $\lambda_2$ the eigenvalues of $\rho$. We write the density matrix and its square as
\begin{align}
    \rho_\text{ancilla} = \begin{pmatrix}
    a & b\\ c & d
    \end{pmatrix} \quad \text{and} \quad
    \rho_\text{ancilla}^2 = \begin{pmatrix}
    a^2+bc & \cdot\\ \cdot & bc+d^2
    \end{pmatrix}
\end{align}
where matrix elements irrelevant for our protocol are denoted with a dot. 
With the Bell state 
\begin{align}
    \ket{\Phi^-} = \frac{1}{\sqrt{2}} \left( \ket{00}-\ket{11}\right)
\end{align}
we construct the Bell measurement operator 
\begin{align}
    \ket{\Phi^-}\bra{\Phi^-} = \frac{1}{2} \begin{pmatrix}
1 & 0 & 0 & -1\\
0 & 0 & 0 & 0\\
0 & 0 & 0 & 0\\
-1 & 0 & 0 & 1
\end{pmatrix}.
\end{align}
The diagonal matrix elements of $\rho_\text{ancilla}^2$ may then be attained by considering a composite system where the second system with controlled backward-time evolution (implementable by changing the sign of $H$). Then, the density matrix of the ancilla of the second system effectively corresponds to the transpose of the density matrix of the first ancilla $\rho_\text{ancilla2}=\rho_\text{ancilla1}^T$. With (Eqn.~\ref{eqn:rho_anc}), the composite system then gives
\begin{align}
    & \rho_\text{ancilla1} \otimes \rho_\text{ancilla2} = \rho_\text{ancilla1} \otimes \rho_\text{ancilla1}^T \\
    =& \begin{pmatrix}
    a^2 & \cdot & \cdot & bc\\
    \cdot & \cdot & \cdot & \cdot\\
    \cdot & \cdot & \cdot & \cdot\\
    bc & \cdot & \cdot & d^2
    \end{pmatrix} = 
    \frac{1}{4}\begin{pmatrix}
    1 & \cdot & \cdot & |\alpha|^2\\
    \cdot & \cdot & \cdot & \cdot\\
    \cdot & \cdot & \cdot & \cdot\\
    |\alpha|^2 & \cdot & \cdot & 1
    \end{pmatrix}.
\end{align}
If $\ket{\psi}$ is in an eigenstate, the density matrix of the ancilla $\rho_\text{ancilla}$ is pure and the expectation of the $\ket{\Phi^-}$ measurement is 
\begin{align}
    \langle\rho_\text{ancilla1} \otimes \rho_\text{ancilla2}\rangle_{\ket{\Phi^-}} = \frac{1}{4} (1-|\alpha|^2) = 0,
\end{align}
\subsection{Choosing suitable time values in the ancilla protocol} \label{App:tau}
\begin{figure}[htbp]
    \centering
    \includegraphics[width=.9\linewidth]{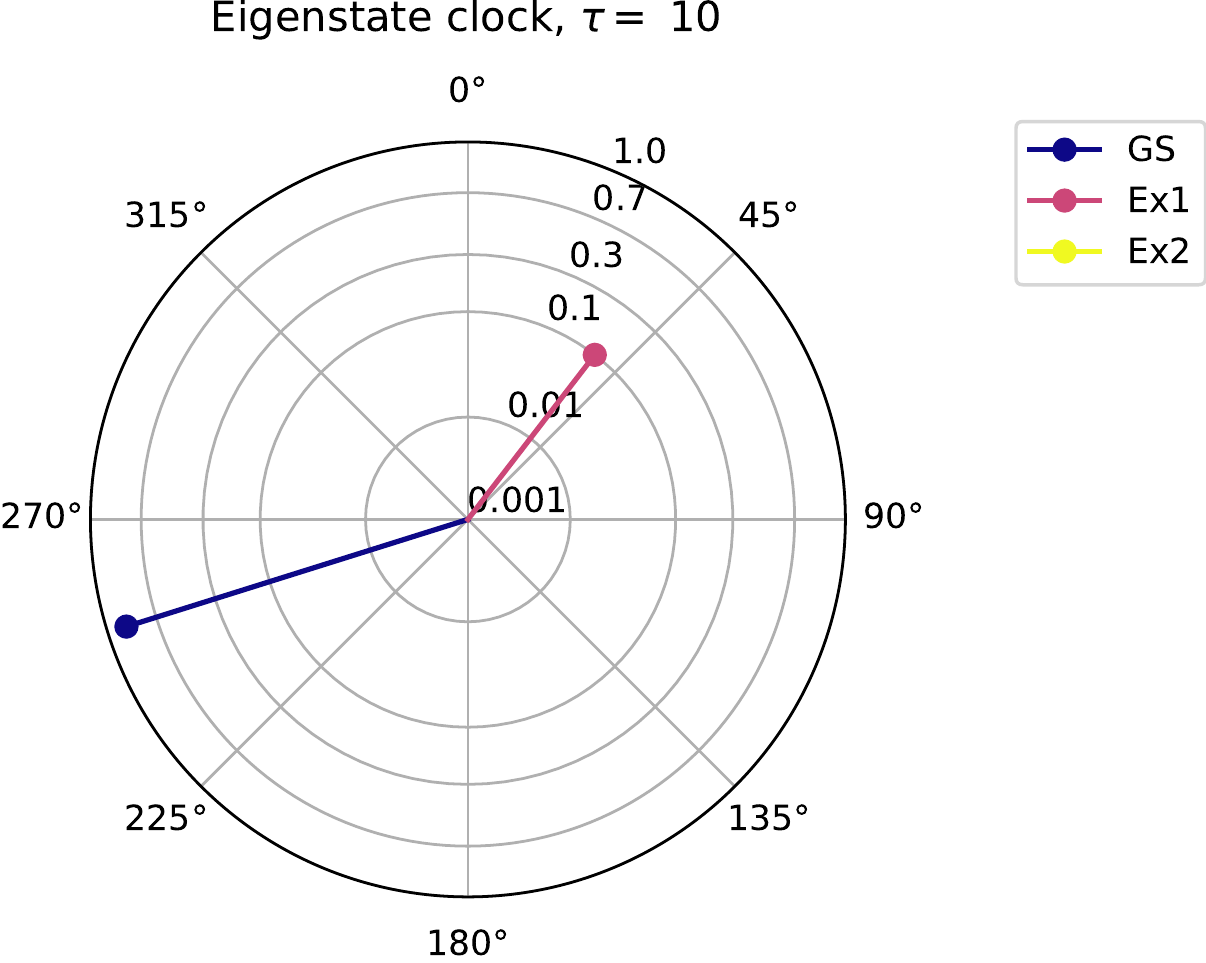}
    \caption{Eigenstate clock for a second larger value of $\tau=10$. For $\tau$=1 the summand corresponding to the ground state and the summand corresponding to the first excited state lie quite closely together in phase which leads to problematic constructive interference (Fig.~\ref{fig:clock}). However, as the pointers rotate with their respective eigenenergy for changing $\tau$, they are well-separated for $\tau=10$. Then, we find destructive interference in the summation which is necessary for the explanatory power of the protocol. Note that in this instance, the summands corresponding to higher excited eigenstates are very small and not visible.}
    \label{fig:clocks}
\end{figure}

We have argued that a measure for the eigenstate closeness is given by
\begin{align}
     \braket{\psi | \psi_\text{evo}} = \braket{\sigma_x - i \sigma_y}_\text{ancilla} = \sum_j |\psi_j|^2 e^{-i E_j \tau}  =: \alpha. 
\end{align}
using the one-ancilla protocol.
For suitable $\tau$, $\ket{\psi_\text{evo}}$ is an eigenstate of the fixed Hamiltonian $H$ only if $|\alpha|=1$. The time $\tau$ needs to be chosen so that the complex summands of $\alpha$ with non-vanishing amplitude do not have approximately equal phases. In this unlikely case of matching phases, we would see constructive interference so that $\alpha=1$ could be true even if $\ket{\psi_\text{evo}}$ is not an eigenstate. Visualizing the summands of $\alpha$ on a complex plane (Fig. \ref{fig:clocks}), this becomes rather intuitive. The choice for $\tau$ is related to the spectrum of $H$. It is reasonable to assume and confirmed in our simulations that a quantum state in our algorithm has the largest overlap with the ground state. As transitions to high energy excited states are extremely rare, for this argument we assume that the overlap with the first excited state was the only other non-vanishing overlap. Then, we simply have 
\begin{align}
    \alpha = |\psi_0|^2 e^{-i E_0 \tau} + |\psi_1|^2 e^{-i E_1 \tau}.
\end{align}
In the case of destructive interference 
\begin{align}
    |\alpha| = \min_\tau |\alpha(\tau)|
\end{align}
which corresponds to
\begin{align}
    e^{-i E_0 \tau} + e^{-i E_1 \tau} \overset{!}{=} 0
    \Leftrightarrow \tau(E_1 - E_0) =& \pi (2k+1) \\
    \Leftrightarrow \tau =& \frac{\pi}{\Delta} (2k+1)
\end{align}
with $k\in\mathbb{Z}$ and $\Delta=E_1-E_0$. An arbitrary $\tau$ would correspond to choosing an $l\in \mathbb{Z}$ in $\tau = \pi l/\Delta$ at random. For $l\gg 1$, the probability of choosing an odd value of $l$ is approximately 1/2. Therefore, by testing several random values of $\tau \in O(\Delta^{-1})$ it is possible to deduce information about the system whether it is in a mixed state or an eigenstate with high confidence.
\subsection{Bound for the single-ancilla protocol} \label{App:Bound}
We extend the discussion about suitable values of $\tau$. As argued above, an answer is generally dependent on the spectral properties of $H$, which are not available a priori.\\
We shall provide a bound for the single ancilla method without any knowledge about the spectrum of $H$ or the populations of the different eigenstates. A byeffect of this generality is that the bound is not tight. This is because in a realistic setting of a (quasi-)adiabatic evolution, the eigenstate populations of higher excites are expected to be very small.\\
We consider the expectation value \footnote{In writing the expectation value, we make use of the (unnormalized) sinc function which is defined as $\sinc (x):= \sin(x) / x$.} of $\alpha(\tau)$
\begin{align}
    & \mathbb{E}_{\tau\: \sim \text{ unif. dist. in } [0, K]} (|\alpha(\tau)|^2) \\
    &= \frac{1}{K} \int_0^K |\alpha(\tau)|^2 d\tau \\
    &= \frac{1}{K} \int_0^K \sum_{i,j} |\psi_i|^2 |\psi_j|^2 e^{-iE_i\tau} e^{-iE_j\tau} d\tau \\
    &= \sum_{i,j} |\psi_i|^2 |\psi_j|^2 \sinc((E_i - E_j)K).
\end{align}
In the limit of large $K$, we obtain
\begin{align}
    \lim_{K\rightarrow\infty} \mathbb{E}_{\tau\: \sim \text{ unif. dist. in } [0, K]} (|\alpha(\tau)|^2)= \sum_i |\psi_i|^4 =: E^2
\end{align}
where the spectral dependence has entirely averaged out. Such an $E^2$ corresponds to what one would observe in the laboratory. We note that $|\psi_0|^2$ was the ground state overlap of $\ket{\psi}$. As $\ket{\psi}$ is normalized, we can write 
\begin{align}
    |\psi_i|^2 \leq 1 - |\psi_0|^2 \quad \forall i > 0,
\end{align}
i.e. for all $i$ which do not correspond to the ground state. Then, multiplying with $|\psi_i|^2_{i>0}$ yields 
\begin{align}
    |\psi_i|^4 \leq \left(1 - |\psi_0|^2\right) |\psi_i|^2 \quad \forall i > 0.
\end{align}
As the latter inequality holds for all $i>0$, we obtain
\begin{align}
    \sum_{i=1} |\psi_i|^4 \leq \left(1 - |\psi_0|^2\right) \sum_{i=1} |\psi_i|^2 = \left(1 - |\psi_0|^2\right)^2.
\end{align}
For the expectation value $E^2$ we can now write down the inequality
\begin{align}
    E^2 = |\psi_0|^4 + \sum_{i=1} |\psi_i|^4 \leq |\psi_0|^4 + \left(1 - |\psi_0|^2\right)^2.
\end{align}
Solving for $|\psi_0|^2$, we obtain
\begin{align}
    |\psi_0|^2 \geq \frac{1}{2} + \frac{1}{2}\sqrt{2E^2-1},\quad \text{or} \quad
    |\psi_0|^2 \leq \frac{1}{2} - \frac{1}{2}\sqrt{2E^2-1}. \label{eqn:ancillainequ}
\end{align}
So far we have not assumed anything about the populations. Letting the ground state population $|\psi_0|^2$ be the largest of the eigenstate populations, we have
\begin{align}
    |\psi_0|^2 \geq \frac{1}{2} + \frac{1}{2}\sqrt{2E^2-1}
\end{align}    
as a lower bound for $|\psi_0|^2$ for $E^2 \geq 1/2$. Also, (Eqn.~\ref{eqn:ancillainequ}) implies that in this case the smaller ground state populations are therefore upper bounded as 
\begin{align}
    |\psi_i|^2 \leq \frac{1}{2} - \frac{1}{2}\sqrt{2E^2-1} \quad \forall i > 0.
\end{align}
For $E^2 < 1/2$, no non-trivial bound for $|\psi_0|^2$ (apart from $|\psi_0|^2 \in [0,1]$) can be given in the limit of very large $K$ (with this approach).
\section{Plots on the number of measurements and noise}
\begin{figure*}[!htbp]
    \centering
    \includegraphics[width=.7\textwidth]{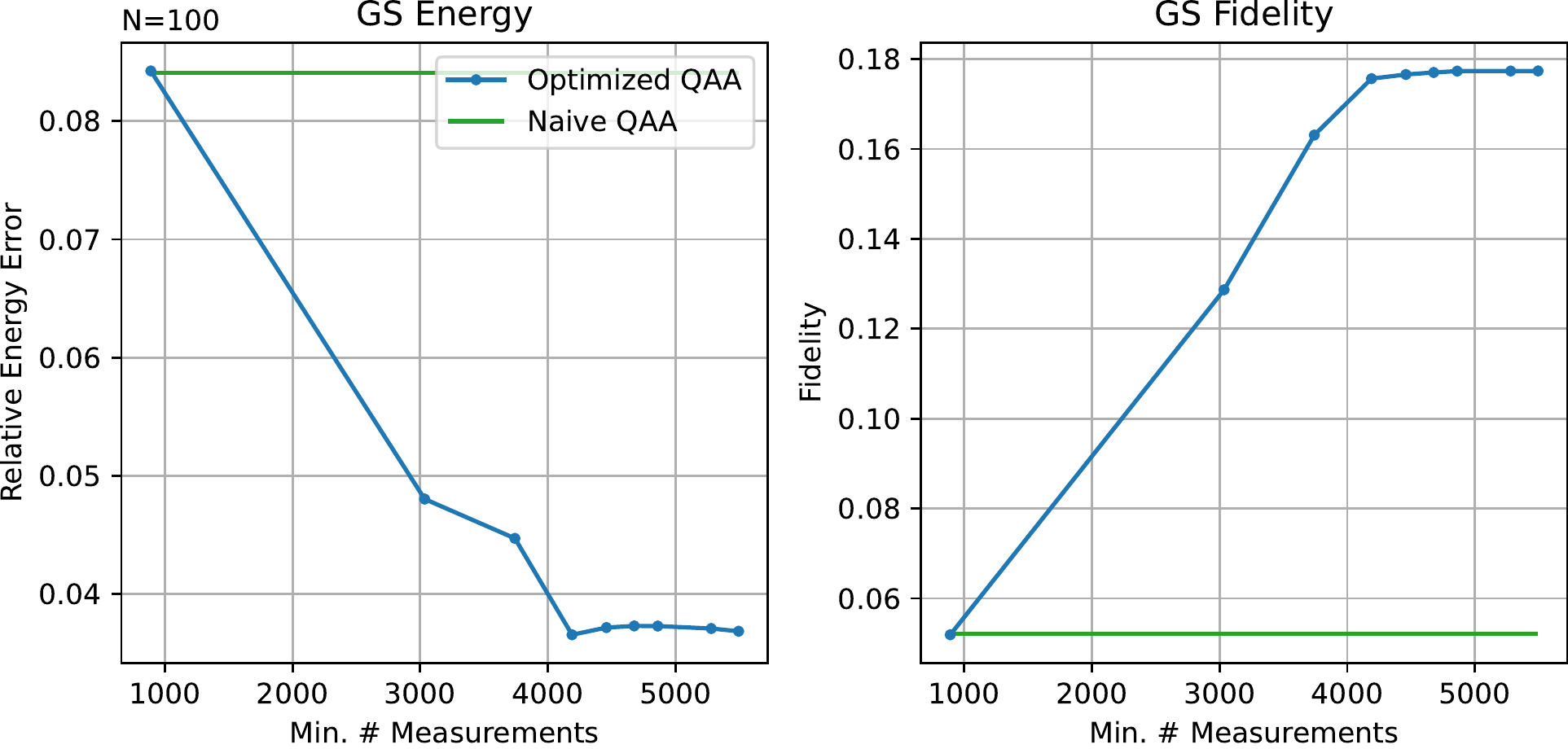}
    \caption{The number of measurements for this explicit instance obtained using the Chernoff-Hoeffding inequality and the assumptions mentioned in the main text. Already for few thousand total measurements, the end result of the adiabatic evolution is significantly improved. The data corresponds to the first eight iterations of an instance of the black box optimizer with five chunks for fixed total time $T=1$ and $N=100$.}
    \label{fig:measment}
\end{figure*}

\begin{figure*}[hbtp]
    \centering
    \subfloat[The energy density of states of the ZZXZ model for $N=12$. Low energy spectral lines are rare for physical instances.]{%
    \includegraphics[width=.9\columnwidth]{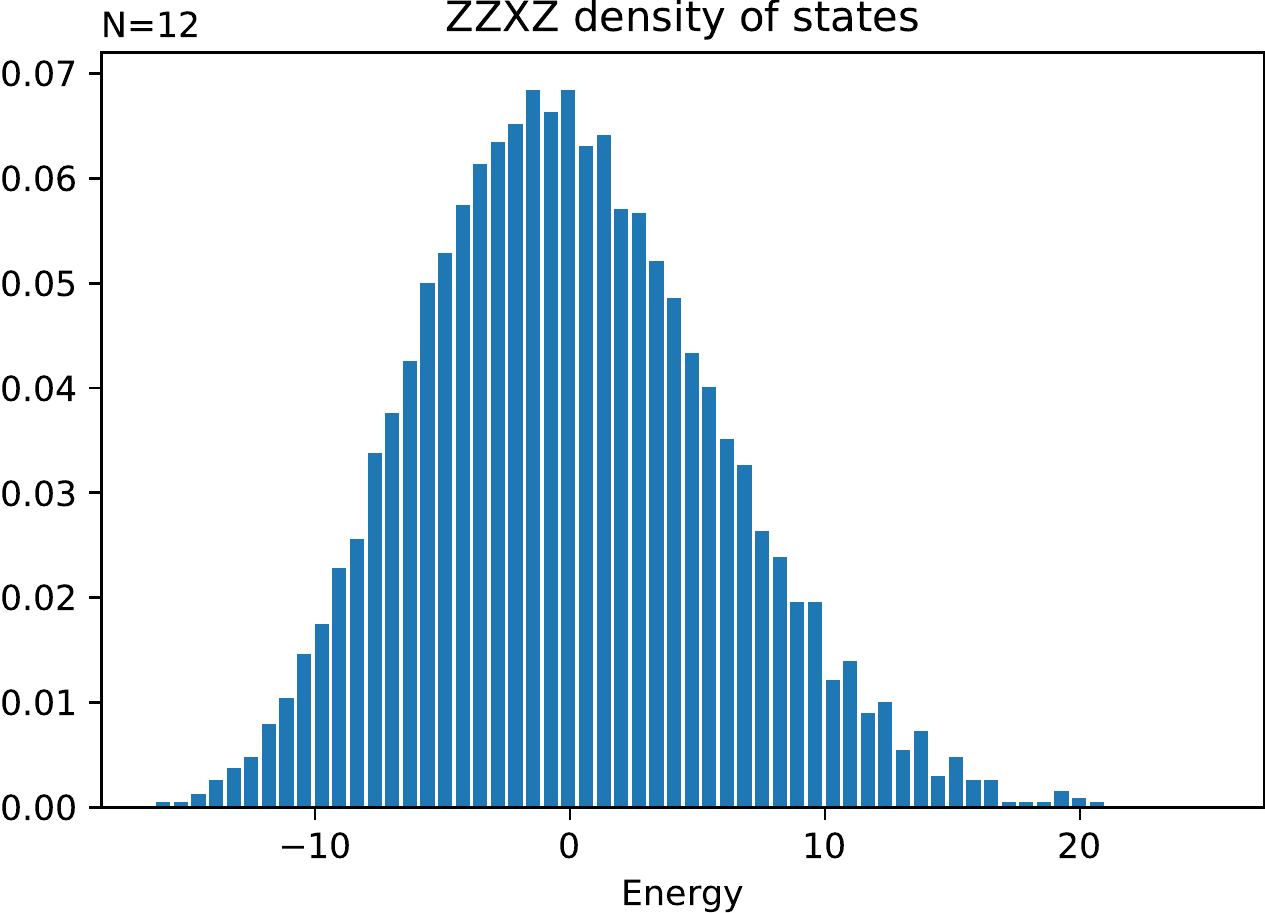}
    \label{fig:DOS}
    }\hspace{2em}
    \subfloat[Relative energy difference with ZZXZ ground state obtained with DMRG for adiabatic ground state preparation. The total time is fixed to $T=1$ corresponding to 16 discrete adiabatic steps. Pauli noise is applied either before the first unitary (step 0) or after each unitary to the center qubit of the spin chain. We observe that there is no qualitative difference at which position in the adiabatic path noise occurs. Note that for a $\sigma^x$ noise gate, there is no difference in the energy at the beginning of the adiabatic path because the initial Hamiltonian $H_0$ commutes with $\sigma^x$.]{%
    \includegraphics[width=.9\columnwidth]{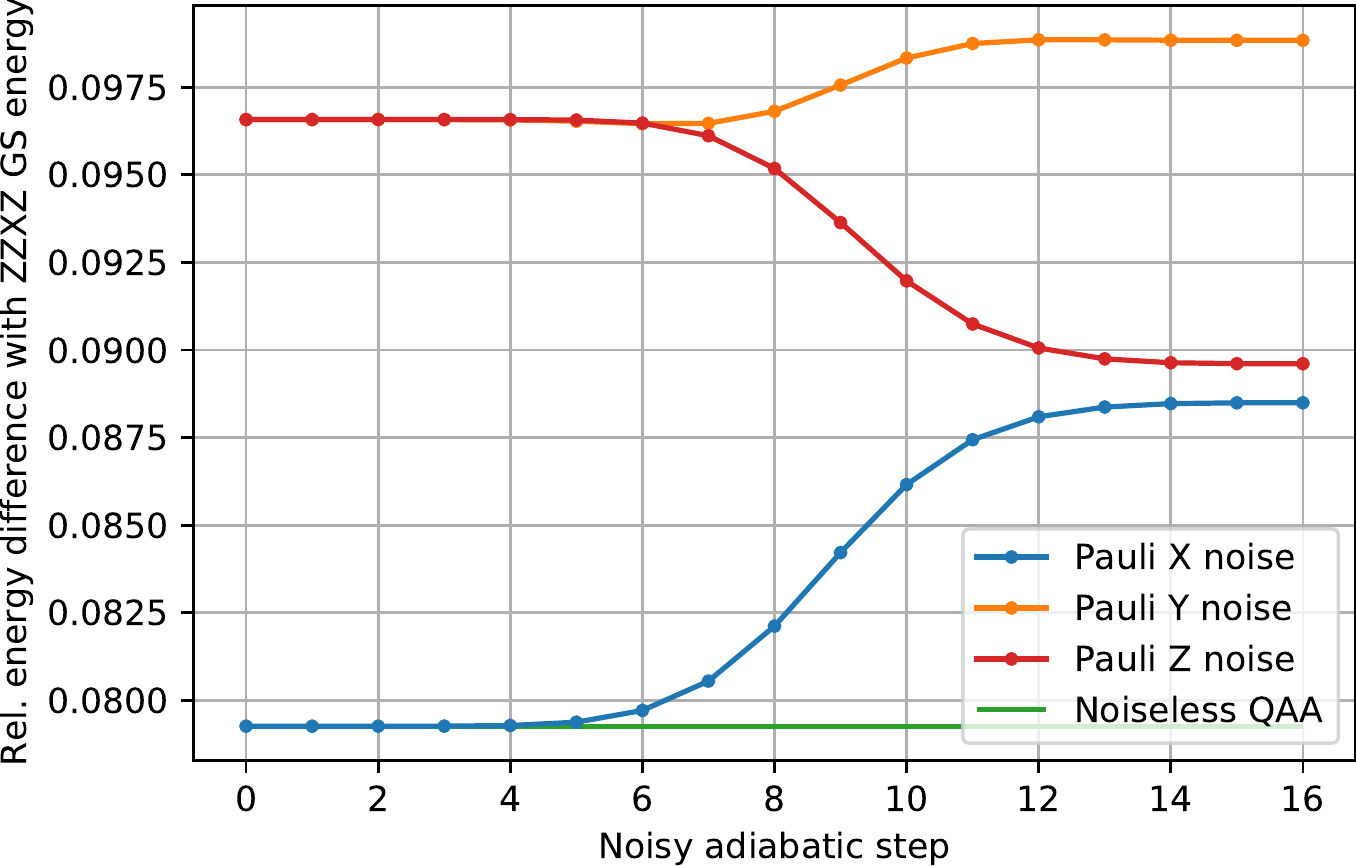}
    \label{fig:adiabnoise}
    }
    \caption{Further plots on the robustness of the adiabatic evolution.}
\end{figure*}

We include plots showing the number of ground state overlaps for an instance of the black box algorithm for $N=100$ and five chunks. Also, we present an estimate of the number of measurements needed using the Chernoff-Hoeffding inequality (Fig.~\ref{fig:measment}). 
\begin{figure*}[!htbp]
    \centering
    \includegraphics[width=.7\textwidth]{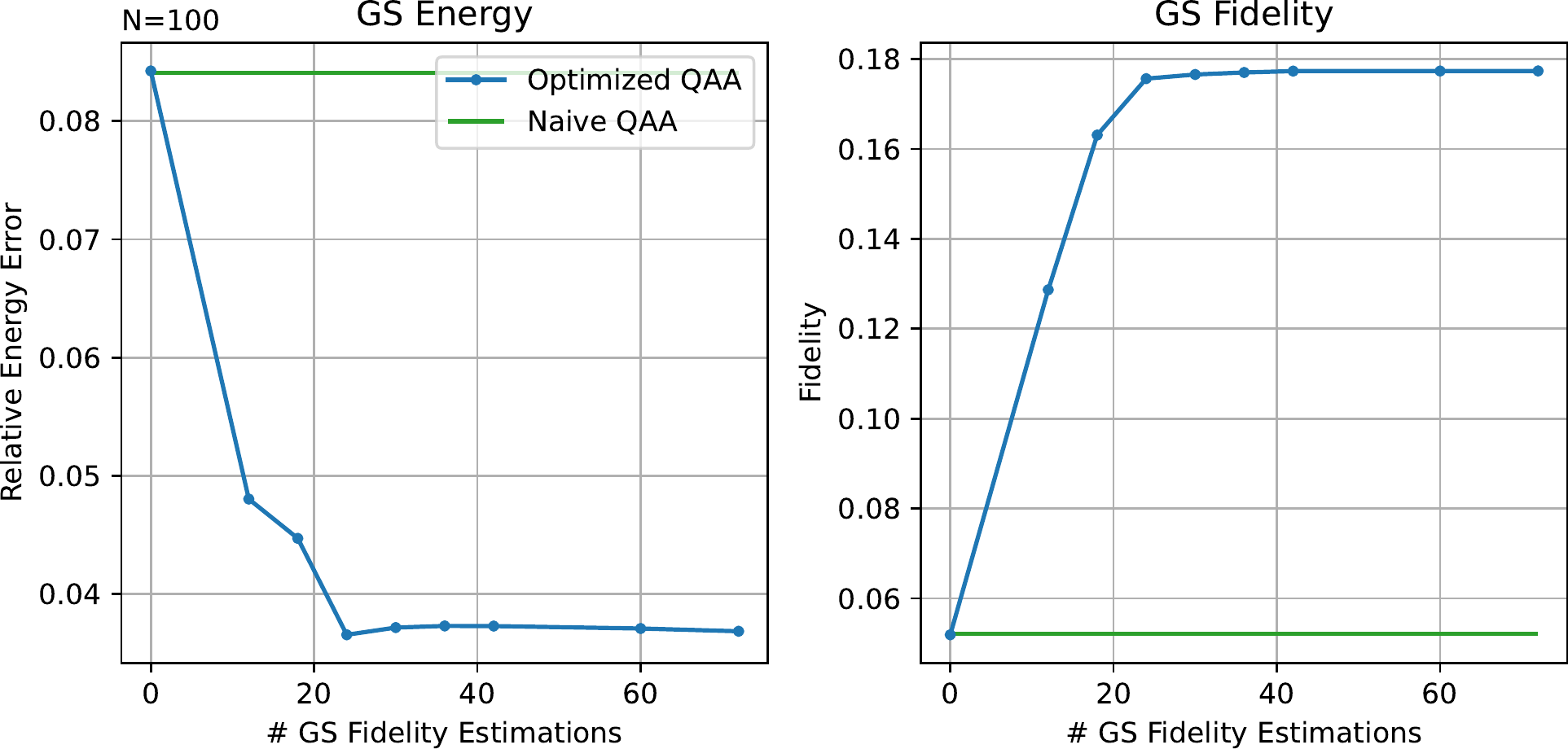}
    \caption{Number of iterations required for this explicit instance. The optimizer algorithm is L-BFGS-B without any further adjustments. Very good results are already obtained after three iterations and less than 25 ground state overlap evaluations. The data corresponds to the first eight iterations of an instance of the black box optimizer with five chunks for fixed total time $T=1$ and $N=100$.}
    \label{fig:measment1}
\end{figure*}
For a discussion of the inherent robustness of adiabatic algorithm, we show a the energy density of states for $N=12$ for the ZZXZ model (Fig.~\ref{fig:DOS}) and an analysis of the relative energy error due to a noisy gate at different positions in the adiabatic evolution (Fig.~\ref{fig:adiabnoise}).
\begin{figure}[htbp]
    \centering
    \includegraphics[width=.9\linewidth]{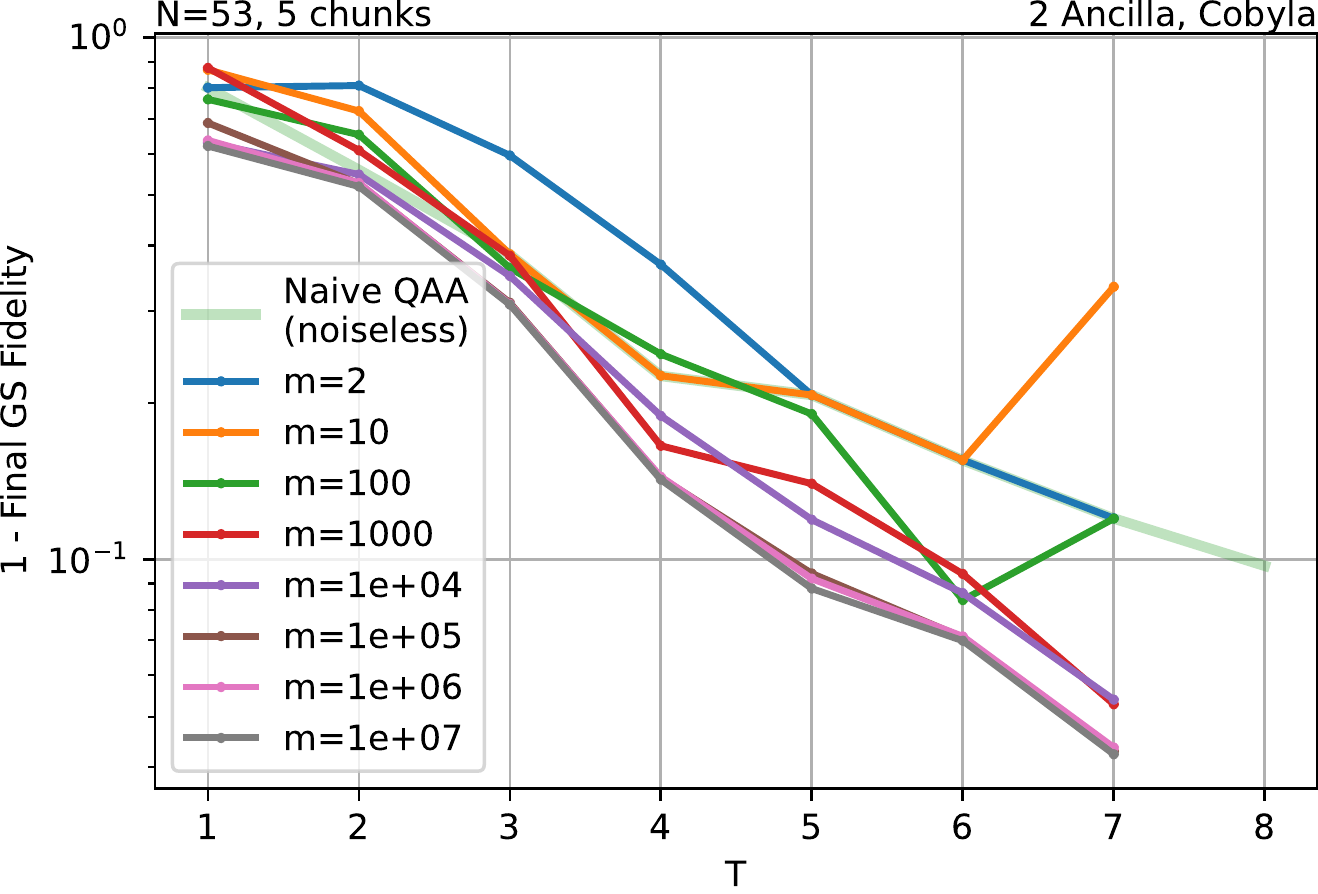}
    \caption{Simulation of shot noise in the black box algorithm for $N=53$, $J=1$. Here, the two-ancilla method is used. The number of measurements taken for every ground state overlap estimation is given by $m$. Note that no extra noise is applied during the circuit. Cobyla was used as the classical optimizer. For $m>10^3$ the results begin to converge and shot noise is sufficiently small.}
    \label{fig:Noise_Shot2}
\end{figure}

\bibliography{biblio}% Produces the bibliography via BibTeX.

%%%%%%%%%% Merge with supplemental materials %%%%%%%%%%
\clearpage
\widetext
\begin{center}
\textbf{\large Supplemental Materials: Adiabatic spectroscopy and a variational quantum adiabatic algorithm}
\end{center}
%%%%%%%%%% Merge with supplemental materials %%%%%%%%%%
%%%%%%%%%% Prefix a "S" to all equations, figures, tables and reset the counter %%%%%%%%%%
\setcounter{equation}{0}
\setcounter{figure}{0}
\setcounter{table}{0}
\setcounter{page}{1}
\makeatletter
\renewcommand{\theequation}{S\arabic{equation}}
\renewcommand{\thefigure}{S\arabic{figure}}
\renewcommand{\bibnumfmt}[1]{[S#1]}
%\renewcommand{\citenumfont}[1]{S#1}
%%%%%%%%%% Prefix a "S" to all equations, figures, tables and reset the counter %%%%%%%%%%
\onecolumngrid
\section{Quantum algorithms}
\subsection{Implementing the Quantum Adiabatic Algorithm}\label{Assec:QAA}
\begin{figure}[htbp]
    \includegraphics[width=.6\linewidth]{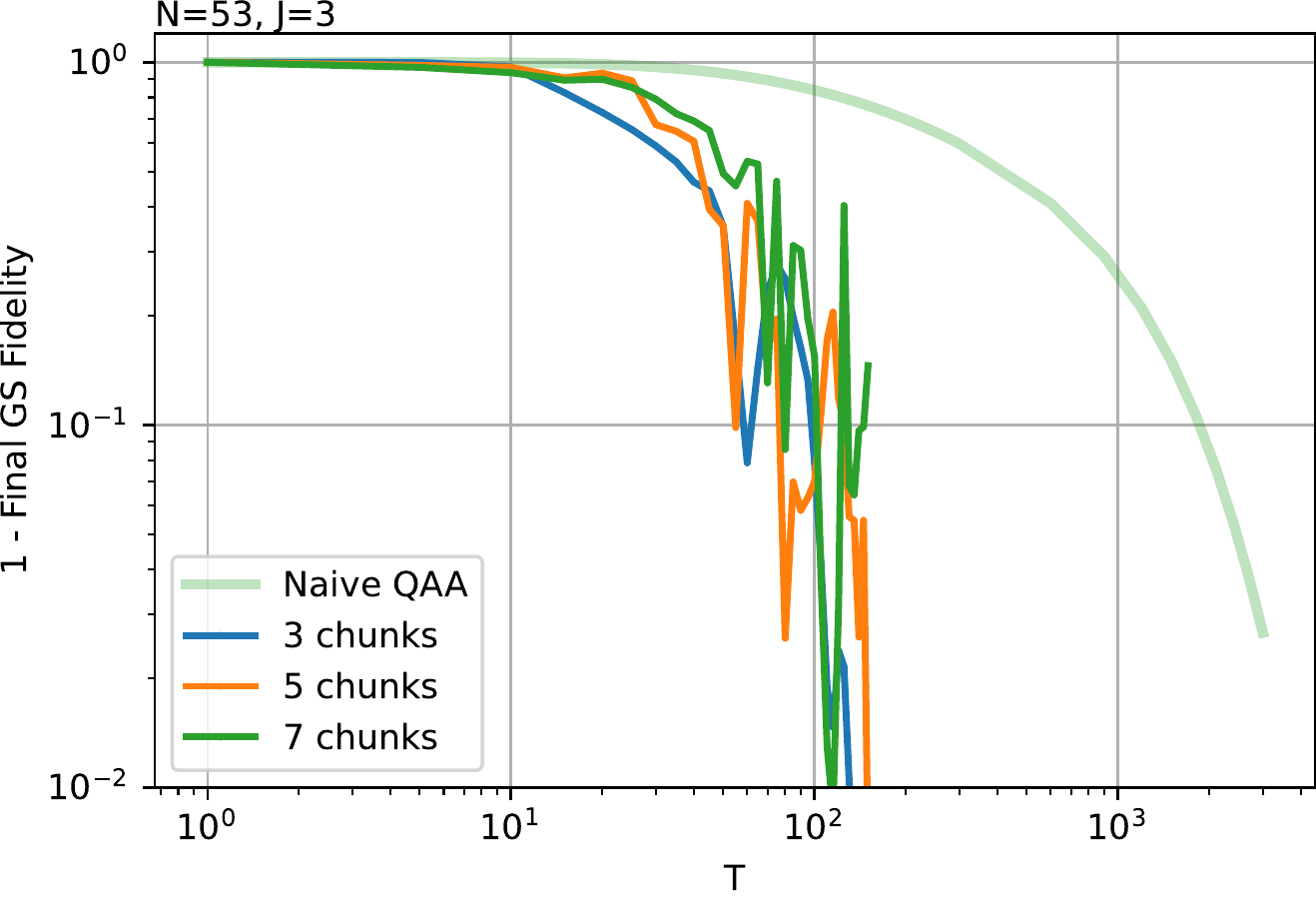}
    \caption{Black box VQAA results for 53 qubits and the ZZXZ model for the case of a crossed phase transition ($J=3$). Here, the adiabatic ramps have been smoothed using the reparametrization $\lambda(s)$. The results for smoothened naive QAA are worse than in Fig.~\ref{fig:BBBPT} leading to a larger separation between VQAA and QAA.}
    \label{fig:BBBPT_smooth}
\end{figure}

For a quantum state $\ket{\psi(t)}$ and a Hamiltonian $H(t)$, the time evolution of the state is given by the Schr\"odinger equation
\begin{align}
    i \frac{\partial}{\partial t} \ket{\psi(t)} = H(t) \ket{\psi(t)},
    \label{eqn:Schroedinger}
\end{align}
where we set the reduced Planck constant to $\hbar=1$. The quantum adiabatic algorithm~\cite{farhi2000quantum} describes how to prepare the unitaries that implement the path
\begin{align}
    H(s) = (1-s) H_0 + s H_T
\end{align}
from a trivial Hamiltonian $H_0$ to the problem Hamiltonian $H_T$. Here, $s$ is parametrized time $s=t/T$ for a total evolution time $T$. From (Eqn.~\ref{eqn:Schroedinger}) we have the unitary time evolution operator $U_{T, 0}$ which implements
\begin{align}
    \ket{\psi(T)}=e^{-i\int_0^T H(t) dt}\ket{\psi(0)} =U_{T, 0}\ket{\psi(0)}.    
\end{align}
On gate-model quantum computers, the adiabatic evolution needs to be approximated by discrete adiabatic steps. We write the unitary operator $U_{T, 0}$ as a product of $M$ unitaries
\begin{align}
    U_{T,0}=U_{T, T-\delta}U_{T-\delta, T-2\delta}\ldots U_{\delta, 0},
\end{align}
with $\delta=T/M$.  Here, we approximately assume the Hamiltonian $H(s)$ to be fixed in time during a time interval $\delta$:
\begin{align}
    U_{(l+1)\delta, l\delta} \approx e^{-i\delta H(l\delta)},
\end{align} 
for $l\in \{0,1,\dots, M-1\}$. In~\cite{farhi2000quantum}, the maximal size of $\delta$ is bounded as $\delta\ll 1/\|H_T-H_0\|$ in order for the approximation to be valid. However, we found a time step $\delta=1/16=0.0625$ to give sufficient accuracy for our simulations (cf.~\cite{banuls2006simulation}).\\
In the adiabatic approximation, higher order corrections can be partially suppressed by using a reparametrization with smooth derivatives (corresponding to the Hamiltonian being in a Gevrey class as a function of time). Such a smooth parametrization which removes discontinuities in the derivatives at the beginning and end of every chunk can be achieved by
\begin{align}
    \lambda(s)=\lambda_0+(\lambda_f-\lambda_0)\sin^2\left(\frac{\pi}{2}\sin^2\left(\frac{\pi s}{2}\right)\right),
\end{align}
where $\lambda_0=s_0$ and $\lambda_f=s_f$ are initial and final time values, respectively of a chunk. Our numerical analyses showed, however, that in the regimes that we probe, this smooth reparametrization gives inferior results than a simple linear ramp. This can be understood by reminding ourselves that a smooth reparametrization $\lambda(s)$ lead to a faster traverse of an avoided level crossing in the middle of the adiabatic path. For a fair comparison of the VQAA, we therefore benchmarked against a linear adiabatic schedule. The main benchmark for $N=53$ qubits with smooth reparametrizations is shown for reference in Fig.~\ref{fig:BBBPT_smooth}.
 \\
In order to apply a time evolution unitary operator $U(\delta, s)$, we make use of the second-order Suzuki-Trotter approximation~\cite{hatano2005finding}
\begin{align}
     U(\delta, s) = e^{-i\delta H(s)} \approx \left(e^{-i\delta H(s)_\text{even}/(2K)}e^{-i\delta H(s)_\text{odd}}e^{-i\delta H(s)_\text{even}/(2K)}\right)^K,
\end{align}
with $H(s)=(1-\lambda(s))H_0+ \lambda(s)H_T$. Here, $K=2$ has proven to be sufficient without any practical decrease in accuracy~\cite{banuls2006simulation}.\\
In our implementation of time evolution, we truncate singular values after every application of a two-site gate and renormalize the singular values in order to keep the MPS normalized. We monitor the loss in accuracy by keeping track of the norm of the MPS throughout the real time evolution. In general, we aim for a norm error of not larger than 1\%. Exact ground states and excited states for benchmarking purposes are obtained using density-matrix renormalization group (DMRG) methods~~\cite{schollwock2011density, banuls2013mass}. The code for the classical simulation was written in Python, building upon the blueprint~\cite{tenpy} as a basic building block.
\subsection{Theoretical aspects of an adiabatic evolution}~\label{sec:adiabatic}
Here we outline the main concepts of a theoretical treatment of adiabaticity. When considering transitions from the ground state to the excited states, the spectral gap between the ground state and the first excited state is of the greatest importance. Notably, the accumulation of relative phases plays a crucial role as well. We derive the differential equations governing the time evolution of a quantum state. Here, we follow the calculations given in~\cite{chruscinski2012geometric}, yet it is instructive to use parametrized time $s=t/T$ instead of real time $t$.\\
Then, the time-dependent and time-independent Sch\"odinger equations with $\hbar=1$ are given by
\begin{align}
    i \frac{\partial}{\partial s} \ket{\psi(s)} = TH(s)\ket{\psi(s)} \label{eqn:tdSchroedinger},\\
    H(s) \ket{k(s)} = E_k(s) \ket{k(s)} \label{eqn:tiSchroedinger},
\end{align}
respectively. The $\ket{k(s)}$, $k \in [0, n]$, are the eigenvectors of $H(s)$ and the eigenenergies $E_n(s)$ are for simplicity of exposition assumed to be non-degenerate and distinct. $E_0(s)$ is the ground state energy at a given $s$. \\
We use the representation for a quantum state
\begin{align}
    \ket{\psi(s)} = \sum_k c_k(s) e^{-i T \int_0^s E_k(\tau)d\tau} \ket{k(s)} = \sum_k c_k(s) e^{-i T \phi_k} \ket{k(s)},
\end{align}
writing  $\phi_k = \int_0^s E_k(\tau)d\tau$, and insert it into the time-dependent Schr\"odinger equation (Eqn.~\ref{eqn:tdSchroedinger})
\begin{align}
    \sum_k \left[ \dot c_k(s) e^{-i T \phi_k} \ket{k(s)} - i T E_k(s) c_k(s) e^{-i T \phi_k} \ket{k(s)} + c_k(s) e^{-i T \phi_k} |\dot k(s)\rangle  \right]\\  =
    -iT H(s) \sum_k c_k(s) e^{-i T \phi_k} \ket{k(s)}, \label{eqn:long}
\end{align}
where the dot denotes the derivative with respect to parametrized time:
\begin{align}
    \dot c(s)=\frac{\partial}{\partial s}c(s).
\end{align}
The second term on the left hand side and the right hand side of (Eqn.~\ref{eqn:long}) vanish as  $\ket{\psi(s)}$ satisfies the time-independent Schr\"odinger equation (Eqn.~\ref{eqn:tiSchroedinger}), which gives us
\begin{align}
    \sum_k \left[ \dot c_k(s) e^{-i T \phi_k} \ket{k(s)} + c_k(s) e^{-i T \phi_k} |\dot k(s)\rangle  \right] = 0. \label{eqn:beforeM}
\end{align}
Multiplying (Eqn.~\ref{eqn:beforeM}) by the state $\bra{m(s)}$ from the left yields
\begin{align}
     \sum_k \left[ \dot c_k(s) e^{-i T \phi_k} \underbrace{\braket{m(s)|k(s)}}_{\delta_{mk}} + c_k(s) e^{-i T \phi_k} \braket{m(s)|\dot k(s)}  \right] &= 0
\end{align}
and we obtain
\begin{align}
     \dot c_m(s) e^{-i T \phi_m} = - \sum_k c_k(s) e^{-i T \phi_k} \braket{m(s)|\dot k(s)}.
\end{align}
Then, the system of differential equations for the coefficients to describe the quantum state is given by
\begin{align}
    \dot c_m(s) = - \sum_k c_k(s) \exp\left(-i T \int_0^s \left(E_k(\tau) - E_m(\tau)\right)d\tau\right) \braket{m(s)|\dot k(s)} \quad \forall m\in [0,n].
    \label{eqn:beforeadiabapprox}
\end{align}
The $i\braket{m(s)|\dot k(s)}$ are also referred to as the Berry connections~\cite{berry1984quantal}.\\
From the time-independent Schr\"odinger equation (cf.~Eqn.~\ref{eqn:tiSchroedinger}), by taking the time-derivative, we obtain
\begin{align}
    \dot H(s) \ket{k(s)} + H(s) \ket{\dot k(s)} = \dot E_k(s) \ket{k(s)} + E_k(s) \ket{\dot k(s)}
\end{align}
and by multiplying with $\bra{m(s)}$ from the left, we get
\begin{align}
    \braket{m(s)|\dot k(s)} = \frac{1}{E_k(s) - E_m(s)}\braket{m(s)|\dot H(s)|k(s)}.
\end{align}
Under the adiabatic approximation we consider the evolution generated by $H(s)$ adiabatic if the changes of $H(s)$ are very slow compared to the time scale of the system, where the time scale $\Delta T_{km}$ is defined as the characteristic time of transition between the $k$th and the $m$th eigenstate:
\begin{align}
    |\braket{m(s)|\dot H(s)|k(s)}| \ll \frac{E_k(s)-E_m(s)}{\Delta T_{km}}.
\end{align}
In the adiabatic limit of infinitely slow change in $H(s)$, we have 
\begin{align}
    \lim_{\Delta T_{km} \rightarrow \infty} \braket{m(s)|\dot k(s)} = 0, \quad m \neq k.
\end{align}
Then, in the adiabatic limit, (Eqn.~\ref{eqn:beforeadiabapprox}) reads
\begin{align}
        \dot c_m(s) = - c_m(s) \braket{m(s)|\dot m(s)}
\end{align}
with $\dot c_m(0)=\delta_{nm}$, i.e. at the beginning of the evolution the whole population is in the $n$th eigenstate and will strictly remain in this eigenstate subspace if the evolution is perfectly adiabatic. For $n=0$ we have the special case of the QAA: if a (perfect) adiabatic evolution starts in the ground state, the initial state will remain in the instantaneous ground state subspace until the target Hamiltonian is reached at $s=1$.
\section{Adiabatic spectroscopy} \label{ssec:sub_adspec}
For better clarity, we outline the simple algorithmic procedure of adiabatic spectroscopy as described in the main text.\\
\begin{algorithm}[H]
    \SetAlgoLined
    Set resolution $r$ and target overlap $O_T$\\
    Obtain set of data points $\{s_i\}$, $i\in \{1,\dots,r\}$ \\
    \For{$i=1$ \KwTo $r$}{
        Probe different $\widetilde T_j$ so that (linear) adiabatic sweep from $s=0$ to $s=s_i$ gives
        instantaneous ground state overlap of $O_T$ at $s_i$ (using search algorithm of choice).\\
        Set $T_i$ with the $\widetilde T_j$ which succeeds best in producing that overlap.
    }
    \uIf{
    Large positive gradient around value $s^*$ }
    {Avoided level crossing detected at $s^*$}
    \caption{Adiabatic spectroscopy}
\end{algorithm}
\subsection{Variational algorithms and parametrized circuits} \label{ssec:vqa}
The concept of a parametrized circuit is to write a trial wave function which approximates the desired quantum state. This trial state is prepared as a product of many unitary operators, of which each depends on a classical variational parameter acting upon an initial trivial state. The operator preparing this state can be written as
\begin{align}
    U(\bm\varphi) = U_L(\bm\varphi_L)\cdots U_2(\bm\varphi_2) U_1(\bm\varphi_1)
\end{align}
where the $\{ \bm \varphi_i\}$, $\forall i\in\{1,\dots, L\}$, are vectors of continuous parameters.
\begin{figure}[htbp]
    \centering
    \begin{tikzpicture}
        \node[scale=1] {
            \begin{quantikz}
                \lstick{$\ket{\psi_\text{init}}$} & \gate{U_1(\bm\varphi_1)}\qwbundle[alternate]{} & \gate{U_{2}(\bm\varphi_{2})} \qwbundle[alternate]{} & \ \ldots \  \qwbundle[alternate]{} & \gate{U_L(\bm\varphi_L)} \qwbundle[alternate]{} & \qwbundle[alternate]{} & \meter{}\qwbundle[alternate]{}
            \end{quantikz}
        };
    \end{tikzpicture}
    \caption{Quantum circuit for a parametrized circuit. Multiple parametrized unitaries act upon a trivial initial state. Our diagrams were created using the Quantikz package~\cite{kay2018tutorial}.}
    \label{fig:paramcircuit}
\end{figure}
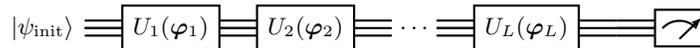
The performance depends strongly on the details of the target Hamiltonian and appropriate parameter initialization can be necessary, albeit is ultimately not sufficient due to the dependence of the algorithm performance on the representative power of the chosen ansatz $U(\bm\varphi)$.
\subsection{The quantum approximate optimization algorithm}
The quantum approximate optimization algorithm~(QAOA)~\cite{farhi2014quantum} is a highly popular example of a parametrized ansatz and much research has been done at investigating the properties of this algorithm~\cite{zhou2018quantum}. We outline QAOA here:\\
For an initial Hamiltonian $H_0=\sum^N_{j=1}\sigma_j^x$, the state $\ket{+}^{\otimes N}$ is usually chosen to be the initial state. However, the ground state of $H_0$ is $\ket{-}^{\otimes N}$ and it would correspond to the initial state of an adiabatic quantum algorithm. Starting with an initial spin state 
\begin{equation}
    \ket{\psi_\text{init}}=\ket{+,...,+}=\frac{1}{\sqrt{2^N}}(\ket{\downarrow}+\ket{\uparrow})^{\otimes N},
\end{equation}
we write down the QAOA ansatz
\begin{equation}
    \ket{\psi_p(\boldsymbol{\gamma},\boldsymbol{\beta})}=e^{-i\beta_p  H_0}e^{-i\gamma_p  H_T}...e^{-i\beta_1  H_0}e^{-i\gamma_1  H_T}\ket{\psi_\text{init}}
\end{equation}
where $\boldsymbol{\beta}=(\beta_1,...,\beta_p)$, $\boldsymbol{\gamma}=(\gamma_1,...,\gamma_p)$ with integer $p\geq1$. $ H_0$ and $ H_T$ are non-commuting Hamiltonians. $ H_0=\sum^N_{j=1}\sigma_j^x$ is also called the mixing Hamiltonian and $H_T$ is the problem Hamiltonian. In the usual case, $H_T$ is diagonal in the computational basis and encodes combinatorial optimization problems, e.g.~MAX-CUT, 3SAT, etc.\\
The energy cost function
\begin{equation}
    E_p(\boldsymbol{\gamma},\boldsymbol{\beta})=\braket{\psi_p(\boldsymbol{\gamma},\boldsymbol{\beta})| H_T|\psi_p(\boldsymbol{\gamma},\boldsymbol{\beta})}
\end{equation}
is determined with a quantum device by repeated measurements in the computational basis. Using a classical computer, the energy is then variationally minimized towards a local minimum $E_p(\boldsymbol{\gamma^*},\boldsymbol{\beta^*})$.

\section{Simulating the adiabatic evolution with matrix product states} \label{sec:MPS}
\subsection{Tensor network techniques}
In order to describe a general $N$-qubit quantum state with $d$ spin degrees of freedom, there are $d^N$ coefficients $c_{i_i, i_2, \ldots,i_N}$ in
\begin{align}
\ket{\psi} = \sum_{0\leq i_1, i_2, \ldots, i_N < d} c_{i_i, i_2, \ldots,i_N}\ket{i_1, i_2, \ldots, i_N}
\label{eqn:quantumstate}
\end{align}
that we need to account for. The $\ket{i_k}_{i=1}^{d}$ are forming an orthonormal basis of local Hilbert space (for $1\leq k<N$). However, and to our great benefit, ground states of gaped local one-dimensional Hamiltonians obey an area law regarding their entanglement entropy~\cite{hastings2007area}. In fact, area law states occupy an exponentially small corner of the many-body Hilbert space only, which makes a more economical description of the quantum state possible. If we consider the coefficients $c_{i_i, i_2, \ldots,i_N}$ being a very large tensor which can be decomposed into a chain of tensors of order 3, we obtain an alternative approach. We describe the quantum state by a matrix product state (MPS) ansatz for open boundary conditions by
\begin{align}
    \ket{\psi_\text{MPS}} = \sum_{0\leq i_1, i_2, \ldots, i_N < d} B_1^{i_i} B_2^{i_2} \ldots B_N^{i_N} \ket{i_1, i_2, \ldots, i_N}.
\end{align}
The complex matrices $B_k^{i_k}$ are $(D \times D)$-dimensional with the exception of the boundary tensors which are vectors. $D$~is the dimension of the virtual bonds in the MPS representation, limiting the amount of entanglement that can be represented by the MPS ansatz. Describing the state as a matrix product state only requires $ND^2d$ parameters which is clearly preferable if a moderate virtual bond dimension $D$ suffices. Therefore, MPS methods are an extremely successful ansatz to describe such quantum states obeying the area law such as aforementioned ground states~\cite{verstraete2006matrix}. In general, MPS are not well-suited to simulate the dynamics of a quantum system, due to the increase of entanglement in the system. However, in our algorithm which describes a (quasi-)adiabatic dynamic, the quantum states are very close to the ground state of the system.  This making MPS a perfect candidate for classically simulating quantum adiabatic algorithms.\\
For the simulation of time evolution we employ the time-evolving block decimation algorithm (TEBD)~\cite{vidal2004efficient} which is the computationally cheapest method of several time-evolution methods for tensor networks and it is also well-suited for Hamiltonians with local interactions. This being said, other powerful time-evolution methods have been developed in recent years, such as MPO~W$^{\text{I,II}}$, TDVP and the global Krylov method~\cite{paeckel2019time}. \\
An appealing property of the ZZXZ model is the suppressed the light cone spreading due to real time confinement~\cite{kormos2017real}. This leads to a very slow increase in entanglement when time evolving the system, making this model a good candidate for simulations with MPS. The maximum bond dimension of the MPS could be kept at modest values while maintaining very good accuracy in the MPS representation. 
\subsection{Simulating noisy quantum circuits} \label{App:NoisyCircuits}
\begin{figure}[htbp]
    \centering
    \begin{quantikz}
        \lstick[wires=4]{$\ket{\psi_{\text{init}}}^{\otimes N}$} & \gate{\mathcal{N}} & \gate[wires=4]{U_1} & \gate{\mathcal{N}} & \gate[wires=4]{U_2} & \gate{\mathcal{N}} & \qw \ \ldots \ \\
        & \gate{\mathcal{N}} & & \gate{\mathcal{N}} & & \gate{\mathcal{N}} & \qw \ \ldots \ \\
        & \gate{\mathcal{N}} & & \gate{\mathcal{N}} & & \gate{\mathcal{N}} & \qw \ \ldots \ \\
        & \gate{\mathcal{N}} & & \gate{\mathcal{N}} & & \gate{\mathcal{N}} & \qw \ \ldots \ 
    \end{quantikz}
    \caption{Quantum circuit including the noise tensors $\mathcal{N}$ which are all distinct from each other. Before and every $N$-site unitary, a noise event $\mathcal{N}$ is applied to the respective tensor $B_i$.}
    \label{fig:NoiseTensor}
\end{figure}
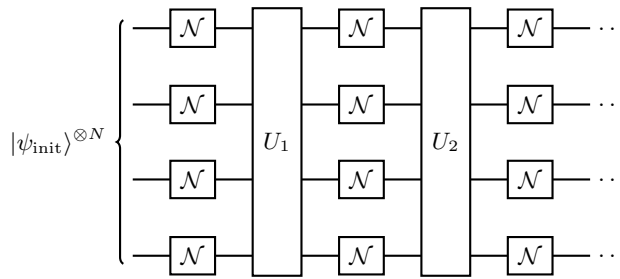

In order to simulate noisy quantum circuits, two main approaches are feasible. A noisy quantum channel can be represented by working with density matrices instead of quantum states. This is because an MPS cannot represent a mixed state. By considering Matrix Product Density Operators instead of MPS, such noisy quantum channels can be efficiently dealt with. We have chosen the second approach where noisy circuits are simulated through statistically averaging over an ensemble of MPS
\begin{align}
    \rho \rightarrow \rho_\text{noisy} = \frac{1}{n} \sum^n_{k=1} \underbrace{\mathcal{N}_k \ket{\psi}}_{|\widetilde{\psi}_k\rangle} \bra{\psi} \mathcal{N}_k^\dag
\end{align}
with the noise tensors $\mathcal{N}$ being unitary matrices. Each of the MPS $|\widetilde{\psi}_k\rangle$ will be different due to the probabilistic nature of noise. In the limit of a large MPS sample size $n$, both approaches will yield the same results.\\
A discrete noise model is considered, where $p$ is the noise strength and probability that a noise event occurs. Such a noise event can be either a bit flip ($\sigma_x$), a phase flip ($\sigma_z$) or a combination of both. The noise channel for discrete noise is
\begin{align}
    \rho \rightarrow \mathcal{N}\rho\mathcal{N}^\dag = (1-p)\rho + \frac{p}{3}(\sigma_x\rho\sigma_x + \sigma_y\rho\sigma_y + \sigma_z\rho\sigma_z).
\end{align}
The distinct noise tensors $\{\mathcal{N}_i\}$ are contracted with respective tensors $\{B_i\}$ for $i\in\{1,\dots,N\}$ before the first and after every $N$-site unitary (Fig.~\ref{fig:NoiseTensor}).
As an example, in the discrete noise model with noise strength of $p=10^{-4}$, $N=100$ qubits, a time evolution for total time $T=5$ and $16$ discrete steps per time unit, we expect on average $10^{-4}\cdot 100 \cdot (1+5) \cdot 16 = 0.96 \approx 1$ noise event in the entire quantum circuit.
\section{Gate count}
We include a short discussion of the one- and two-qubit gates required to implement the algorithms described here. Given the different experimental platforms, we focus on three exemplary use-cases which shall stand representative of others. One straightforward way to implement a variational quantum adiabatic routine is by making use of the ancilla-free method on an \emph{analogue quantum simulator}. As quantum simulators do not require any Trotter overhead for the time evolution, the cost of the optimization procedure is directly given by the overhead in evolution time. Our numerical estimations in (Fig.~\ref{fig:measment1}) show that around 20-30 ground state fidelity estimations can suffice for a significant improvement of the performance of the adiabatic algorithm. In the  case with the highest cost, the ancilla-free method is used only at the end of the adiabatic evolution. The backward-time evolution leads to an additional factor of 2 in the cost of this method.\\
There are different approaches to implementing the time evolution in the VQAA on a \emph{gate-model} architecture. One straightforward estimate of the cost shall be given here by counting the gates required to implement the trotterized evolution as in the TEBD for the numerical simulation for a large one-dimensional quantum system. The time evolution on the qubit chain with $N=53$ sites is decomposed into two-qubit unitaries. We chose $M=16$, $K=2$ and the Suzuki-Trotter order to be 2 (cf.~Suppl.) and every unitary can be decomposed into 3 CNOTs and 7 arbitrary single-qubit gates~\cite{shende2004minimal}. Thus, per unit of time, around 120 CNOTs/$N$ and 270 single-qubit gates / $N$ are required.\\
Next, we would like to count the gates required for the single-ancilla method and focus on the number of CNOTs. For the ancilla method, every two-qubit unitary from our consideration above is now controlled by the auxiliary qubit. Hence, we require $3\cdot6=18$ CNOTs for the three Toffoli gates~\cite{barenco1995elementary} and at most $7\cdot3=21$ CNOTs to control the single qubit rotations, resulting in around 1530 CNOTs/$N$ per unit of time as an upper bound.\\
As the gate count for the ancilla method depends crucially on the value of $\tau$, we make a remark on how the ancilla method would be used in practice. As explained in (App.~\ref{App:tau}), the time $\tau$ for the controlled time evolution depends on the spectral gap of the respective Hamiltonian $H(s)$. While one might be interested in the instantaneous ground state overlap along the adiabatic path, the ground state overlap at $s=1$ is of most interest. Note that while the spectral gap $\Delta$ along the path might be very small for hard problem instances, this is not necessarily the case at the end of the path. The ZZXZ model is a good example (Fig.~\ref{fig:DMRG}) where $\Delta$ is of the order of one at $s=1$. \\
A chain of trapped ions are an experimental system corresponding to our numerical analyses where no SWAP gates are required to realize the circuit. \emph{Superconducting}-qubit devices, however, usually do not have an all-to-all connectivity and rather feature a two-dimensional architecture. We provide a worst-case-connectivity estimate for the number of SWAPs by doing the count for a chain of $N$ qubits where only neighbouring qubits are connected. The ancilla qubit can be moved through the chain next to the two-site unitary to be applied with $N-2$ SWAPs per brick-wall layer. Also, three extra SWAPs are necessary per three-qubit controlled unitary so that the single qubit rotation to be controlled is next to the ancilla qubit. All-in-all, the overhead due to SWAP gates is at 580 additional CNOTs/$N$ for swapping operations on this chain topology per unit of time.\\
We stress again that these estimates are concerned with counting the CNOTs that would have been required to implement the gates used in the numerical simulation on a quantum device. The actual gate count will be significantly lower if the circuit is optimized with regard to the experimental realization instead.
\section{Bayesian inference in hypothesis testing} \label{sec:hypo}
\begin{figure}[htbp]
    \centering
    \subfloat[
        Beta distribution for $a=10$ and $b=2$ as a prior Bayesian inference hypothesis testing. The left and right $\alpha$-errors are indicated for $H_0=0.9$ and $\epsilon=0.05$. When more samples are obtained, the posterior likelihood distribution is updated.
        ]{%
        \centering
        \includegraphics[width=.45\columnwidth]{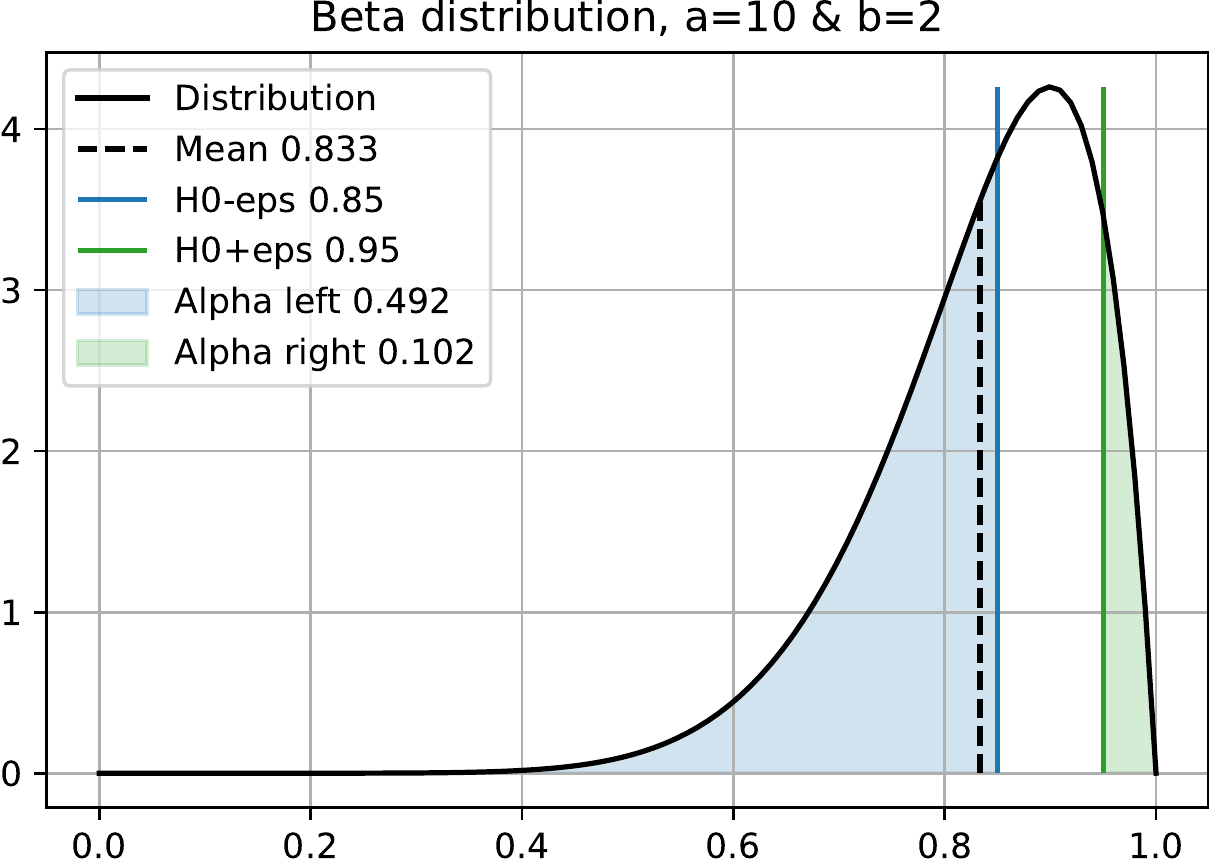}
    }\hspace{4em}
    \subfloat[
        Development of the $\alpha$-errors for different total sample number $N$ and actual probability of a 1-measurement of $p=0.99$ much larger than $H_0=0.9$. The left $\alpha$-error decreases exponentially fast and drops below the stopping criterion of 0.05 already for under 25 samples. For the left and right $\alpha$-error, the shaded areas indicate the region of uncertainty of one standard deviation.
        ]{%
        \centering
        \includegraphics[width=.45\columnwidth]{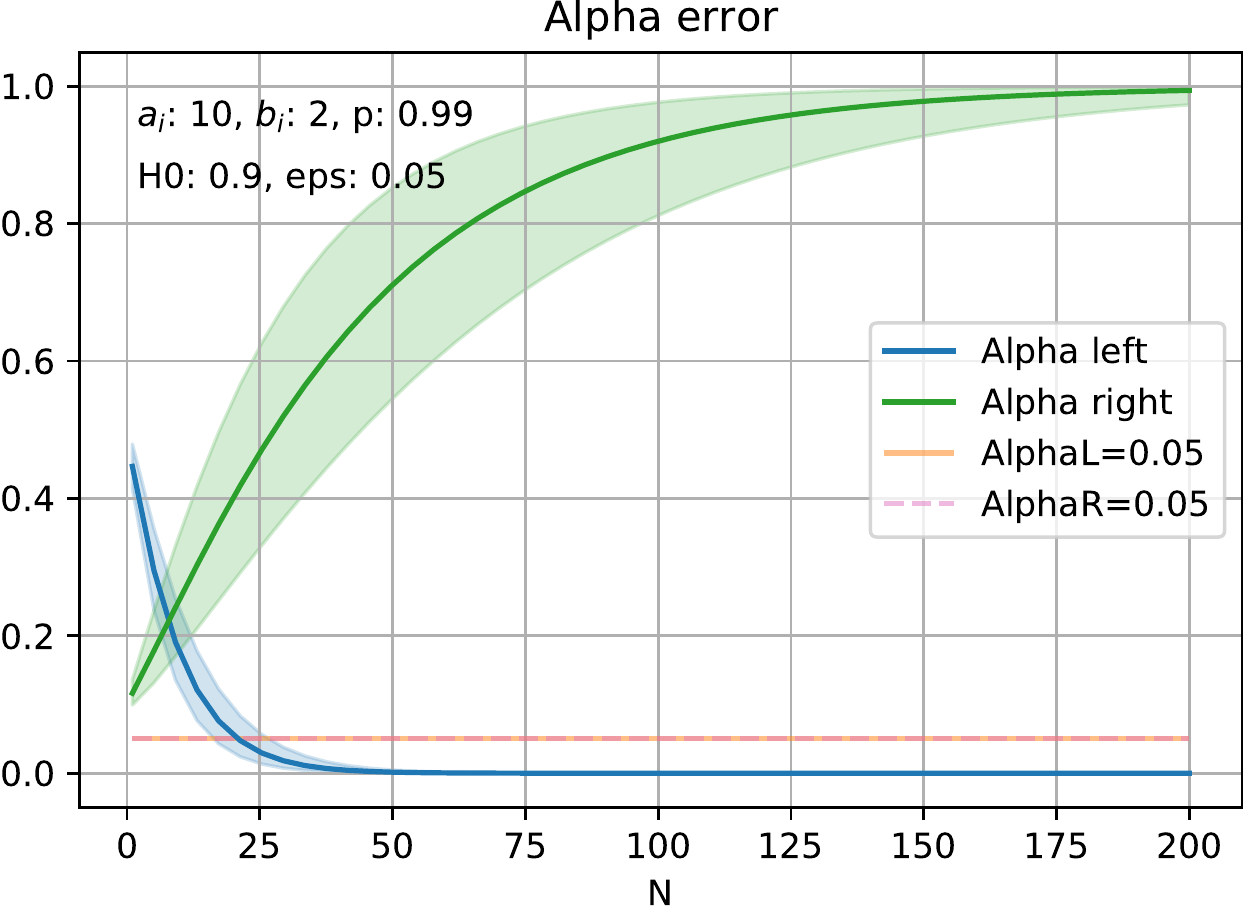}
    }
    
    \subfloat[
        The total number of samples needed in the worst case scenario of $H_0=p=0.9$. The parameter $\epsilon$ determines the size of the interval (2$\epsilon$) where no decision can be made. For very small epsilon, the number of samples necessary diverges.
        ]{%
        \centering
        \includegraphics[width=.45\columnwidth]{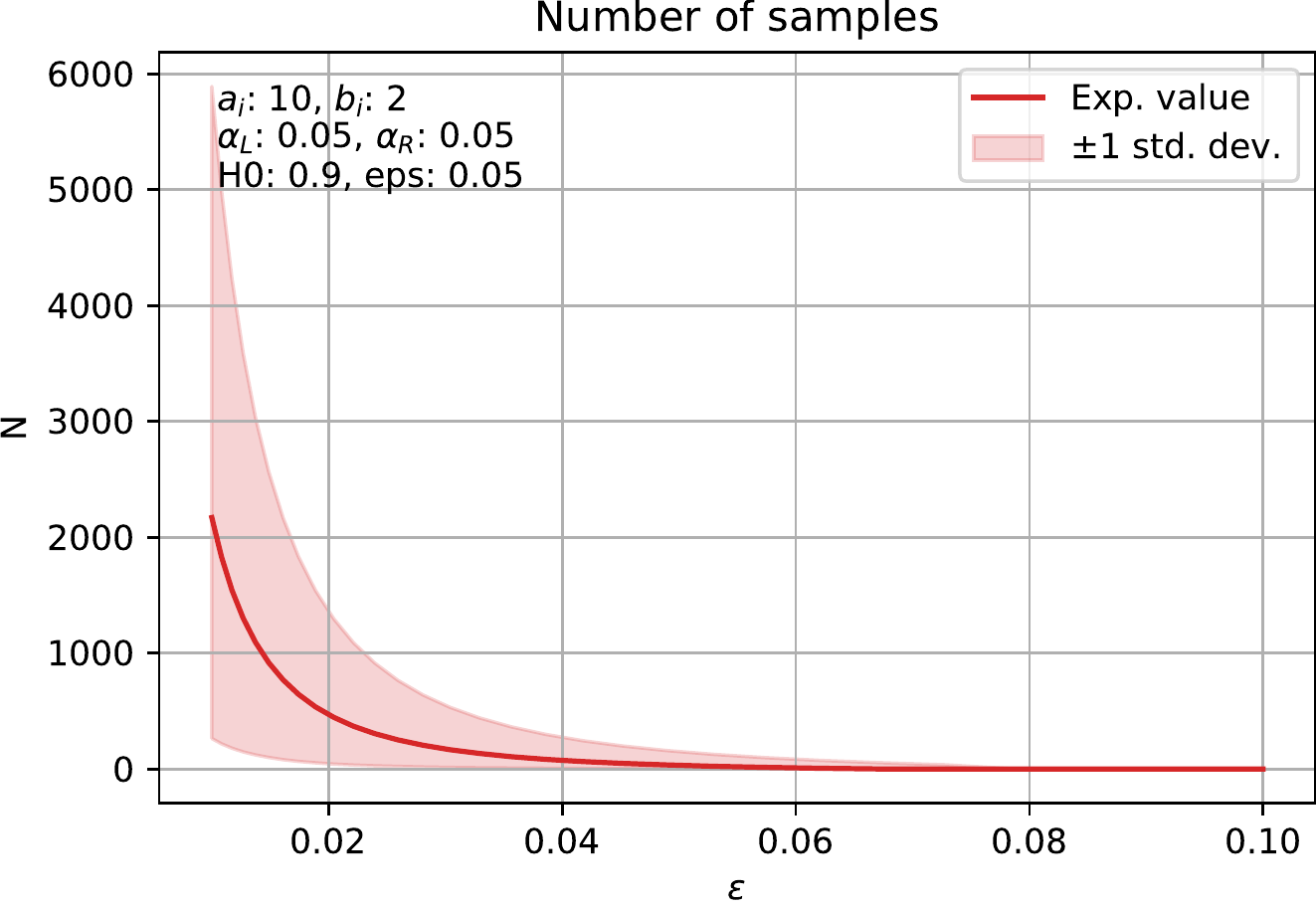}
    }\hspace{4em}
    \subfloat[
        Given a value for $H_0=0.9$, the expected number of necessary samples is shown depended on the actual probability $p$ to measure a 1.
        ]{%
        \centering
        \includegraphics[width=.45\columnwidth]{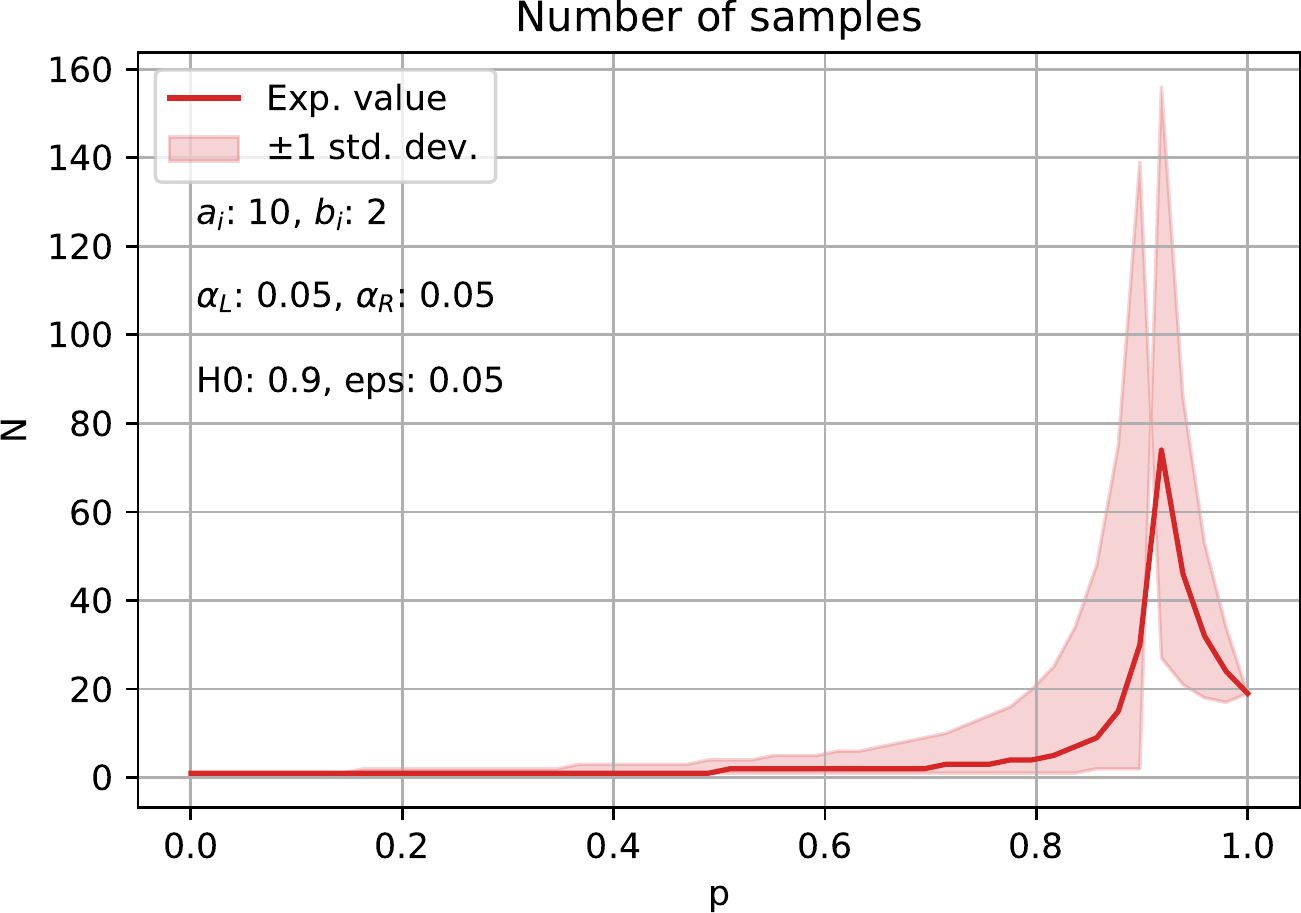}
    }
  \caption{Example for Bayesian hypothesis testing.}
  \label{fig:hypo}
\end{figure}

We give a brief summary on how to use Bayesian inference in order to do highly efficient hypothesis testing using our entangled-ancillas protocol (Sec.~\ref{sec:entangled-ancilla}) to decide whether the ground state overlap is larger than a given threshold value. In an experimental setup, we would like to only obtain as few samples as needed. Therefore, after each measurement, we would like make use of a stopping criterion to decide whether we need to continue sampling or are able to terminate. Unfortunately, stopping rules are quite cumbersome to deal with in a frequentist approach. And fortunately, Bayesian inference turns out to be a most convenient tool.\\
The measurement outcomes of a quantum expectation value such as the ground state overlap with our protocol are assumed to behave as an independent and identically distributed (i.i.d.)~random Bernoulli variable. The total measurement data is $X = \{x_1, ..., x_N\}$ with a Beta prior $\theta = \text{Beta}(a_i, b_i)$. The prior prevents overfitting for few data points. The Beta distribution is normalized and given as 
\begin{align}
    \text{Beta}(\mu | a, b) = \frac{\Gamma(a+b)}{\Gamma(a) \Gamma(b)} \mu^{a-1}(1-\mu)^{b-1}
\end{align}
with 
\begin{align}
    \Gamma(x) = \int_{0}^{\infty} {\mu}^{x-1} e^{-\mu} \, d\mu.
\end{align}
The mean and the variance of the Beta distribution are
\begin{align}
    \mathbb{E}[X] =& \frac{a}{a+b}, \\
    \text{var}[X] =& \frac{ab}{(a+b)^2(a+b+1)}.
\end{align}
The Beta distribution is the conjugated prior for the Bernoulli distribution, implying that the posterior after one measurement is again a Beta distribution with new parameters \cite{gundersen_2020}:
\begin{align}
    p(\theta|X) \propto& \prod_{n=1}^N p(x_n|\theta)p(\theta) \\
    =& \prod_{n=1}^N \theta^{x_n}(1-\theta)^{1-x_n}\frac{1}{\text{Beta(a,b)}}\theta^{a-1}(1-\theta)^{b-1} \\
    \propto& \; \theta^{\sum_{i=1}^N x_i + a - 1}(1-\theta)^{N-\sum_{i=1}^N x_i + b - 1}.
\end{align}
When sampling in the experiment, our posterior becomes the prior for the next measurement until our stopping criterion is fulfilled. We merely need to keep track of the 0's and 1's or spin-ups and spin-downs that are measured.
\begin{align}
    p(\theta | X) =& \;\text{Beta}(a_N, b_N), \\
    a_N =& \sum_{i=1}^N x_i + a,\\
    b_N =& \;N - \sum_{i=1}^N x_i + b.
\end{align}
If we are expecting mostly zeros when close to the ground state, we might initialize our prior with $a=10$ and $b=2$. If the algorithm asks for a ground state overlap greater than $H_0$ (Zero hypothesis: $|\braket{\psi | \psi_\text{evo}}| > H_0)$, samples are gathered until a stopping criterion is met and $H_0$ is either accepted or rejected. A decision upon the hypothesis is hard when $H_0$ is very close to the mean as many samples will be necessary to distinguish them. Setting up an interval $[H_0-\epsilon, H_0+\epsilon]$ where a decision cannot be made overcomes this problem. Then, $H_0$ can be updated to a different value and the sampling is restarted. The following outcomes of hypothesis testing are possible:\\
\begin{algorithm}[H]
    \SetAlgoLined
    Set $\alpha$-threshold \\
    \For{$i$ \KwTo Maximum number of samples}{
        Take sample number $i$\\
        Update left and right $\alpha$-error \\
        \uIf{Left $\alpha$-error $ < \alpha$-threshold}{Accept hypothesis}
        \uElseIf{Right $\alpha$-error $ < \alpha$-threshold}{Reject hypothesis}
    }
    \uIf{
    Hypothesis can be neither accepted nor rejected}
    {Update $\alpha$-threshold and repeat}
    \caption{Decision algorithm for hypothesis testing}
\end{algorithm}
\noindent Identifying a 0-measurement (perfect ground state overlap) with increasing $a$ by 1 and a 1-measurement with increasing $b$ by 1, the $a_N$ and $b_N$ contain the sampling data after $N$ samples. When close to a ground state, we expect mainly $a$'s to be counted and a distribution strongly biased towards $p=1$. A suitable stopping criterion might be an alpha error probability of under 5\%. A sample case is shown in (Fig.~\ref{fig:hypo}).

\section{Carbon footprint}
\begin{table}[h]
    \centering
    \begin{tabular}[b]{l r}
        \hline
        \textbf{Numerical simulations} & \\
        \hline
        Estimated project-related cluster kernel hours & 255,000\\
        Est.~thermal design power per kernel [W]& 6.25\\
        Total energy consumption of simulations [kWh] & 1,60\\
        Est.~specific emissions in Germany [kgCO$_2$eq/kWh]& 0.5\\
        Total CO$_2$eq emissions for numerical simulations [kg] & 800\\
        \hline
        \textbf{Transport} & \\
        \hline
        No substantial project-related CO$_2$eq emissions for transport. & \\
        \hline
        \textbf{Total} & \\
        \hline
        CO$_2$eq emissions [kg] & 800\\
        \emph{Emissions have been offset with \href{https://www.atmosfair.de}{atmosfair.de}.} & \\
        \hline
        \hline
    \end{tabular}
    \caption{Table of project-related CO$_2$eq emissions. While some figures can only be estimated approximately, this information is intended to increase transparency about the climate impact of this project. Disclosing emissions of scientific research can help to raise awareness for the climate crisis. See also  \href{https://scientific-conduct.github.io}{scientific-conduct.github.io}.}
\end{table}
\end{document}